\documentclass[pra,aps,amsmath,amssymb,superscriptaddress,longbibliography]{revtex4-1}

\usepackage[sc,osf]{mathpazo}
\usepackage{physics}
\usepackage{outlines}

\usepackage{tikz}
\usetikzlibrary{fit}
\usepackage{nicefrac}

\usepackage{slashbox}
\usepackage{graphicx}
\usepackage{tikz-network}
\usetikzlibrary{patterns,decorations.pathreplacing}

\newcommand{\blu}{\color{blue}}
\newcommand{\red}{\color{red}}

\newcommand{\bla}{\color{black}}

\makeatletter
\renewcommand*\env@matrix[1][*\c@MaxMatrixCols c]{%
  \hskip -\arraycolsep
  \let\@ifnextchar\new@ifnextchar
  \array{#1}}
\makeatother

\newtheorem{definition}{Definition}
\newtheorem{thm}{Theorem}
\newtheorem{prop}[thm]{Proposition}

\usepackage{hyperref}
\usepackage{bm}
\usepackage{amsmath}
\usepackage{amssymb}
\usepackage{graphicx}
\usepackage{tabularx}
\usepackage{tabu}
\usepackage{longtable}
\usepackage{multirow}
\usepackage{color} 
\newcommand{\btp}{\begin{tikzpicture}}
\newcommand{\etp}{\end{tikzpicture}}

\definecolor{color1}{RGB}{230,230,250}
\definecolor{color2}{RGB}{255,255,224}
\definecolor{green2}{RGB}{0,100,0}

\newcommand{\newc}{\newcommand}
\newc{\beq}{\begin{equation}}
\newc{\eeq}{\end{equation}}
\newc{\kt}{\rangle}
\newc{\br}{\langle}
\newc{\beqa}{\begin{eqnarray}}
\newc{\eeqa}{\end{eqnarray}}
\newc{\longra}{\longrightarrow}
\newc{\Ob}{{\mathcal O}}
\newc{\uta}{U^{T_{1}}}
\newc{\utad}{U^{T_{1}\dagger}}
\newc{\utb}{U^{T_{2}}}
\newc{\utbd}{U^{T_{2}\dagger}}
\newc{\utap}{U'^{T_{1}}}
\newc{\utapd}{U'^{T_{1}\dagger}}
\newc{\utbp}{U'^{T_{2}}}
\newc{\utbpd}{U'^{T_{2}\dagger}}

\begin{document}

\title{From dual-unitary  to quantum Bernoulli circuits: Role of the entangling power in constructing a quantum ergodic hierarchy }
\author{S. Aravinda}
\email{aravinda@physics.iitm.ac.in}
%\affiliation{Department of Physics, Indian Institute of Technology
%Madras, Chennai, India~600036}
\author{Suhail Ahmad Rather}
\email{suhailmushtaq@physics.iitm.ac.in}
%\affiliation{Department of Physics, Indian Institute of Technology
%Madras, Chennai, India~600036}
\author{Arul
Lakshminarayan}
\email{arul@physics.iitm.ac.in} 
\affiliation{Department of Physics, Indian Institute of Technology
Madras, Chennai, India~600036}

\begin{abstract}
Deterministic classical dynamical systems have an ergodic hierarchy, from ergodic through mixing, to Bernoulli systems that are ``as random as a coin-toss". Dual-unitary  circuits have been recently introduced as solvable models of many-body nonintegrable quantum chaotic systems  having a hierarchy of ergodic properties. We extend this to include the apex of a putative quantum ergodic hierarchy which 
is Bernoulli, in the sense that correlations of single and two-particle observables vanish at space-time separated points. We derive a condition based on the entangling power $e_p(U)$ of the basic two-particle unitary building block, $U$, of the circuit, that guarantees mixing, and when maximized, corresponds to Bernoulli circuits. Additionally we show, both analytically and numerically, how local-averaging over random realizations of the single-particle unitaries, $u_i$ and $v_i$ such that the building block is $U^\prime = (u_1 \otimes u_2 ) U (v_1 \otimes v_2 )$ leads 
to an identification of the average mixing rate as being determined predominantly by the entangling power $e_p(U)$. Finally we provide several, both analytical and numerical, ways to construct dual-unitary operators  
covering the entire possible range of entangling power. We construct a coupled quantum cat map which is dual-unitary  for all local dimensions and a 2-unitary or perfect tensor for odd local dimensions, and can be used to build Bernoulli circuits. 
\end{abstract}

\maketitle
\tableofcontents

\section{Introduction}

The complexity of interacting many-body systems, classical or quantum, matches their importance and prevalence that one cannot easily overstate. 
This is hardly surprising, given that even low-dimensional systems can display chaos that originates from the lack of sufficient constants of motion or nonintegrability. However, low-dimensional chaos is reasonably well-understood in terms of phase-space structures or spectral properties. Seemingly abstract toy models have played a crucial role in this process. The iconic one-dimensional logistic map has a long history with attributions to John von Neumann and Ulam from 1940 through early computer studies in the 1950's \cite{ulam1947combination}. It came to a head with the more modern and definitive studies of May \cite{may1976simple}, Li, Yorke \cite{LiYorke}, Feigenbaum \cite{feigenbaum1978quantitative} and others, culminating in the discovery of universality in one-dimensional maps. 

Closer to Hamiltonian systems, two-dimensional area preserving maps were studied as early as 1937 when Hopf introduced \cite{hopf}  what is now called the bakers map as an example of a mixing system \cite{Ott_2002}, and we find the so-called Arnold cat map in a book by Arnold and Avez, published in 1968 \cite{arnold1968ergodic}. The quantum avatars of the cat and bakers maps were introduced respectively by Berry and Hannay in 1980 \cite{hannay1980quantization}, and by Balazs and Voros in 1987 \cite{balazs1987quantized,balazs1989quantized}. Such quantum maps \cite{berry1979quantum} did provide vast simplifications over say a hydrogen atom in crossed electric and magnetic fields \cite{LesHouches91}. They were useful in understanding semiclassical methods and identifying classical structures in quantum states \cite{Saraceno1990,keating1991cat,Almeida1991,ConnorSteveHeller,Nonnen_cat,qmaps_book}. However, they were arguably not nearly as successful in being simple enough to penetrate the veils
of quantum chaos, something which their classical counterparts did with ease.

Random unitary circuit models involving random two-body local interactions have been extensively studied in the recent past \cite{Nahum2018, nahum2017quantum,khemani2018operator,von2018operator,chan2018solution}
as minimal models of many-body quantum nonintegrable systems. In such systems, the growth of bipartite enatnglement is found to be linear using a ``minimal spacetime membrane" approach and is validated in the limit of  large local Hilbert space dimension \cite{nahum2017quantum}. It is notable that quantum supremacy was purportedly demonstrated with a realization of such
circuits in 2-D with superconducting junctions \cite{QSupreme}. 
In addition, the introduction of a special class of unitary circuits of  solvable lattice models that are typically nonintegrable is
remarkable and hold potential as toy models of many-body quantum chaos \cite{Akila2016,Bertini2019,gutkin2020exact,BraunPRE2020}. The key aspect that make these models solvable is an underlying spacetime duality that is realized for certain interactions, hence
these are referred to as  dual-unitary circuits \cite{Bertini2019}.  Quantities such as the two-point correlation functions, operator entanglements, spectral form factors, have become largely analytically accessible in these models \cite{Bertini2018,Bertini2019,bertini2019entanglement,gutkin2020exact,
bertini2019operator,kos2020chaos,lerose2020influence,garratt2020many,flack2020statistics,bertini2019operatorII}. Using matrix product states as initial states, time evolution under dual-unitary  circuits has been analyzed, including the evaluation of block entanglement entropy \cite{piroli2020exact}. The OTOC are analytically tractable for dual-unitary  circuits \cite{claeys2020maximum} and for random dual-unitary circuits \cite{bertini2020scrambling}. The effect of perturbations that break the dual nature of the circuits is studied in \cite{kos2020correlations} where the structural stability of these models is motivated as being analogous to that of classically chaotic systems \cite{Ott_2002}. Computational power of the circuits built from one and two dimensional dual-unitary operators are also studied recently and shows the quantum computational supremacy of these circuits \cite{suzuki2021computational}.

\subsection{Correlations and the Ergodic hierarchy}
The dual-unitary  circuits are defined in discrete time and can be interpreted as Floquet systems with a time-periodic forcing.
In fact the dual kicked chain model is a many-body Floquet system and the circuit model may be thought of as being equivalent to it \cite{gutkin2020exact}. 
They have been shown to be capable of displaying, in the thermodynamic limit, a range of properties that could be non-ergodic, ergodic or mixing. 
The mixing rates of these circuits can be found from a simple contractive channel that is a completely positive map, such as
implemented in open quantum systems. The channel depends only on the two-body interaction and governs the spread of correlations 
across space-time \cite{Bertini2019}. These properties are inferred from the behavior of correlations such as $C_{AB}(t;x,y)=\br A^y(t) B^x\kt -\br A^y \kt \br  B^x\kt$. Here operators $A^y(0)$ and $B^x$ are localized  initially at  $y$ and $x$ sites respectively, and $A^y(t)$ is time-evolved under the many-body dynamics.

For comparison, we recall the ergodic hierarchy of classical dynamical systems, which may be stated, largely, in terms of correlations \cite{sep-ergodic-hierarchy, Ergodic_Hierarchy, Ott_2002}.
Let $f(x)$ and $g(x)$ be two functions on phase space $\Omega$, $x=(q,p)\in \Omega$ and $d\mu(x)$ be an invariant measure under the dynamics $x \mapsto x_t= \phi^t(x)$. For Hamiltonian systems the relevant measure is simply the Lebesgue measure or volume on phase space $dq dp$, in which case $d\mu(\phi^{-t} x)=d\mu(x)$ from Liouville's theorem. Consider the correlation
\beq
C(t;f,g)=\left \br f(\phi^t x) g(x)\right \kt -\left \br f(x) \right\kt \left \br g(x) \right\kt,
\eeq
where for any function $h(x)$, $\br h(x) \kt = \int_{\Omega} h(x) d\mu(x)$, is its phase space average. Denote the time-average of any function $h'(t)$ 
as $\overline{h'(t)}=\lim_{T \rightarrow \infty}\int_{0}^T h'(t) dt /T$, or a corresponding sum for discrete time systems.

The ergodic hierarchy is
\[
\begin{split}
\text{Ergodic}\;\supset \; &\text{Weak mixing} \;\supset \;  \text{Mixing}\;\supset\;\\ &\text{Kolmogorov}\;\supset\; \text{Bernoulli}
\end{split}, 
\]
where $ A \supset B \supset C$ indicates that the property $B$ implies $A$ but not $C$.
The properties may then be identified as follows: ergodic systems are those for which $\overline{C(t;f,g)}=0$, weakly mixing ones are those for which 
$\overline{|C(t;f,g)|}=0$, while mixing implies that the correlations vanish without time averaging: $\lim_{t \rightarrow \infty} C(t;f,g)=0$.

Kolmogorov (or K-) and Bernoulli (or B-) systems need a more elaborate setup. While K-systems can be identified from nonintuitive properties
of correlations, they are usually identified as systems with a positive metric (or Kolmogorov-Sinai) entropy, or simply from the 
principal Lyapunov exponent being positive. Chaotic systems are typically K, and imply exponential divergence of trajectories. At the apex of the hierarchy are Bernoulli systems for which there exists a partition of phase-space such that for characteristic functions $f$ and $g$ on these partitions and their images, $C(t;f,g)=0$ for $t\neq 0$, that is the correlations vanish instantly \cite{Ott_2002,Ergodic_Hierarchy}. These give rise to a symbolic dynamics based on these partitions with a full left-shift dynamics, and a one-to-one relationship between a stochastic Bernoulli process such as coin-tosses and deterministic itineraries of orbits.

Returning to dual-unitary  circuits in $1+1$ \cite{Bertini2019,Akila2016}, a crucial consequence of duality is that all correlations between initially  localized operators vanish inside the light cone $|y-x|<t$, and also enables the explicit calculation of correlation functions along the light cone $y-x =\pm t$ in terms of the bipartite dual-unitary  operators. It has been noted before that the correlations along the light cone can either not vanish with time (non-ergodic), or can average to $0$ (ergodic) or can tend to zero exponentially fast (mixing). The exponential fast mixing subsumes the case of K-systems. While this quantum classification has not singled out an equivalent of Bernoulli systems, we show that when the local systems are larger than $2$ dimensional, that is qutrits or above, it is possible for the dual-unitary  circuit to 
become Bernoulli in the sense that $C_{AB}(t;x,y)=C_0 \delta_{t,0}\delta_{x,y}$, it vanishes at {\it all} spacetime separated points. 

The ergodic nature of dual-unitary  circuits can depend on the extent or support of the local operators used and most of the work thus far has been relevant to single-particle operators. While in this work we also largely confine ourselves to such observables, in the Bernoulli case it is not hard to see that even extended observables with two-particle support, have vanishing correlations. Hence these ``Bernoulli circuits" represent extreme cases 
presumably at the apex of a putative quantum ergodicity hierarchy. We must also emphasize that these statements are strictly valid for all time only in the
thermodynamic limit. For finite systems, the statements are valid till half the particle number. A previous discussion of a quantum ergodic hierarchy also 
relies on correlation functions \cite{Quantum_Ergodic_Hierarchy}, however the examples provided are, to our understanding, not rigorous at these levels. 

\subsection{The role of the entangling power of the building blocks}

The building blocks of these circuits, the two-particle unitary operators $U$ determine to a large extent the nature
of the circuits themselves. The dual-unitary circuits are constructed from $U$ that maximize their operator entanglement \cite{Gopalakrishnan2019,SAA2020}. 
The Bernoulli circuits considered in this paper are special cases that are constructed from $U$ that are 2-unitary operators (as defined in \cite{Goyeneche2015}) that are distinguished by having maximal entangling power.
In fact the entangling power of $U$, denoted herein as $e_p(U)$, plays a central role in the mixing properties of the circuit and forms the principal
topic in this paper. While precise definitions are given in the next section, the entangling power
measures the average entanglement that is produced when $U$ acts on all product, unentangled, states \cite{Zanardi2000,Zanardi2001}. The {\sc swap} operator $S$ which acts by interchanging the subsystems:
\beq
\label{eq:swap}
\begin{split}
S\ket{\phi_1}\ket{\phi_2} &= \ket{\phi_2}\ket{\phi_1}, \quad \forall \ket{\phi_1},\ket{\phi_2}, \\
S(u_1\otimes u_2)S &= u_2\otimes u_1, \forall u_1,  u_2, 
\end{split}
\eeq
is an example of a dual-unitary  operator, however
its entangling power is zero as it clearly engenders no entanglement. It sits diametrically opposite to 2-unitary operators and 
consequently dual operators in the vicinity of the {\sc swap} have the smallest mixing properties. Previously, 2-unitary operators in 
circuits have been used as perfect-tensors in networks to demonstrate that the tripartite mutual information, a measure of scrambling, 
takes the largest negative value possible, already indicating the special status of these systems \cite{Hosur2016}.

The entangling power acts as a foil for an interaction strength among the particles. It has been shown to determine global operator entanglement
in bipartite time evolution \cite{Bhargavi2017,Bhargavi2019} as well to some extent in Floquet spin chains \cite{PalLak2018}.  
The role of the entangling power of the basic two-particle unitary $U$ in determining the ergodic nature of a dual-unitary  circuit built from it is not evident due to its local unitary invariance (LUI), as measures of entanglement are required to be. That is, if 
\beq
\label{eq:LU}
U^\prime = (u_1 \otimes u_2 ) U (v_1 \otimes v_2 ),
\eeq
where $u_i$ and $v_i$ are single-particle operators, $U$ and $U'$ are local unitary (LU) equivalent, and  $e_p(U)=e_p(U')$  \cite{Zanardi2000,Zanardi2001}. The dual-unitary  property is preserved under local unitaries, thus $U'$ is also dual-unitary. However, the dynamical properties of circuits constructed from $U$ and $U'$ are very different. For example, the two-point correlation functions and the mixing rates, depend on the local unitaries. 
The qualitative nature of the circuit changes with the use of local operators or ``fields". This is not surprising as for example the Heisenberg model on the ring is integrable, however adding a nonhomogeneous local magnetic field can drive it to an ergodic phase. Similarly, local unitaries can enhance entangling power to typical values if sandwiched between nonlocal operators \cite{Bhargavi2017,Bhargavi2019}.
 
Nevertheless we show in this paper that the entangling power plays a central role in two respects: (1)  
it is shown that there exists a threshold such that if $e_p(U)>e_p^*$ the dual-unitary circuit built from it is guaranteed to be  mixing, and (2)
the {\it local unitary averaged} mixing properties of dual circuits, when the single particle operators $u_i$ and $v_i$ are sampled uniformly, are to a very good approximation solely determined by $e_p(U)$. 
 The threshold $e_p^*=(q^2-2)/(q^2-1)$ is a simple function of the local, single-particle, Hilbert space dimension $q$.  For $e_p(U)\leq e^*_p$ the circuit
could be mixing, ergodic or non-ergodic. If $e_p(U)=0$ then the circuit cannot be mixing, and if and only if $e_p(U)=1$, the circuit is Bernoulli.
Thus $e_p^*$ signals a strong enough entangling power that no matter what the local fields are, the circuit is mixing. For qubits $q=2$, it turns
out that $e_p(U)$ cannot exceed $e_p^*=2/3$, hence Bernoulli circuits cannot be constructed from qubits.
However, for all $q>2$ the entangling power can reach $e_p(U)=1$. The only case where it was not known
if this was the case till very recently was $q=6$, but an explicit construction of 2-unitary operators in this dimension has settled
that question \cite{SRatherAME46}. This is summarized in the following rudimentary phase diagram. 
\[
\btp [scale=0.65]
\draw [ultra thick,violet] (0,0) -- (5,0);
%\draw [ultra thick] (5,0) -- (8,0);
\draw [thick, fill=blue] (0,0) circle (0.2cm);
%\draw [thick, fill=red] (8,0) circle (0.2cm);
\draw [thick, fill=violet] (5,0) circle (0.2cm);
%\draw [->,dashed] (2,-2) .. controls (2,-1) and (0,-1) .. (0,-0.6);
%\draw [->,dashed] (6,-2) .. controls (6,-1) and (8,-1) .. (8,-0.6);

\node at (-3,-.5){$q=2$};
%\node[text=red] at (8,-1) {(2-unitary)};
%\node[text=red] at (9,0.5){Bernoulli};
%\node[text=red] at (8,-0.5) {$\mathbf{1}$};
\node[text=blue] at (0,-1) {({\sc swap})};
\node[text=blue] at (-1,0.5){Non-mixing};
\node[text=blue] at (0,-0.5) {$\mathbf{0}$};
\node[text=violet] at (2.8,0.5) {$\supseteq$ Mixing}; 
%\node at (6.5,0.5) {Mixing};  
\node at (5,-0.6) {$\mathbf{e_p^*=\frac{2}{3}}$};
%\node at (4,-2) {Entangling power $e_p(U)$}; 
%\draw [->] (7.0,-2) -- (7.8,-2); 

\etp 
\] 
 
\[
\btp [scale=0.65]
\draw [ultra thick,violet] (0,0) -- (5,0);
\draw [ultra thick] (5,0) -- (8,0);
\draw [thick, fill=blue] (0,0) circle (0.2cm);
\draw [thick, fill=red] (8,0) circle (0.2cm);
\draw [thick, fill=violet] (5,0) circle (0.2cm);
%\draw [->,dashed] (2,-2) .. controls (2,-1) and (0,-1) .. (0,-0.6);
%\draw [->,dashed] (6,-2) .. controls (6,-1) and (8,-1) .. (8,-0.6);

\node at (-3,-.5){$q>2$};
\node[text=red] at (8,-1) {(2-unitary)};
\node[text=red] at (9,0.5){Bernoulli};
\node[text=red] at (8,-0.5) {$\mathbf{1}$};
\node[text=blue] at (0,-1) {({\sc swap})};
\node[text=blue] at (-1,0.5){Non-mixing};
\node[text=blue] at (0,-0.5) {$\mathbf{0}$};
\node[text=violet] at (2.8,0.5) {$\supseteq$ Mixing}; 
\node at (6.5,0.5) {Mixing};  
\node at (5,-0.6) {$\mathbf{e_p^*=\frac{q^2-2}{q^2-1}}$};
\node at (4,-2) {Entangling power $e_p(U)$}; 
\draw [->] (7.0,-2) -- (7.8,-2); 

\etp 
\]
The special value of the entangling power, $e_p^*$, appears as a natural border  for various families of dual-unitaries, including qubits. We show explicit
examples, including a quantum cat map, wherein $e_p(U)=e_p^*$ and the circuit is not mixing, therefore the 
bound for guaranteed mixing is tight.
 In fact $e_p^*$ is the final threshold of a series of entangling powers that are multiples of $1/(q^2-1)$ that when breached ensure a minimum number of modes remain mixing whatever the local unitaries maybe. If $e_p(U)=1$, the $U$ has maximum possible entangling power, and is 2-unitary by the definition in \cite{Goyeneche2015}, then so is $U'$ and the Bernoulli property of circuits is local unitary invariant.

\subsection{Various constructions of dual-unitary and 2-unitary operators}

Using an ensemble with the Fourier gate (a dual-unitary  operator) and different local unitaries, the dynamical properties of dual circuits were studied in \cite{gutkin2020local}. Similarly, \cite{claeys2020ergodic} constructed the dual-unitary  operators from diagonal unitaries and the {\sc swap} operator,
to exhibit mixing circuits. However, we show below that the entangling power of such dual-unitaries do not explore the whole possible range. 
In \cite{SAA2020}, we constructed a non-linear iterative algorithm to produce ensembles of unitaries which are arbitrarily close to dual-unitaries
and in some cases 2-unitaries and these provide examples with large entangling powers. Thus we are able to cover a wide dynamical range and indeed construct various families of dual operators and find their entangling powers. 

%The range of possible dynamic behaviors from non-ergodic to mixing, when $0 <e_p(U) \leq e_p^*$ is precisely due to the freedom of the local unitaries. This prompts us to define local unitary 
%averages, with random single particle unitaries, of quantities such as the spectral gap and the mixing rates. We show, both analytically and numerically, that such average rates have a very close one-to-one connection with the entangling power, and that this relationship seems universal for local dimension larger than $2$. Universality here is in the sense that it is independent of the structure of the dual operator or the local dimension. 
%
%For local dimension $2$, or qubits, there are also some exact results connecting the entangling power and mixing rates. 
% For arbitrary circuits, we derive analytically the maximal mixing rate under arbitrary local single particle unitaries. 
%It implies that as long as $U$ is not local, that is $e_p(U)>0$, however small it may be, there
%are also local operators that will render the circuit mixing. We motivate that this fact remains true in all dimensions as well.
%It maybe noted that for qubits, the entangling power cannot exceed $e_p^*$ and hence there are no Bernoulli circuits. For any other local dimension, if the entangling power exceeds $e_p^*$, the local unitaries are not capable of changing the mixing nature of the circuit, although it 
%may still change the rates.

Except for local dimension $2$, the parameterization of dual-unitary  operators is unknown. There are examples that have 
been constructed and studied in the past based on the Fourier transform (its duality has been discussed via its operator entanglement in \cite{Bhargavi2019,Tyson2003}), complex Hadamard matrices, and diagonal unitaries.
However, the systematic exploration of these must also consider their entangling power as this has a crucial impact on the
nature of the circuit as we have motivated above. In this paper we provide various constructions of families of dual-unitary  operators
that are explicit and analytical and yield ensembles with a range of possible entangling powers. 

We also provide analytical examples of 2-unitary operators that are to our knowledge new,
and related to higher dimensional quantum cat maps. Such 2-unitaries imply the existence of corresponding absolutely maximally entangled (AME) states, wherein 
every bipartite cut is maximally mixed, and these are equivalent to  perfect tensors with 4 indices  \cite{Goyeneche2015,Bhargavi2019}. The fact that there are no Bernoulli circuits for qubits is then related to the well-known fact from quantum information that there are no absolutely maximally entangled states of $4$ qubits \cite{Higuchi2000}. Thus the present work  ties in quantum information theoretic measures  with dual-unitary  circuits and it is conceivable that it generalizes for other special circuits as well.

\subsection{Structure of the paper}

The structure of this paper is as follows. In Sec. (\ref{sec:termi}) dual-unitary  circuits are recalled and the roles of the matrix realignment and partial
transpose--two central operations in quantum information, are developed. 
The Bernoulli circuits are constructed in Sec. (\ref{sec:Bernoulli}) and necessary and sufficient conditions for this  property
is discussed. It is also pointed out that correlations between operators with support on two sites retain the property of being Bernoulli. 
The explicit relation between entangling power and mixing in dual-unitary  circuits is developed in Sec.(\ref{sec:MixingRateAnalytics}). The mixing rates are defined and the sufficiency condition for the dual-unitary  circuit to be mixing is derived in Sec. (\ref{sec:mixmeas}). Local, single-particle, averaged spectral radius, which is related to the  measure of mixing, is calculated and found to be a simple function of entangling power in Sec. (\ref{ref:genavg}).  Explicit analytical calculations of mixing rates for the case of two qubit unitary operators is carried in Sec. (\ref{sec:qubit}).

 Dual-unitary  ensembles are constructed and their corresponding CPTP maps, involved in the calculation of correlation functions, are studied in Sec. (\ref{sec:DualConstructions}). The dual-unitaries constructed from block-diagonal unitary operators are in Sec. (\ref{sec:blockdiag}), the ``dual-CUE" resulting from an iterative map acting on the circular unitary ensemble (CUE) is in Sec. (\ref{sec:dualcue}). The designs from permutation matrices and the role of Latin squares is discussed in Sec. (\ref{sec:permu}) and finally a quantized coupled cat map is constructed in Sec. (\ref{sec:cat}). The cat map provides an analytically simple construction of dual-unitary  operators in all local dimensions and 2-unitaries or perfect tensors in any odd local dimension. The latter is to our knowledge the first such construction of 2-unitaries valid in all odd-dimensions.  
Finally we  summarize and discuss some possible directions in Sec. (\ref{sec:discu}).    

\section{Dual-unitary  to Bernoulli circuits via operator entanglements\label{sec:termi}}

In this section we review the dual-unitary  circuits for being self-contained and 
motivate how this naturally gives rise to their being cast in information theoretic language.
We point out that the two important rearrangements of matrix elements in quantum information theory that are used to detect bipartite separability of states--the realignment \cite{chen2002matrix,rudolph2003some} and the partial transpose \cite{peres1996separability} show up prominently in the dual-unitary  circuits.
This allows us to introduce the entangling power that plays an important role subsequently. These considerations lead 
naturally to the definition of Bernoulli circuits as being at the apex of an quantum ergodic hierarchy. 

\subsection{Dual-unitary  circuits, Realignment and partial transpose \label{sec:defn}}

Consider a periodic lattice of length $2L$ and each site, located for convenience at half-integer values, is occupied by a particle whose internal degree of freedom is a state in a Hilbert space $\mathcal{H}^q$
of dimension $q$. The unitary $U$ is a two-particle operator in $\mathcal{H}^q \otimes \mathcal{H}^q$ and  acts on all pairs of nearest neighbor sites. 
It is convenient to use the diagrammatic approach as explained in \cite{Bertini2019}. Unitary operators $U$ and $U^\dagger$ acting on bipartite Hilbert space  are diagrammatically represented as  

\begin{equation}
\br i \alpha |U|j \beta\kt = 
\begin{tikzpicture}[baseline=(current  bounding box.center),scale = 0.5]
\draw [thick,fill=teal,rounded corners=2.2pt] (1,1) rectangle (2,2); 
\node at (0.3,0.3) {$j$};
\node at (2.8,0.3) {$\beta$};
\node at (0.3,2.8) {$i$};
\node at (2.8,2.8) {$\alpha$};
\draw [thick] (2,2) -- (2.5,2.5);
\draw [thick] (1,2) -- (0.5,2.5);
\draw [thick] (1,1) -- (0.5,0.5);
\draw [thick] (2,1) -- (2.5,0.5);
\end{tikzpicture},  \qquad  U^{\dagger} = 
\begin{tikzpicture}[baseline=(current  bounding box.center),scale = 0.5]
\draw [thick,fill=purple,rounded corners=2.2pt] (1,1) rectangle (2,2); 
\draw [thick] (2,2) -- (2.5,2.5);
\draw [thick] (1,2) -- (0.5,2.5);
\draw [thick] (1,1) -- (0.5,0.5);
\draw [thick] (2,1) -- (2.5,0.5);
\end{tikzpicture}.
\nonumber 
\end{equation}

Connecting the legs is equivalent to  contracting  the indices, hence the unitarity of $U$, 
$\br i \alpha |UU^\dagger|j \beta \kt =\delta_{ij} \delta_{\alpha \beta}$
is diagrammatically  
\begin{equation}
\label{eq:unidia1}
\begin{tikzpicture}[baseline=(current  bounding box.center),scale = 0.4]
\node at (0.3,0.3) {$j$};
\node at (2.8,0.3) {$\beta$};
%\node at (0.6,2.6) {$k$};
%\node at (2.3,2.6) {$\gamma$};
\node at (0.3,4.7) {$i$};
\node at (2.8,4.7) {$\alpha$};
\draw [thick,fill=purple,rounded corners=2.2pt] (1,1) rectangle (2,2); 
\draw [thick] (1,2) to[out=110, in=-110] (1,3);
\draw [thick] (2,2) to[out=70, in=-70] (2,3);
\draw [thick,fill=teal,rounded corners=2.2pt] (1,3) rectangle (2,4);
\draw [thick] (1,4) -- (0.5,4.5);
\draw [thick] (2,4) -- (2.5,4.5);
\draw [thick] (1,1) -- (0.5,0.5);
\draw [thick] (2,1) -- (2.5,0.5);
\end{tikzpicture} 
=
\begin{tikzpicture}[baseline=(current  bounding box.center),scale = 0.4]
\node at (2.8,0.3) {$j$};
\node at (5.4,0.3) {$\beta$};
\node at (2.8,4.7) {$i$};
\node at (5.4,4.7) {$\alpha$};
\draw [thick] (3.5,1) -- (3.5,4);
\draw [thick] (4.5,1) -- (4.5,4);
\draw [thick] (3.5,1) -- (3.0,0.5);
\draw [thick] (3.5,4) -- (3,4.5);
\draw [thick] (4.5,4) -- (5,4.5);
\draw [thick] (4.5,1) -- (5,0.5);
\end{tikzpicture}, \;\;
U^\dagger U= \mathbb{1} \equiv 
\begin{tikzpicture}[baseline=(current  bounding box.center),scale = 0.4]
\draw [thick,fill=teal,rounded corners=2.2pt] (1,1) rectangle (2,2); 
\draw [thick] (1,2) to[out=110, in=-110] (1,3);
\draw [thick] (2,2) to[out=70, in=-70] (2,3);

\draw [thick,fill=purple,rounded corners=2.2pt] (1,3) rectangle (2,4);
\draw [thick] (1,4) -- (0.5,4.5);
\draw [thick] (2,4) -- (2.5,4.5);

\draw [thick] (1,1) -- (0.5,0.5);
\draw [thick] (2,1) -- (2.5,0.5);
\end{tikzpicture}
=  
\begin{tikzpicture}[baseline=(current  bounding box.center),scale = 0.4] 

\draw [thick] (3.5,1) -- (3.5,4);
\draw [thick] (4.5,1) -- (4.5,4);
\draw [thick] (3.5,1) -- (3.0,0.5);
\draw [thick] (3.5,4) -- (3,4.5);
\draw [thick] (4.5,4) -- (5,4.5);
\draw [thick] (4.5,1) -- (5,0.5);

\end{tikzpicture}. 
\end{equation}

The one-time evolution operator 
\beq 
\label{eq:UU}
\mathbf{U}=\mathbf{T}_{2L}U^{\otimes L} \mathbf{T}^\dagger_{2L}U^{\otimes L}
\eeq can be thought of as having two stages, in the first the $2L$ particle state is evolved under the unitary $U^{\otimes L}$, acting on pairs $(i,i+\nicefrac{1}{2})$, and in the other on $(i+\nicefrac{1}{2},i+1)$. Here $\mathbf{T}_{2L} \ket{k_1} \otimes \ket{k_2} \otimes  \cdots \ket{k_{2L}} = \ket{k_2} \otimes \ket{k_3} \otimes  \cdots \ket{k_{2L}} \otimes \ket{k_1}$ is the site translation operator. The evolution operator over time $t$ is simply $\mathbf{U}(t) = \mathbf{U}^t$. For example with $L=4$, corresponding to $8$ particles, $\mathbf{U}^2$  is diagrammatically  
\begin{equation}  
\mathbf{U}^2 =  
\btp[baseline=(current  bounding  box.center),scale=0.35] 
\foreach \i in {0,...,3} %for four legs
{
\draw [thick,fill=teal,rounded corners=2.2pt] (1+4*\i,1) rectangle (2+4*\i,2);
\draw [thick,fill=teal,rounded corners=2.2pt] (3+4*\i,3) rectangle (4+4*\i,4);
\draw [thick,fill=teal,rounded corners=2.2pt] (1+4*\i,5) rectangle (2+4*\i,6);
\draw [thick,fill=teal,rounded corners=2.2pt] (3+4*\i,7) rectangle (4+4*\i,8);
\draw [thick] (1+4*\i,1) -- (0.5+4*\i,0.5);  %small legs
\draw [thick] (3+4*\i,8) -- (2.5+4*\i,8.5); %small legs
\draw [thick] (4+4*\i,8) -- (4.5+4*\i,8.5); %small legs
\draw [thick] (2+4*\i,1) -- (2.5+4*\i,0.5); %small legs
\draw [thick] (2+4*\i,2) -- (3+4*\i,3);
\draw [thick] (2+4*\i,6) -- (3+4*\i,7);
\draw [thick] (3+4*\i,4) -- (2+4*\i,5);
} 

\foreach \i in {0,...,2} %for three legs
{
\draw [thick] (4+4*\i,4) -- (5+4*\i,5);
\draw [thick] (4+4*\i,3) -- (5+4*\i,2);
\draw [thick] (4+4*\i,7) -- (5+4*\i,6);
}

\foreach \i in {0,...,1} %for two legs
{
\draw [thick] (16,3+4*\i) -- (16.5,2.5+4*\i);
\draw [thick] (1,2+4*\i) -- (0.5,2.5+4*\i);
}
\draw [thick] (1,5) -- (0.5,4.5); %single leg
\draw [thick] (16,4) -- (16.5,4.5); %single leg 

\node at (8.5,-0.2) {\small{$0$}};
\node at (10.5,-0.2) {$\frac{1}{2}$};
\node at (12.5,-0.2) {\small{$1$}};
\node at  (14.5,-0.2){\small{$\frac{3}{2}$}};
\node at (6.5,-0.2) {\small{$-\frac{1}{2}$}};
\node at (4.5,-0.2) {\small{$-1$}};
\node at ( 2.5,-0.2) {\small{$-\frac{3}{2}$}};
\node at (0.5,-0.2) {\small{$-2$}};
\node at (0.1,0.5) {\small{$0$}};
\node at (0.1,2.7){\small{$\frac{1}{2}$}};
\node at  (0.1,4.5){\small{$1$}};
\node at  (0.1,6.7){\small{$\frac{3}{2}$}};
\node at (0.1,8.5){\small{$2$}};

\node at (18,-1.2) {$x$};
\node at (-0.5,10){$t$};
\draw [->,thick] (-0.8,-1) -- (17.2,-1);
\draw [->,thick] (-.8,-1) -- (-.8,9);

\etp. 
\label{eq:unidia} 
\nonumber 
\end{equation} 

Let $\{a_j\}_{j = 0}^{q^2 -1}$ be an orthonormal operator basis in the single-particle local Hilbert space $\mathcal{H}^q$ under the Hilbert-Schmidt inner product $\tr(a^{\dagger}_j a_k) = q \delta_{j,k}$. If we choose $a_0 = \mathbb{1}_q$, it follows that $\tr(a_j) = 0,$ for $j> 0$. The dynamical correlation function of single-particle operators in the infinite temperature state can be found from the following quantities (where $ i,j \neq 0$) \cite{Bertini2019}:
\begin{equation}
\label{eq:dyncor}  
\begin{split}
& D^{ij}(x,y,t) \equiv \frac{1}{q^{2L}} \tr[a_j^x \mathbf{U}^{-t} a_i^y \mathbf{U}^{t}]=\\
&\frac{1}{q^{2L}}
\btp[baseline=(current  bounding  box.center),scale=0.35] 
%Top picture 
\foreach \i in {0,...,3} %for four legs
{
\draw [thick,fill=purple,rounded corners=2.2pt] (1+4*\i,11) rectangle (2+4*\i,12);
\draw [thick,fill=purple,rounded corners=2.2pt] (3+4*\i,9) rectangle (4+4*\i,10);
\draw [thick,fill=purple,rounded corners=2.2pt] (1+4*\i,15) rectangle (2+4*\i,16);
\draw [thick,fill=purple,rounded corners=2.2pt] (3+4*\i,13) rectangle (4+4*\i,14);
\draw [thick] (3+4*\i,9) -- (2.5+4*\i,8.5); %small legs
\draw [thick] (4.5+4*\i,8.5) -- (4+4*\i,9); %small legs
\draw [thick] (1+4*\i,16) -- (0.5+4*\i,16.5);  %small legs
\draw [thick] (2+4*\i,16) -- (2.5+4*\i,16.5);  %small legs
\draw [thick] (3+4*\i,10) -- (2+4*\i,11);  
\draw [thick] (3+4*\i,14) -- (2+4*\i,15);  
\draw [thick] (2+4*\i,12) -- (3+4*\i,13); 
} 
\foreach \i in {0,...,2} %three legs 
{
\draw [thick] (4+4*\i,10) -- (5+4*\i,11);
\draw [thick] (4+4*\i,14) -- (5+4*\i,15);
\draw [thick] (5+4*\i,12) -- (4+4*\i,13);
}
\foreach \i in {0,...,1} %for two legs
{
\draw [thick] (16,10+4*\i) -- (16.5,10.5+4*\i);
\draw [thick] (1,11+4*\i) -- (0.5,10.5+4*\i);
} 
\draw [thick] (16,13) -- (16.5,12.5); %single leg
\draw [thick] (1,12) -- (0.5,12.5); %single leg
%lower picture
\foreach \i in {0,...,3} %for four legs
{
\draw [thick,fill=teal,rounded corners=2.2pt] (1+4*\i,1) rectangle (2+4*\i,2);
\draw [thick,fill=teal,rounded corners=2.2pt] (3+4*\i,3) rectangle (4+4*\i,4);
\draw [thick,fill=teal,rounded corners=2.2pt] (1+4*\i,5) rectangle (2+4*\i,6);
\draw [thick,fill=teal,rounded corners=2.2pt] (3+4*\i,7) rectangle (4+4*\i,8);
\draw [thick] (1+4*\i,1) -- (0.5+4*\i,0.5);  %small legs
\draw [thick] (3+4*\i,8) -- (2.5+4*\i,8.5); %small legs
\draw [thick] (4+4*\i,8) -- (4.5+4*\i,8.5); %small legs
\draw [thick] (2+4*\i,1) -- (2.5+4*\i,0.5); %small legs
\draw [thick] (2+4*\i,2) -- (3+4*\i,3);
\draw [thick] (2+4*\i,6) -- (3+4*\i,7);
\draw [thick] (3+4*\i,4) -- (2+4*\i,5);
} 
\foreach \i in {0,...,2} %for three legs
{
\draw [thick] (4+4*\i,4) -- (5+4*\i,5);
\draw [thick] (4+4*\i,3) -- (5+4*\i,2);
\draw [thick] (4+4*\i,7) -- (5+4*\i,6);
}
\foreach \i in {0,...,1} %for two legs
{
\draw [thick] (16,3+4*\i) -- (16.5,2.5+4*\i);
\draw [thick] (1,2+4*\i) -- (0.5,2.5+4*\i);
}
\draw [thick] (1,5) -- (0.5,4.5); %single leg
\draw [thick] (16,4) -- (16.5,4.5); %single leg 
\draw[thick, fill=black] (6.5,8.5) circle (0.1cm); 
\draw[thick, fill=black] (10.5,16.5) circle (0.1cm);
\Text[x=6,y=8]{$a^y_i$}
\Text[x=10.2,y=17]{$a^x_j$}
\etp. \end{split}
\end{equation}

By reshaping of the matrix elements of $U$, its realignment $R_1$ (denoted by $\tilde{U}$ in \cite{Bertini2019}) is defined as the permutation 
\begin{equation}
\br \beta \alpha|U^{R_1}| j i\kt =\br i \alpha|U|j \beta\kt.
\label{eq:r1} 
\end{equation} 
This can be used to define the  evolution in the ``spatial direction" by the transfer operator 
\begin{equation}
\mathbf{U}_R =  \mathbf{T}_{2t} (U^{R_1})^{\otimes t} \mathbf{T}^\dagger_{2t} (U^{R_1})^{\otimes t}. 
\label{eq:spaevo}   
\end{equation}
With periodic boundary conditions, these models satisfy $\tr\left(\mathbf{U}^t\right) = \tr\left(\mathbf{U}_R^L\right)$ and are in a sense dual to each other \cite{Akila2016}. It has been shown in \cite{Bertini2019} that if $U^{R_1}$ is also unitary, the correlations in Eq.~(\ref{eq:dyncor}) vanish inside the light cone $|y-x|<t$ and the resultant circuit has been called dual-unitary . These circuits are built of dual-unitary  operators.

\begin{definition}[Dual-unitary  \cite{Bertini2019} ]\label{def:dual}
A unitary operator $U$ is dual-unitary  if $U^{R_1}$ is also unitary. 
\end{definition}
This is represented diagrammatically as  
\begin{equation}
\label{eq:dual2}
\begin{tikzpicture}[baseline=(current  bounding box.center),scale = 0.5]
\draw [thick,fill=teal,rounded corners=2.2pt] (1,1) rectangle (2,2); 
\draw [thick] (1,2) to[out=110, in=-110] (1,3);
\draw [thick] (2,3) -- (2.3,2.7);
\draw [thick] (2,2) -- (2.3,2.3);
\draw [thick] (0.5,0.5) to[out=110, in=-110] (0.5,4.5);

\draw [thick,fill=purple,rounded corners=2.2pt] (1,3) rectangle (2,4);
\draw [thick] (1,4) -- (0.5,4.5);
\draw [thick] (2,4) -- (2.5,4.5);

\draw [thick] (1,1) -- (0.5,0.5);
\draw [thick] (2,1) -- (2.5,0.5);
\end{tikzpicture}
= 
\begin{tikzpicture}[baseline=(current  bounding box.center),scale = 0.5]
\draw [thick] (5,2) to[out=110, in=-110] (5,3);
\draw [thick] (5,2) -- (6,2);
\draw [thick] (5,3) -- (6,3);
\draw [thick] (6,3) -- (6.3,2.7);
\draw [thick] (6,2) -- (6.3,2.3);

\draw [thick] (4.5,4.5) -- (5,4) -- (6,4) -- (6.5,4.5);
\draw [thick] (4.5,0.5) -- (5,1) -- (6,1) -- (6.5,0.5);

\draw [thick]  (4.5,0.5) to[out=110, in=-110] (4.5,4.5);
\end{tikzpicture},\;\;\;\;
\begin{tikzpicture}[baseline=(current  bounding box.center),scale = 0.5]
\draw [thick,fill=teal,rounded corners=2.2pt] (1,1) rectangle (2,2); 
\draw [thick] (1,2) -- (0.5,2.3);
\draw [thick] (2,2) to[out=70, in=-70] (2,3);
\draw [thick] (0.5,0.5) -- (1,1);
\draw [thick] (1,3) -- (0.5,2.7);
\draw [thick] (2.5,0.5) to[out=70, in=-70] (2.5,4.5);

\draw [thick,fill=purple,rounded corners=2.2pt] (1,3) rectangle (2,4);
\draw [thick] (1,4) -- (0.5,4.5);
\draw [thick] (2,4) -- (2.5,4.5);

\draw [thick] (1,1) -- (0.5,0.5);
\draw [thick] (2,1) -- (2.5,0.5);
\end{tikzpicture} =
\begin{tikzpicture}[baseline=(current  bounding box.center),scale = 0.5]
\draw [thick] (4.5,4.5) -- (5,4) -- (6,4) -- (6.5,4.5);
\draw [thick] (4.5,0.5) -- (5,1) -- (6,1) -- (6.5,0.5);
\draw [thick] (6.5,0.5) to[out=70, in=-70] (6.5,4.5);
\draw [thick] (4.5,2.7) -- (5,3) -- (6,3); 
\draw [thick] (4.5,2.3) -- (5,2) -- (6,2);
\draw [thick] (6,2) to[out=70, in=-70] (6,3);
\end{tikzpicture}.
\end{equation}

%The entanglement properties of unitary operators play an important role in understanding the quantum dynamics from the perspective of entanglement it generates \cite{Zanardi2000,Zanardi2001,Nielsen2003}. 

The realignment in Eq. (\ref{eq:r1}) is intimately related to bipartite operator entanglement \cite{Zyczkowski2004,Bengtsson2007}.
It has been pointed out \cite{SAA2020} that the unitary $U$ is dual iff the operator entanglement of $U$ is maximum.  Besides this, the realignment in Eq. (\ref{eq:r1}) has important applications in  investigating the complete positivity of quantum maps \cite{oxenrider1985matrix,sudarshan1961stochastic}, and in characterizing the  entanglement of mixed states \cite{chen2002matrix,rudolph2003some}. To visualize the realignment $R_1$ in Eq. (\ref{eq:r1}, consider a $4 \times 4 $ matrix defined with the elements in the product basis as $X_{i\alpha,j\beta} = \mel{i\alpha}{X}{j\beta}$. Each $2 \times 2 $ block of  $4 \times 4 $ matrix $X$ is vectorized and arranged as a column in the $X^{R_1}$ matrix. This can be visualized easily as: 
\begin{equation}
\label{eq:matX}
X  = \begin{pmatrix}[cc|cc] 
 \blu  X_{00,00} & \blu X_{00,01} & X_{00,10} & X_{00,11} \\
  \blu X_{01,00} & \blu X_{01,01}  & X_{01,10} & X_{01,11} \\
  \hline 
  \red X_{10,00} & \red X_{10,01} & \color{green2}X_{10,10} & \color{green2} X_{10,11} \\
  \red X_{11,00} & \red X_{11,01} & \color{green2}X_{11,10} & \color{green2} X_{11,11}
 \end{pmatrix} 
\end{equation}
\begin{equation}
X^{R_1} = \begin{pmatrix}[c|c|c|c]
           \blu X_{00,00} & \red X_{10,00} & X_{00,10} & \color{green2}X_{10,10} \\
           \blu X_{01,00} & \red X_{11,00} & X_{01,10} & \color{green2}X_{11,10} \\
           \blu X_{00,01} & \red X_{10,01} & X_{00,11} & \color{green2}X_{10,11} \\
           \blu X_{01,01} & \red X_{11,01} & X_{01,11} & \color{green2}X_{11,11}
          \end{pmatrix} 
\end{equation}

To see the connection to entanglement consider the four-party pure state on $\otimes^4 \mathcal{H}^q$
\beq
\label{eq:4party}
|\psi\kt = \frac{1}{q}\sum_{i\alpha j \beta} U_{i \alpha; j \beta}|i \alpha j \beta\kt_{1234} = (U\otimes \mathbb{1}) |\Phi^+\kt_{13} |\Phi^+\kt_{24},
\nonumber 
\eeq
where
\beq
\label{eq:iso}
|\Phi^+\kt =\frac{1}{\sqrt{q}}\sum_{j=1}^q|jj\kt
\eeq
is a maximally entangled state, and $U$ acts on the $12$ subsystem.
 The partition $12|34$ is maximally entangled as the reduced density matrix $\rho_{12}=U U^{\dagger}/q^2=\mathbb{1}_{q^2}/q^2$, due to 
 the unitarity of $U$. The $31|42$ partition has the state $\rho_{31}=U^{R_1} U^{R_1\,\dagger}/q^2$, and if $U$ is dual-unitary  this 
 partition is also maximally entangled \cite{Bhargavi2019}. 

The operator Schmidt decomposition of $U$ is 
\begin{equation}
U = \sum_{j=0}^{q^2-1} \sqrt{\gamma_j} \; X_{j} \otimes Y_{j},
\label{eq:sch}
\end{equation}
such that $\tr (X_{j}^\dagger X_{k}) = \br X_j|X_k\kt= \tr(Y_{j}^\dagger Y_{k}) = \br Y_j|Y_k\kt= \delta_{jk}$, that is the operators $X$ and $Y$ form orthonormal bases. The vectors $|X_j\kt$ are columnwise vectorization of the $q\times q$ matrices $X_k$, that is $\br j|X_k|i \kt =\br ij|X_k \kt$.
It is easy to see that $(X\otimes Y)^{R_1}=|Y\kt \br X^*|$, and hence 
\beq
\rho_{31}=\frac{1}{q^2}U^{R_1} U^{R_1\,\dagger}=\frac{1}{q^2}\sum_{j=0}^{q^2-1} \gamma_j|Y_j\kt \br Y_j|.
\eeq

Hence, the eigenvalues of $\rho_{31}$ are the Schmidt coefficients $\gamma_j \geq 0$ which determines the entanglement of $U$: iff only 
one of them is nonvanishing, the operator is of product form. As is clear from the Schmidt decomposition, the $\gamma_j$ are LU invariants under local unitary transformations of the form Eq.~(\ref{eq:LU}), and may be used to define a variety of entropies of entanglement. As $\sum_{j = 0}^{q^2-1} \gamma_j/q^2 = 1$, $\{\gamma_j/q^2, 0 \leq j \leq q^2-1\}$ defines a discrete probability distribution, in terms of which the entropies maybe defined.
For our purposes, the linear entropy, the \textit{operator entanglement} \cite{Zanardi2001} defined as 
\begin{equation}
	E(U) = 1 - \frac{1}{q^4} \tr[\left(U^{R_1}U^{R_1\dagger}\right)^2]=1 - \frac{1}{q^4} \sum_{j=0}^{q^2-1} \gamma_j^2,
	\label{eq:EUdefn}
\end{equation}
where $0 \leq E(U) \leq E(S)= 1 - \frac{1}{q^2}$, is useful. Here $S$ is the swap operator in Eq.~(\ref{eq:swap}).
From this, the operator entanglement $E(U)$ is maximum iff $U^{R_1}$ is also unitary,
and all $\gamma_j=1$. Thus the maximal operator entanglement is equivalent to the unitary operator being dual-unitary, a property that 
is LU invariant as $E(U')=E(U)$.  The OTOC for bipartite systems \cite{Ravi_PRB_2020, Yan_2020}, averaged over all local unitary operators was shown to  be related to the Loschmidt Echo \cite{Yan_2020}. The averaged OTOC was related to the operator entanglement of bipartite unitary operators \cite{styliaris2020information}, and provides another operational meaning to the operator entanglement by relating it to the measures of chaos in bipartite systems. 

Returning to the circuit, using the fact that $\mathbf{U}$ is invariant under two-site shifts, $\mathbf{U} \mathbf{T}^2_{2L} = \mathbf{T}^2_{2L} \mathbf{U}$, to find the dynamical correlation function in Eq. (\ref{eq:dyncor}), it is sufficient to consider the pair of functions 
\beq
\begin{split}
C_+^{ij}(x,t) &= D^{ij}\left(x,0,t\right), \\
C_-^{ij}(x,t) &= D^{ij}(x+\textstyle \frac{1}{2},\frac{1}{2},t).
\end{split}
\nonumber 
\eeq 
These two correlations are when the single-particle operators are to the "left" or "right" leg of a two-particle $U$. 
The only non-zero correlation for dual-unitary  circuits lie along the light cone and was shown, quite remarkably, in \cite{Bertini2019} to be given in terms of a completely positive trace preserving (CPTP) map constructed only from the two-particle unitary operator $U$:
\beq
C^{ij}_{\pm}(\pm t,t) = \frac{1}{q} \tr[\mathcal{M}_{\pm}^{2t}(a^i)a^j].  
\label{eq:Map1}
\eeq 
This can be seen from the diagrammatic evaluation of the correlation functions  $C^{ij}_+ (t,t) =$
\begin{equation} 
 \frac{1}{q^{2t+1}}
\btp [baseline=(current  bounding  box.center),scale=0.35] 
\foreach \i in {0,...,3} %for four legs
{
\draw [thick,fill=purple,rounded corners=2.2pt] (2*\i,1+2*\i) rectangle (1+2*\i,2+2*\i);
\draw [thick,fill=teal,rounded corners=2.2pt] (2*\i,-2-2*\i) rectangle (1+2*\i,-1-2*\i);  
\draw [thick] (1+2*\i, 2+2*\i) -- (2+2*\i, 3+2*\i);
\draw [thick] (1+2*\i, -2-2*\i) -- (2+2*\i, -3-2*\i);
} 

\foreach \i in {0,...,2} %for four legs
{
\draw [thick] (3+2*\i,3+ 2*\i) -- (3.5+2*\i,2.5+2*\i) -- (3.5+2*\i,-2.5-2*\i) -- (3+2*\i,-3-2*\i); 
\draw [thick] (2+2*\i,4+ 2*\i) -- (1.5+2*\i,4.5+2*\i) -- (-1.5-2*\i,4.5+2*\i) -- (-2-2*\i,4+2*\i) -- (-2-2*\i,-4-2*\i) -- (-1.5-2*\i,-4.5-2*\i) -- (1.5+2*\i,-4.5-2*\i) -- (2+2*\i,-4-2*\i) ; 
}

\draw [thick] (8,9) -- (8,-9); 
\draw[thick, fill=black] (8,9) circle (0.1cm);
\Text[x=8,y=9.3]{$a_j$}

\draw [thick] (0,-2) -- (-0.5,-2.5) ; 
\draw [thick] (0,2) -- (-0.5,2.5) ;
\draw [thick]  (-0.5,-2.5) to[out=110, in=-110] (-0.5,2.5);
\draw [thick]  (0,-1) to[out=110, in=-110] (0,1);
\draw [thick]  (1,-1) to[out=70, in=-70] (1,1);

\draw[thick, fill=black] (-0.2,0) circle (0.1cm);
\Text[x=-0.6,y=0.3]{$a_i$}
\etp .
\label{eq:corrndia}
\end{equation} 

The CPTP maps  $\mathcal{M}_{\pm}$  are 
\begin{equation}
\mathcal{M}_{+}(a) = \frac{1}{q} \tr_1 \left[ U^{\dagger}(a \otimes \mathbb{1})U\right] = \frac{1}{q}
\begin{tikzpicture}[baseline=(current  bounding  box.center),scale = 0.6]
\draw [thick,fill=teal,rounded corners=2.2pt] (1,1) rectangle (2,2); 
\draw [thick] (1,2) to[out =110, in = -110] (1,3);
\draw [thick] (2,2) to[out=70, in=-70] (2,3);
\draw [thick] (0.5,0.5) -- (1,1);
\draw[thick, fill=black] (0.9,2.5) circle (0.1cm);

\draw [thick] (0.5,0.5) to[out=110, in=-110] (0.5,4.5);

\draw [thick,fill=purple,rounded corners=2.2pt] (1,3) rectangle (2,4);
\draw [thick] (1,4) -- (0.5,4.5);
\draw [thick] (2,4) -- (2.5,4.5);

\draw [thick] (1,1) -- (0.5,0.5);
\draw [thick] (2,1) -- (2.5,0.5);
\Text[x=0.6,y=2.6]{$a$} 
\end{tikzpicture},
\label{eq:mpdia}
\end{equation} 
\begin{equation}
\mathcal{M}_{-}(a) = \frac{1}{q} \tr_2 \left[ U^{\dagger}( \mathbb{1}\otimes a )U\right] = \frac{1}{q}
\begin{tikzpicture}[baseline=(current  bounding  box.center),scale = 0.6]
\draw [thick,fill=teal,rounded corners=2.2pt] (1,1) rectangle (2,2); 
\draw [thick] (1,2) to[out =110, in = -110] (1,3);
\draw [thick] (2,2) to[out=70, in=-70] (2,3);
\draw [thick] (0.5,0.5) -- (1,1);
\draw [thick] (2.5,0.5) to[out=70, in=-70] (2.5,4.5);

\draw [thick,fill=purple,rounded corners=2.2pt] (1,3) rectangle (2,4);
\draw [thick] (1,4) -- (0.5,4.5);
\draw [thick] (2,4) -- (2.5,4.5);

\draw [thick] (1,1) -- (0.5,0.5);
\draw [thick] (2,1) -- (2.5,0.5);

\draw[thick, fill=black] (2.1,2.5) circle (0.1cm);
\Text[x=2.5,y=2.7]{$a$}
\end{tikzpicture}. 
\label{eq:mmdia}
\end{equation} 

These linear maps take density matrices to density matrices and are unital (preserve the identity).
Let the action $\mathcal{M}_{+}(\rho)=\rho'$ be denoted as $M_{+}[U]|\rho\kt = |\rho'\kt$, where 
$|\rho\kt$ is the $q^2$ dimensional column vector formed from its row-vectorization, and the dependence on the 
unitary $U$ is made explicit. That is, $\br jl|\rho\kt \equiv \br j |\rho|l \kt$ and  $ M_{\pm}[U]\in \mathbb{C}^{q^2 \times q^2}$, where $M_{-}[U]$ is similarly defined. Consider 
\beq
\begin{split}
q \br \alpha|\mathcal{M}_{+}&(\rho)|\beta \kt = \sum_{i} \br i \alpha| U^{\dagger}(\rho \otimes \mathbb{1})U|i  \beta \kt \\
%&=\sum_{ij\gamma l \delta} \br i \alpha | U^{\dagger}|j \gamma \kt \br j \gamma |(\rho \otimes \mathbb{1})|l \delta\kt \br l \delta |U|i \beta \kt \\
&=\sum_{ij\gamma l} \br i \alpha | U^{\dagger}|j \gamma \kt  \br l \gamma |U|i \beta \kt \br j |\rho|l \kt\\
&=\sum_{jl i \gamma } \br i \gamma|\utbd|j \alpha\kt \br l \beta|\utb|i \gamma \kt \br jl |\rho\kt\\
%&=\sum_{jl} \br l \beta |\utb \utbd|j \alpha \kt \br jl |\rho\kt\\
&=\sum_{jl} \br \alpha \beta |\left(\utb \utbd\right)^{R_1}|j l \kt \br jl |\rho\kt.
\end{split}
%\nonumber
\eeq
However, as $ \br \alpha|\mathcal{M}_{+}(\rho)|\beta \kt=\br \alpha|\rho'|\beta\kt =\br \alpha \beta|\rho'\kt$ we identify  
\beq
\label{eq:M+UT}
M_{+}[U]=\frac{1}{q}\left(\utb \utbd \right)^{R_1}.
\eeq
Similarly, we get 
\beq 
\label{eq:M-UT}
M_{-}[U]=\frac{1}{q} \left( \utbd \utb \right)^{R_2} .
\eeq 
The reordering of matrix elements used in the above equations are defined as follows: 
\begin{subequations}
\label{eq:T1T2R2}
\begin{align}
&\text{Partial transpose}\; T_1:  &\br i \alpha|X|j \beta\kt = \br j \alpha|X^{T_1}| i \beta\kt,\\
&\text{Partial transpose}\; T_2:  &\br i \alpha|X|j \beta\kt = \br i \beta|X^{T_2}| j \alpha\kt,\\
&\text{Realignment}\; R_2: &\br i \alpha|X|j \beta\kt = \br ij|X^{R_2}| \alpha \beta \kt.
\end{align}
\end{subequations}
To visualize these operations, consider a $4\times 4$ matrix with matrix elements $X_{i\alpha,j\beta} = \mel{i\alpha}{X}{j\beta}$ as given in Eq.~(\ref{eq:matX}). 
\begin{equation}
\begin{split} 
 X^{R_2} &= \begin{pmatrix}
            \blu X_{00,00} & \blu X_{00,01} & \blu X_{01,00} & \blu X_{01,01} \\
            \hline 
            X_{00,10} & X_{00,11} & X_{01,10} & X_{01,11} \\
            \hline
            \red X_{10,00} & \red X_{10,01} & \red X_{11,00} &\red X_{11,01} \\
            \hline 
            \color{green2}X_{10,10} &\color{green2} X_{10,11} & \color{green2}X_{11,10} & \color{green2} X_{11,11} 
           \end{pmatrix} \\
X^{T_1} & = \begin{pmatrix}[cc|cc] 
  \blu X_{00,00} & \blu X_{00,01} & {\red X_{10,00}}  & {\red X_{10,01}} \\
  \blu X_{01,00} & \blu X_{01,01} & {\red X_{11,00}} & {\red X_{11,01}}   \\
  \hline 
    X_{00,10} &  X_{00,11} & \color{green2} X_{10,10} & \color{green2} X_{10,11} \\
    X_{01,10} &  X_{01,11}  & \color{green2} X_{11,10} & \color{green2} X_{11,11}
 \end{pmatrix} \\
X^{T_2} & = \begin{pmatrix}[cc|cc] 
  \blu X_{00,00} & \blu  {\bf X_{01,00}} & X_{00,10} & {\bf X_{01,10}}  \\
  \blu {\bf X_{00,01}}  & \blu  X_{01,01} & {\bf X_{00,11}} & X_{01,11} \\
  \hline 
  \red X_{10,00} & \red {\bf X_{11,00}} & \color{green2} X_{10,10} & \color{green2}{\bf X_{11,10}} \\
  \red {\bf X_{10,01}} & \red X_{11,01} & \color{green2}{\bf X_{10,11}} &\color{green2} X_{11,11}
 \end{pmatrix} 
\end{split} 
\end{equation}

Note also that 
\beq
\label{eq:M_pm_relation}
M_+[SUS]=M_-[U],
\eeq
which follows from the definitions in Eq.~(\ref{eq:M+UT}), ~(\ref{eq:M-UT}) and the fact that $\tr_1(SAS)=\tr_2(A)$. Identifying these CPTP maps
in terms of the well-known partial transpose operation--another rearrangement of matrix elements, devised to detect bipartite entanglement \cite{peres1996separability},
has some immediate consequences that we presently investigate.

Along with the swap operator $S$ , 
and the usual transpose operation $A^T$, the realignment and partial transpose operations  satisfy the following, easily verifiable, but useful identities, which we use frequently. 
\begin{subequations}
\begin{align}
(A^{T_1})^{T_2}=(A^{T_2})^{T_1}=A^T, &\;\;(A^{T_1})^T =(A^T)^{T_1}, \label{eq:IdentityT1T2a}\\
%SA^{T_1}S=(SAS)^{T_2},& \;\; SA^{T_2}S=(SAS)^{T_1},\\
(A^{R_1})^{R_2}=&(A^{R_2})^{R_1}=S A^T S,\\
A^{R_1}&=(SA^{R_2} S)^T, \label{eq:IdentityR1R2}\\
(SA)^{R_2}=SA^{T_1},& \;\; (AS)^{R_1}=A^{T_1}S, \label{eq:IdentityT1R2}\\
(SA)^{T_2}=SA^{R_1}, & \;\; (AS)^{T_1}=A^{R_1}S, \label{eq:IdentityR1T2}\\
(AS)^{T_2}=A^{R_2}S,&\;\; (SA)^{T_1}= S A^{R_2}. \label{eq:IdenityAST2}
\end{align}
\label{eq:Idenities}
\end{subequations}
The effect of local operators is as follows:
\begin{subequations}
\begin{align}
[ (u_1 \otimes u_2) A (u_3 \otimes u_4)]^{R_1}&=(u_4^T \otimes u_2) A^{R_1} (u_3 \otimes u_1^T),\\
[ (u_1 \otimes u_2) A (u_3 \otimes u_4)]^{R_2}&=(u_1 \otimes u_3^T) A^{R_2} (u_2^T \otimes u_4),\\
[ (u_1 \otimes u_2) A (u_3 \otimes u_4)]^{T_1}&=(u_3^T \otimes u_2) A^{T_1} (u_1^T \otimes u_4),\\
[ (u_1 \otimes u_2) A (u_3 \otimes u_4)]^{T_2}&=(u_1 \otimes u_4^T) A^{T_2} (u_3 \otimes u_2^T).
\end{align}
\label{eq:LocalIdentities}
\end{subequations}

First, we recall some basic properties of the channel. $M_\pm$ is a Liouvillian superoperator representation of the channel $\mathcal{M}_\pm$. $M_\pm$ is a non-Hermitian operator whose eigenvalues $\{\lambda_j\}$ are bounded $\abs{\lambda_j} \leq 1$ \cite{Bertini2019}. The channel $\mathcal{M}_\pm$ is unital: it fixes the maximally mixed state $\rho=\mathbb{1}_q/q$, thus there is always one eigenvalue $=1$, the ``trivial eigenvalue" corresponding to the (unnormalized) eigenvector $|\rho\kt=|\mathbb{1}_q\kt/q=|\Phi^{+}\kt/\sqrt{q}$,
where  $|\Phi^+\kt$ is the maximally entangled state from Eq.~(\ref{eq:iso}).  We note that $\mathcal{M}_+(a)$ is {\it complementary} \cite{devetak2005capacity, holevo2007complementary,smaczynski2016selfcomplementary} to the channel 
\[
\mathcal{E}_+(a) =   \frac{1}{q} \tr_2 \left[ U^{\dagger}(a \otimes \mathbb{1})U\right],
\] where the subsystem initially in the most mixed state is traced out, which is the traditional process of tracing out a
complex environment. The channel $\mathcal{E}_+(a)$ is obtained by replacing $U$ in  $\mathcal{M}_+(a)$ by $US$. The Liouvillian superoperator of the channel $\mathcal{E}_+(a)$ is $E_+[U] = M_+[US] = (1/q) \tr \left((US)^{T_2} (US)^{T_2 \dagger} \right)^{R_1} =  \left(U^{R_2}U^{R_2 \dagger} \right)^{R_1} /q$ as noted in \cite{Zyczkowski2004}, and if $U$ is dual-unitary, this channel is $\mathbb{1}_{q^2}^{R_1}/q= \dyad{\Phi_+} $. Here the result from Eq.~\ref{eq:IdenityAST2} is used.

The spectrum of $M_+$ arranged in increasing order of its magnitude is used below as
$\lambda_0=1 \geq |\lambda_1| \geq |\lambda_2| \geq  \cdots \geq |\lambda_{q^2-1}|$. Generically the eigenvalues of both $M_{\pm}$ 
will be referred to as $\lambda_k$ below.   
The authors of Ref. (\cite{Bertini2019}), classified the ergodic properties of the dual-unitary  circuits from the nontrivial eigenvalues ($k>0$) of $M_{\pm}$ as 
\begin{description}
\item[Non-interacting] All  $2(q^2-1)$ eigenvalues $\lambda_k = 1$, the correlations remain constant.  
\item[Non-ergodic (Interacting and nonintegrable)] There are $0 <  k < 2(q^2-1) $ nontrivial eigenvalues $\lambda_j = 1$, then the correlations in the $k$ modes remain constant.
\item[Ergodic and non-mixing] All nontrivial eigenvalues $\lambda_k \neq 1$ but there exists atleast one eigenvalue with unit modulus $\abs{\lambda_k} = 1$.   
\item[Ergodic and mixing] All $2(q^2-1)$ eigenvalues lie within the unit circle, $\abs{\lambda_k} < 1$. All the correlations decay.  
\end{description}
following the discussion of Bernoulli systems in the introduction, we add 
%\blu (Bernoulli intro copied from inside section) \bla 
%
%There is a well-established hierarchy in deterministic classical dynamical systems, with Bernoulli systems at its apex. 
%They are not just mixing, but are isomorphic to random processes such as a coin-toss. 
%There exists a partition, the generating partition, of the phase space such that the correlations built from the characteristic function on these, decay instantaneously and hence they are always independent.  The quantum analogs of such Bernoulli systems can be constructed from the dual circuits if the all the nontrivial eigenvalues  $\lambda_k, \; k >0$ vanishes. 
\begin{description}
\item[Bernoulli] All $2(q^2-1)$ nontrivial eigenvalues vanish: $\lambda_k=0$.
\end{description}
Bernoulli implies mixing, but not the other way around, and we will motivate below that this is
indeed a unique class by itself and lies at the apex of a possible quantum ergodic hierarchy.

\subsection{Bernoulli circuits, entangling power and 2-unitarity \label{sec:Bernoulli}} 

To begin, note that the correlations on the lightcone vanishes  if Eq. (\ref{eq:mpdia}) and (\ref{eq:mmdia}) reduces as follows 
\begin{equation*}
\mathcal{M}_{+}(a) =  \frac{1}{q}
\begin{tikzpicture}[baseline=(current  bounding  box.center),scale = 0.4]
\draw [thick] (1,2) to[out =110, in = -110] (1,3);
\draw [thick] (2,2) to[out=70, in=-70] (2,3);
\draw [thick] (0.5,0.5) -- (1,1);
\draw[thick, fill=black] (0.9,2.5) circle (0.1cm);

\draw [thick] (0.5,0.5) to[out=110, in=-110] (0.5,4.5);

\draw [thick] (1,4) -- (0.5,4.5);
\draw [thick] (2,4) -- (2.5,4.5);

\draw [thick] (1,1) -- (0.5,0.5);
\draw [thick] (2,1) -- (2.5,0.5);
\Text[x=0.6,y=2.6]{$a$} 

\draw [thick] (1,2) -- (2,2) ;
\draw [thick] (1,1) -- (2,1) ;
\draw [thick] (1,3) -- (2,3) ;
\draw [thick] (1,4) -- (2,4) ;
\end{tikzpicture}
= \frac{\tr(a)}{q} \mathbb{1}_q =  \;\;
\mathcal{M}_{-}(a) = \frac{1}{q}
\begin{tikzpicture}[baseline=(current  bounding  box.center),scale = 0.4]
\draw [thick] (1,2) to[out =110, in = -110] (1,3);
\draw [thick] (2,2) to[out=70, in=-70] (2,3);
\draw [thick] (0.5,0.5) -- (1,1);
\draw [thick] (2.5,0.5) to[out=70, in=-70] (2.5,4.5);

\draw [thick] (1,4) -- (0.5,4.5);
\draw [thick] (2,4) -- (2.5,4.5);

\draw [thick] (1,1) -- (0.5,0.5);
\draw [thick] (2,1) -- (2.5,0.5);

\draw[thick, fill=black] (2.1,2.5) circle (0.1cm);
\Text[x=2.5,y=2.7]{$a$}

\draw [thick] (1,2) -- (2,2) ;
\draw [thick] (1,1) -- (2,1) ;
\draw [thick] (1,3) -- (2,3) ;
\draw [thick] (1,4) -- (2,4) ;

\end{tikzpicture}, 
\label{eq:mmdia2}
\end{equation*} 
as $\tr(a)=0$.
This motivates the introduction of ``T-duality" based on the $T_2$ partial transpose, see Eq.~(\ref{eq:T1T2R2}). Although we do not imply
any evident duality as such, it is convenient to refer to this in parity to the definition 
of dual-unitarity.
\begin{definition}[T-dual unitary] \label{def:Tdual} 
If $U$ and $U^{T_2}$ are both unitary, $U$ is T-dual.
\end{definition}
Diagrammatically, 
\begin{equation}  
\begin{tikzpicture}[baseline=(current  bounding box.center),scale = 0.5]
\draw [thick,fill=teal,rounded corners=2.2pt] (1,1) rectangle (2,2); 
\draw [thick] (1,2) -- (0.5,2.3);
\draw [thick] (2,2) to[out=70, in=-70] (2,3);
\draw [thick] (0.5,0.5) -- (1,1);
\draw [thick] (1,3) -- (0.5,2.7);
\draw [thick] (0.5,0.5) to[out=110, in=-110] (0.5,4.5);

\draw [thick,fill=purple,rounded corners=2.2pt] (1,3) rectangle (2,4);
\draw [thick] (1,4) -- (0.5,4.5);
\draw [thick] (2,4) -- (2.5,4.5);

\draw [thick] (1,1) -- (0.5,0.5);
\draw [thick] (2,1) -- (2.5,0.5);
\etp = 
\begin{tikzpicture}[baseline=(current  bounding box.center),scale = 0.5]
\draw [thick] (1,2) -- (0.5,2.3);
\draw [thick] (2,2) to[out=70, in=-70] (2,3);
\draw [thick] (0.5,0.5) -- (1,1);
\draw [thick] (1,3) -- (0.5,2.7);
\draw [thick] (0.5,0.5) to[out=110, in=-110] (0.5,4.5);

\draw [thick] (1,4) -- (0.5,4.5);
\draw [thick] (2,4) -- (2.5,4.5);

\draw [thick] (1,1) -- (0.5,0.5);
\draw [thick] (2,1) -- (2.5,0.5);

\draw [thick] (1,2) -- (2,2) ;
\draw [thick] (1,1) -- (2,1) ;
\draw [thick] (1,3) -- (2,3) ;
\draw [thick] (1,4) -- (2,4) ;

\etp, \;\;\;\; 
\begin{tikzpicture}[baseline=(current  bounding box.center),scale = 0.5]
\draw [thick,fill=teal,rounded corners=2.2pt] (1,1) rectangle (2,2); 
\draw [thick] (0.5,0.5) -- (1,1);
\draw [thick] (2.5,0.5) to[out=70, in=-70] (2.5,4.5);
\draw [thick] (2,3) -- (2.3,2.7);
\draw [thick] (2,2) -- (2.3,2.3);
\draw [thick] (1,2) to[out=110, in=-110] (1,3);

\draw [thick,fill=purple,rounded corners=2.2pt] (1,3) rectangle (2,4);
\draw [thick] (1,4) -- (0.5,4.5);
\draw [thick] (2,4) -- (2.5,4.5);

\draw [thick] (1,1) -- (0.5,0.5);
\draw [thick] (2,1) -- (2.5,0.5);
\end{tikzpicture} = 
\begin{tikzpicture}[baseline=(current  bounding box.center),scale = 0.5] 
\draw [thick] (0.5,0.5) -- (1,1);
\draw [thick] (2.5,0.5) to[out=70, in=-70] (2.5,4.5);
\draw [thick] (2,3) -- (2.3,2.7);
\draw [thick] (2,2) -- (2.3,2.3);
\draw [thick] (1,2) to[out=110, in=-110] (1,3);

\draw [thick] (1,4) -- (0.5,4.5);
\draw [thick] (2,4) -- (2.5,4.5);

\draw [thick] (1,1) -- (0.5,0.5);
\draw [thick] (2,1) -- (2.5,0.5);

\draw [thick] (1,2) -- (2,2) ;
\draw [thick] (1,1) -- (2,1) ;
\draw [thick] (1,3) -- (2,3) ;
\draw [thick] (1,4) -- (2,4) ;
\end{tikzpicture}.
\label{eq:tdual2}
\end{equation}

Using $U^{T_2}$, similar to defining $E(U)$ in Eq. (\ref{eq:EUdefn}) it can be shown \cite{Bhargavi2019} that 
\begin{equation}
	E(US) = 1 - \frac{1}{q^4} \tr[\left(U^{T_2}U^{T_2\dagger}\right)^2].
	\label{eq:EUS}
\end{equation}
$E(US)$ is maximum iff $U^{T_2}$ is also unitary, that is $U$ is T-dual. In the context of the 4-party state in Eq.~(\ref{eq:4party}), 
$\rho_{14}=U^{T_2}U^{T_2\dagger}/q^2$ is the state of the $14|32$ partition and $E(US)$ is the linear entropy of entanglement of this
bipartite split.

The quantity that in a way brings together measures of duality and T-duality is the entangling power $e_p(U)$.
This is defined as the average entanglement produced by the action of a unitary operator $U$ on an ensemble of  pure product states distributed according to the Haar measure, 
\begin{equation}
	e_p(U) = C_q\, \overline{\mathcal{E}(U \ket{\phi_1} \otimes \ket{\phi_2})}^{\ket{\phi_1},\ket{\phi_2}},
	\nonumber 
\end{equation}
where $\mathcal{E}(\cdot)$ can be any entanglement measure and $C_q$ is  scaling factor. By considering the linear entropy as an entanglement measure, the entangling power $e_p(U)$ of $U$  has been related to the operator entanglements $E(U)$ and $E(US)$ \cite{Zanardi2001}. The normalized entangling power $0\leq e_p(U) \leq 1$, with $C_q=(q+1)/(q-1)$ is then:
\begin{equation}
	e_p(U) = \frac{1}{E(S)}  \left[ E(U) + E(US) - E(S)\right],
	\label{eq:ep}
\end{equation}
where $E(S) = \frac{q^2-1}{q^2}$ has already been introduced. Note that this choice of scaling is different from that used in \cite{Zanardi2001}, but is convenient to have the maximum value be $1$.
The entangling power $e_p(U)$  is maximum {\it iff} both $U^{R_2}$ and $U^{T_2}$ of $U$ is unitary. Thus {\it 2-unitary operators} have been previously defined \cite{Goyeneche2015}.
\begin{definition}[2-unitary] \label{def:2-uni}
	If  $U$ is dual-unitary  and T-dual then it is 2-unitary. 
\end{definition}

Note that if $U$ and $U^{R_1}$ is unitary then $U^{R_2}$ is also unitary and vice versa. Similarly, for T-dual property, if $U$ and $U^{T_1}$ is unitary then $U^{T_2}$ is also unitary and vice versa. Hence there is no need to distinguish between the two realignments for defining the dual property, or to differentiate the two partial transpositions in defining the T-dual property. 
The two classes of dual-unitary  operators are connected by the swap operator. In particular
if $U$ is $\text{T-dual}$ (dual), then $US$ and $SU$ are both dual ($\text{T-dual}$) unitary operators.
This follows as from Eq.~(\ref{eq:IdentityT1R2}) $(SA)^{R_2}=S A^{T_1}$, hence if $A$ is T-dual, $SA$ is dual-unitary, as the swap is unitary. Similarly, $AS$ is also dual-unitary . As simple examples, the identity is T-dual, and  swap $S$ is dual, {\sc cnot} is T-dual and {\sc dcnot}$=$ $S$ $\times$ {\sc cnot} is dual. The operator entanglements $E(U), E(US)$ and the entangling power $e_p(U)$  are all local unitary invariants (LUI).

\begin{prop}
A dual-unitary   quantum circuit is Bernoulli iff $U$ is 2-unitary.
\end{prop}
If $U$ is 2-unitary, then $\utb$ is also unitary and hence from Eq.~(\ref{eq:M+UT})
\beq
\label{eq:M+2uni1}
M_+[U]= \frac{1}{q}\mathbb{1}_{q^2}^{R_1} = |\Phi^+\kt \br \Phi^+|. 
\eeq
Hence the nontrivial part of the spectrum of $M_+[U]$ vanishes, that is $\lambda_k=0$ for $k>0$,
and the correlations decay instantaneously. Local unitary transformations do not alter the situation as 
2-unitarity is LUI and hence $M_+[U']=M_+[U]$. Note that in this case the spectrum of $M_-[U]$, from Eq.~(\ref{eq:M-UT}) is 
also identical with the only non-zero eigenvalue corresponding to the trivial mode. Bernoulli circuits are not ``chiral".

Now assuming that $\lambda_k=0$ for $k>0$, we show that $U^{T_2}$ is unitary. The condition on the eigenvalues of $M_+[U]$ implies that it is rank 1, and hence we can assume a form 
$M_+[U]=|\phi_2\kt \br \phi_1|$, where $|\phi_2\kt$ is the (right) eigenvector
corresponding to the eigenvalue $1$ if $\br \phi_1|\phi_2 \kt =1$. However, this is the trivial eigenvalue and the corresponding normalized eigenvector from the fact that the map is unital is $|\mathbb{1}_q\kt/\sqrt{q}$, where $|\mathbb{1}_q \kt=|\Phi^+\kt$. Considering the left eigenvector corresponding to the eigenvalue $1$, we conclude that $|\phi_1\kt$ is  the same as well. Hence $M_+[U]=|\Phi_+\kt \br \Phi_+|=\mathbb{1}^{R_1}/q$ and we get
from Eq.~(\ref{eq:M+UT}) that $U^{T_2}U^{T_2 \dagger}=\mathbb{1}_{q^2}$. Similar considerations of $M_-[U]$, assuming that its only nonzero eigenvalue is the trivial one, leads to $\utbd \utb=\mathbb{1}_{q^2}$ and hence $U^{T_2}$ is unitary as was to be shown. As $U$ is assumed to be dual, it is in fact 2-unitary. 

If $U$ is 2-unitary the 4-party state in Eq.~(\ref{eq:4party})
is such that all the bipartite splits are maximally entangled, in other words it is an absolutely maximally entangled state
(AME) \cite{HCLRL12,Goyeneche2015}. It is denoted as $\text{AME}(4,q)$ to indicate that there are $4$ particles each of $q$ dimensions. It is well-known
that there are no $\text{AME}(4,2)$ states: 4 qubits cannot form an AME state \cite{Higuchi2000}. This implies that there are no dual-unitary circuits of qubits that are Bernoulli. It is also known that $\text{AME}(4,q)$ exists for all $q$,  with the possible exception of $q=6$. This has been 
demonstrated explicitly by constructing such 2-unitaries as permutations based on orthogonal Latin squares \cite{Clarisse2005}. We will
return to this when constructing dual operators from permutations further below.

%\subsection{Extended observables in Bernoulli circuits \label{sec:Bernoulli}}

The classification of a circuit based on the maps $\mathcal{M}_{\pm}$ is valid for single-particle observables, but may not
be in the same class for observables with larger support over a few particles, or ``extended observables". While a
study of such correlations in dual-unitary  circuits is, to our knowledge, absent, at least in some limited way Bernoulli circuits seem straightforward
examples where correlations of two-particle observables continue to vanish. 
To illustrate this, the dynamical correlation function between the operators defined on the two sites is shown in Eq.~(\ref{eq:Bern1}).
As any two-particle operator can be written as a linear combination of such product operators $a_j^{x_1} \otimes a_k^{x_2}$, it is sufficient
to consider the correlations from such pairs. Again, as for single particle, we adopt an orthonormal basis with identity as one of them 
and consider the correlation function
\begin{equation}
C^{ijkl} (x_1,x_2,t) = \frac{1}{q^{2L}} \tr \left[ a^{x_1}_ka^{x_2}_l \mathbf{U}^{-t} a^{y_1}_i a^{y_2}_j \mathbf{U}^{t} \right],
\label{eq:corext}  
\end{equation}
with $y_1 = 0, y_2 = 1/2$. Thus $C^{ijkl} (x_1,x_2,t)=$ 
\begin{equation}
\label{eq:Bern1} 
\frac{1}{q^{4t}}
\btp [baseline=(current  bounding  box.center),scale=0.3] 
\foreach \i in {0,...,4} %right v shaped 
{
\draw [thick,fill=purple,rounded corners=2.2pt] (2*\i,1+2*\i) rectangle (1+2*\i,2+2*\i);
}

\foreach \i in {0,...,3} %left v shaped 
{
\draw [thick,fill=purple,rounded corners=2.2pt] (-2-2*\i,3 + 2*\i) rectangle (-1-2*\i,4 + 2*\i);
}

\foreach \i in {0,...,2} %left v shaped 
{
\draw [thick,fill=purple,rounded corners=2.2pt] (2*\i,5 + 2*\i) rectangle (1+2*\i,6 + 2*\i);
}

\foreach \i in {0,...,1} %left v shaped 
{
\draw [thick,fill=purple,rounded corners=2.2pt] (-2-2*\i,7 + 2*\i) rectangle (-1-2*\i,8 + 2*\i);
}

\draw [thick,fill=purple,rounded corners=2.2pt] (0,9) rectangle (1,10);

%lower figure 
\foreach \i in {0,...,4} %right v shaped 
{
\draw [thick,fill=teal,rounded corners=2.2pt] (2*\i,-1-2*\i) rectangle (1+2*\i,-2*\i);
}

\foreach \i in {0,...,3} %left v shaped 
{
\draw [thick,fill=teal,rounded corners=2.2pt] (-2-2*\i,-3 - 2*\i) rectangle (-1-2*\i,-2 - 2*\i);
}

\foreach \i in {0,...,2} %left v shaped 
{
\draw [thick,fill=teal,rounded corners=2.2pt] (2*\i,-5 - 2*\i) rectangle (1+2*\i,-4 - 2*\i);
}

\foreach \i in {0,...,1} %left v shaped 
{
\draw [thick,fill=teal,rounded corners=2.2pt] (-2-2*\i,-7 - 2*\i) rectangle (-1-2*\i,-6 - 2*\i);
}

\draw [thick,fill=teal,rounded corners=2.2pt] (0,-9) rectangle (1,-8);

%lines now 
\foreach \i in {0,...,3}
{
\draw [thick] (3+2*\i, -2 -2*\i) -- (3.5 + 2*\i, -1.5 - 2*\i) -- (3.5+2*\i,2.5+2*\i) -- (3+2*\i,3+2*\i);
\draw [thick] (-2-2*\i, -2 -2*\i) -- (-2.5 - 2*\i, -1.5 - 2*\i) -- (-2.5-2*\i,2.5+2*\i) -- (-2-2*\i,3+2*\i);
\draw [thick] (1+2*\i, 2+2*\i) -- (2+2*\i,3+2*\i);
\draw [thick] (1+2*\i, -1-2*\i) -- (2+2*\i,-2-2*\i);
\draw [thick] (-2*\i, 2+2*\i) -- (-1-2*\i,3+2*\i);
\draw [thick] (-2*\i, -1-2*\i) -- (-1-2*\i,-2-2*\i);
}

\foreach \i in {0,...,2}
{
\draw [thick] (2 + 2*\i, 4 +2*\i) -- (1+2*\i,5+2*\i);
\draw [thick] (-1 - 2*\i, 4 +2*\i) -- (-2*\i,5+2*\i);
\draw [thick] (2+2*\i, -3 -2*\i) -- (1+2*\i,-4-2*\i);
\draw [thick] (-1-2*\i, -3 -2*\i) -- (-2*\i,-4-2*\i); 
}

\foreach \i in {0,...,1}
{
\draw [thick] (1 + 2*\i, 6 +2*\i) -- (2+2*\i,7+2*\i);
\draw [thick] ( -2*\i, 6 +2*\i) -- (-1-2*\i,7+2*\i);
\draw [thick] (1 + 2*\i, -5 -2*\i) -- (2+2*\i,-6-2*\i);
\draw [thick] ( -2*\i, -5 -2*\i) -- (-1-2*\i,-6-2*\i);
}

\draw [thick] (2,8) -- (1,9);
\draw [thick] (-1,8) -- (0,9); 
\draw [thick] (2,-7) -- (1,-8);
\draw [thick] (-1,-7) -- (0,-8); 

\foreach \i in {0,...,4}
{
\draw [thick] (-8+4*\i,10) -- (-8.5+4*\i,10.5);
\draw [thick] (-7+4*\i,10) -- (-6.5+4*\i,10.5);
\draw [thick] (-8+4*\i,-9) -- (-8.5+4*\i,-9.5);
\draw [thick] (-7+4*\i,-9) -- (-6.5+4*\i,-9.5);
}

\draw [thick] (0,0) to[out =110, in = -110] (0,1);
\draw [thick] (1,0) to[out=70, in=-70] (1,1); 

\draw[thick, fill=black] (-0.1,0.5) circle (0.1cm);
\draw[thick, fill=black] (1.1,0.5) circle (0.1cm);
\Text[x=-0.8,y=0.5]{$a_i$}
\Text[x=1.8,y=0.5]{$a_j$}

\draw[thick, fill=black] (3.5,10.5) circle (0.1cm);
\draw[thick, fill=black] (5.5,10.5) circle (0.1cm);
\Text[x=3.5,y=11.2]{$a_k$}
\Text[x=5.5,y=11.2]{$a_l$}

\etp. 
\end{equation} 
By using the duality conditions, Eq.~(\ref{eq:dual2}), the correlations vanish inside the lightcone and is shown in the Appendix~(\ref{app:extende}). 

For the correlations along the lightcone, $x_1 = t-\frac{1}{2}, x_2 = t$, the repeated application of unitarity from Eq. (\ref{eq:unidia1}),  gives
$C^{ijkl} (x_1,x_2,t) =$ 
\begin{equation} 
 \frac{1}{q^{4t}}
\btp [baseline=(current  bounding  box.center),scale=0.28] 
\foreach \i in {0,...,4} %right v shaped 
{
\draw [thick,fill=purple,rounded corners=2.2pt] (2*\i,1+2*\i) rectangle (1+2*\i,2+2*\i);
}

%lower figure 
\foreach \i in {0,...,4} %right v shaped 
{
\draw [thick,fill=teal,rounded corners=2.2pt] (2*\i,-1-2*\i) rectangle (1+2*\i,-2*\i);
}

%lines now 
\foreach \i in {0,...,3}
{
\draw [thick] (3+2*\i, -2 -2*\i) -- (3.5 + 2*\i, -1.5 - 2*\i) -- (3.5+2*\i,2.5+2*\i) -- (3+2*\i,3+2*\i);
\draw [thick] (-2-2*\i, -2 -2*\i) -- (-2.5 - 2*\i, -1.5 - 2*\i) -- (-2.5-2*\i,2.5+2*\i) -- (-2-2*\i,3+2*\i);
\draw [thick] (1+2*\i, 2+2*\i) -- (2+2*\i,3+2*\i);
\draw [thick] (1+2*\i, -1-2*\i) -- (2+2*\i,-2-2*\i);
\draw [thick] (-2*\i, 2+2*\i) -- (-1-2*\i,3+2*\i);
\draw [thick] (-2*\i, -1-2*\i) -- (-1-2*\i,-2-2*\i);
}

\foreach \i in {0,...,2}
{
\draw [thick] (2 + 2*\i, 4 +2*\i) -- (1+2*\i,5+2*\i);
\draw [thick] (-1 - 2*\i, 4 +2*\i) -- (-2*\i,5+2*\i);
\draw [thick] (2+2*\i, -3 -2*\i) -- (1+2*\i,-4-2*\i);
\draw [thick] (-1-2*\i, -3 -2*\i) -- (-2*\i,-4-2*\i); 
}

\foreach \i in {0,...,1}
{
\draw [thick] (1 + 2*\i, 6 +2*\i) -- (2+2*\i,7+2*\i);
\draw [thick] ( -2*\i, 6 +2*\i) -- (-1-2*\i,7+2*\i);
\draw [thick] (1 + 2*\i, -5 -2*\i) -- (2+2*\i,-6-2*\i);
\draw [thick] ( -2*\i, -5 -2*\i) -- (-1-2*\i,-6-2*\i);
}

\draw [thick] (2,8) -- (1,9);
\draw [thick] (-1,8) -- (0,9); 
\draw [thick] (2,-7) -- (1,-8);
\draw [thick] (-1,-7) -- (0,-8); 

\foreach \i in {0,...,4}
{
\draw [thick] (-8+4*\i,10) -- (-8.5+4*\i,10.5);
\draw [thick] (-7+4*\i,10) -- (-6.5+4*\i,10.5);
\draw [thick] (-8+4*\i,-9) -- (-8.5+4*\i,-9.5);
\draw [thick] (-7+4*\i,-9) -- (-6.5+4*\i,-9.5);
}

\draw [thick] (0,0) to[out =110, in = -110] (0,1);
\draw [thick] (1,0) to[out=70, in=-70] (1,1);

\draw[thick, fill=black] (-0.1,0.5) circle (0.1cm);
\draw[thick, fill=black] (1.1,0.5) circle (0.1cm);
\Text[x=-0.8,y=0.5]{$a_i$}
\Text[x=1.8,y=0.5]{$a_j$}

\foreach \i in {0,...,3}
{
\draw [thick] (-8+4*\i,9) -- (-8+4*\i,10); 
\draw [thick] (-7+4*\i,9) -- (-7+4*\i,10);
\draw [thick] (-8+4*\i,-8) -- (-8+4*\i,-9); 
\draw [thick] (-7+4*\i,-8) -- (-7+4*\i,-9);
}

\foreach \i in {0,...,2}
{
\draw [thick] (-6+4*\i,7) -- (-6+4*\i,8); 
\draw [thick] (-5+4*\i,7) -- (-5+4*\i,8);
\draw [thick] (-6+4*\i,-6) -- (-6+4*\i,-7);
\draw [thick] (-5+4*\i,-6) -- (-5+4*\i,-7);
}

\foreach \i in {0,...,1}
{
\draw [thick] (-4+4*\i,5) -- (-4+4*\i,6); 
\draw [thick] (-3+4*\i,5) -- (-3+4*\i,6);
\draw [thick] (-4+4*\i,-4) -- (-4+4*\i,-5); 
\draw [thick] (-3+4*\i,-4) -- (-3+4*\i,-5);
}

\draw [thick] (-2,3) -- (-2,4);
\draw [thick] (-1,3) -- (-1,4);
\draw [thick] (-2,-2) -- (-2,-3);
\draw [thick] (-1,-2) -- (-1,-3);

\draw[thick, fill=black] (7.5,10.5) circle (0.1cm);
\draw[thick, fill=black] (9.5,10.5) circle (0.1cm);
\Text[x=7.5,y=11.2]{$a_k$}
\Text[x=9.5,y=11.2]{$a_l$}

\draw [ultra thick,rounded corners=2.2pt] (6,7) rectangle (7,8); 
\draw [ultra thick,rounded corners=2.2pt] (6,-7) rectangle (7,-6);
\draw [ultra thick, red] (5.5,10.5) -- (5,10) -- (5,9) -- (6,8); 
\draw [ultra thick, red] (5.5,-9.5) -- (5,-9) -- (5,-8) -- (6,-7);
\draw [ultra thick, blue] (7,7) -- (7.5,6.5) -- (7.5,-5.5) -- (7,-6);

\etp. 
\label{eq:Bern_circ}
\end{equation} 
Applying the  T-duality condition Eq.~(\ref{eq:tdual2}) to the highlighted part in Eq.~(\ref{eq:Bern_circ})
and repeating its application, the vanishing of the correlation function is evident, as $C^{ijkl} (x_1,x_2,t) =$ 
\begin{equation} 
\frac{1}{q^{4t}}
\btp [baseline=(current  bounding  box.center),scale=0.28] 

%lines now 
\foreach \i in {0,...,3}
{
\draw [thick] (3+2*\i, -2 -2*\i) -- (3.5 + 2*\i, -1.5 - 2*\i) -- (3.5+2*\i,2.5+2*\i) -- (3+2*\i,3+2*\i);
\draw [thick] (-2-2*\i, -2 -2*\i) -- (-2.5 - 2*\i, -1.5 - 2*\i) -- (-2.5-2*\i,2.5+2*\i) -- (-2-2*\i,3+2*\i);
\draw [thick] (1+2*\i, 2+2*\i) -- (2+2*\i,3+2*\i);
\draw [thick] (1+2*\i, -1-2*\i) -- (2+2*\i,-2-2*\i);
\draw [thick] (-2*\i, 2+2*\i) -- (-1-2*\i,3+2*\i);
\draw [thick] (-2*\i, -1-2*\i) -- (-1-2*\i,-2-2*\i);
}

\foreach \i in {0,...,2}
{
\draw [thick] (2 + 2*\i, 4 +2*\i) -- (1+2*\i,5+2*\i);
\draw [thick] (-1 - 2*\i, 4 +2*\i) -- (-2*\i,5+2*\i);
\draw [thick] (2+2*\i, -3 -2*\i) -- (1+2*\i,-4-2*\i);
\draw [thick] (-1-2*\i, -3 -2*\i) -- (-2*\i,-4-2*\i); 
}

\foreach \i in {0,...,1}
{
\draw [thick] (1 + 2*\i, 6 +2*\i) -- (2+2*\i,7+2*\i);
\draw [thick] ( -2*\i, 6 +2*\i) -- (-1-2*\i,7+2*\i);
\draw [thick] (1 + 2*\i, -5 -2*\i) -- (2+2*\i,-6-2*\i);
\draw [thick] ( -2*\i, -5 -2*\i) -- (-1-2*\i,-6-2*\i);
}

\draw [thick] (2,8) -- (1,9);
\draw [thick] (-1,8) -- (0,9); 
\draw [thick] (2,-7) -- (1,-8);
\draw [thick] (-1,-7) -- (0,-8); 

\foreach \i in {0,...,4}
{
\draw [thick] (-8+4*\i,10) -- (-8.5+4*\i,10.5);
\draw [thick] (-7+4*\i,10) -- (-6.5+4*\i,10.5);
\draw [thick] (-8+4*\i,-9) -- (-8.5+4*\i,-9.5);
\draw [thick] (-7+4*\i,-9) -- (-6.5+4*\i,-9.5);
}

\draw [thick] (0,0) to[out =110, in = -110] (0,1);
\draw [thick] (1,0) to[out=70, in=-70] (1,1); 

\draw [thick,fill=purple,rounded corners=2.2pt] (0,1) rectangle (1,2);
\draw [thick,fill=teal,rounded corners=2.2pt] (0,-1) rectangle (1,0);

\draw[thick, fill=black] (-0.1,0.5) circle (0.1cm);
\draw[thick, fill=black] (1.1,0.5) circle (0.1cm);
\Text[x=-0.8,y=0.5]{$a_i$}
\Text[x=1.8,y=0.5]{$a_j$}

\foreach \i in {0,...,3}
{
\draw [thick] (-8+4*\i,9) -- (-8+4*\i,10); 
\draw [thick] (-7+4*\i,9) -- (-7+4*\i,10);
\draw [thick] (-8+4*\i,-8) -- (-8+4*\i,-9); 
\draw [thick] (-7+4*\i,-8) -- (-7+4*\i,-9);
}

\foreach \i in {0,...,2}
{
\draw [thick] (-6+4*\i,7) -- (-6+4*\i,8); 
\draw [thick] (-5+4*\i,7) -- (-5+4*\i,8);
\draw [thick] (-6+4*\i,-6) -- (-6+4*\i,-7);
\draw [thick] (-5+4*\i,-6) -- (-5+4*\i,-7);
}

\foreach \i in {0,...,1}
{
\draw [thick] (-4+4*\i,5) -- (-4+4*\i,6); 
\draw [thick] (-3+4*\i,5) -- (-3+4*\i,6);
\draw [thick] (-4+4*\i,-4) -- (-4+4*\i,-5); 
\draw [thick] (-3+4*\i,-4) -- (-3+4*\i,-5);
}

\draw [thick] (-2,3) -- (-2,4);
\draw [thick] (-1,3) -- (-1,4);
\draw [thick] (-2,-2) -- (-2,-3);
\draw [thick] (-1,-2) -- (-1,-3);

\draw[thick, fill=black] (7.5,10.5) circle (0.1cm);
\draw[thick, fill=black] (9.5,10.5) circle (0.1cm);
\Text[x=7.5,y=11.2]{$a_k$}
\Text[x=9.5,y=11.2]{$a_l$}

\foreach \i in {0,...,3}
{
\draw [thick] (2+2*\i,3+2*\i) -- (3+2*\i,3+2*\i);
\draw [thick] (2+2*\i,4+2*\i) -- (3+2*\i,4+2*\i);
\draw [thick] (2+2*\i,-2-2*\i) -- (3+2*\i,-2-2*\i);
\draw [thick] (2+2*\i,-3-2*\i) -- (3+2*\i,-3-2*\i);
}

\draw [thick,fill=purple,rounded corners=2.2pt] (8,9) rectangle (9,10);
\draw [thick,fill=teal,rounded corners=2.2pt] (8,-9) rectangle (9,-8);

\etp .  
\end{equation} 
The vanishing correlation function for the case $x_1 = x_2 = t-\frac{1}{2}$ is similar and shown in Appendix (\ref{app:extende}). 

More general cases of extendable observables need to be studied, but we do not expect them to vanish in general. This is also
in line with classical Bernoulli systems where the Bernoulli partition is singled out for independence \cite{Ergodic_Hierarchy}. 
There exists unitaries even with $q=2$ such that $M^k_+[U] =|\Phi^+\kt \br \Phi^+|$ for some $k>1$, so that all single-particle  correlations decay to zero beyond $t=k$. However, unlike the Bernoulli case, local transformations can change the nature of this correlation and it need not continue to hold even for two-site observables. 
We will construct examples of Bernoulli and such cases using cat maps, among other possibilities. Finally we observe that if $U$ is not 
a dual-unitary  operator, but is T-dual alone, the correlations along the lightcone still vanish for $t \neq 0$.

\section{The spectral gap, mixing rate, and entangling power} 
\label{sec:MixingRateAnalytics}

\subsection{Rates, and a sufficiency condition for  dual circuits to be mixing\label{sec:mixmeas}}

To separate the trivial eigenvalue so that the largest eigenvalue is a nontrivial one, we define
\beq
\label{eq:MplusTilde}
\tilde{M}_+[U]=M_+[U]-|\Phi^+\kt \br \Phi^+|.
\eeq
 The eigenvalues
of $\tilde{M}_+[U]$ are the same as that of $M_+[U]$, except that the trivial eigenvalue
is replaced by $0$. This follows as $\br \Phi^+|$ is the left eigenvector with 
eigenvalue $1$ as may also be verified directly using the form Eq.~(\ref{eq:M+UT}) for example.
Therefore we can peel off the projector on the maximally entangled state $|\Phi^+\kt \br \Phi^+|$ from
a spectral or Jordan decomposition of $M_+[U]$. The remaining states are orthogonal or can be chosen 
to be orthogonal to $|\Phi^+\kt$ and  $\tilde{M}_+[U]||\Phi^+\kt=0$.

%For, let $M_+[U]|\lambda\kt =\lambda|\lambda\kt$ be an eigenmode of
%eigenvalue $\lambda$, and let $|\lambda\kt = \sum_{j=0}^{q^2-1} c_j |a_j\kt$, where $|a_j \kt$
%is the vectorization of the orthogonal single-particle operator basis with $a_0=\mathbb{1}_q$,
%so that $|a_0\kt =\sqrt{q}|\Phi^+\kt$.
%Then $M_+[U]|\lambda\kt =c_0 \sqrt{q} |\Phi^+\kt+\sum_{j=1}^{q^2-1} c_j M_+[U]|a_j\kt=c_0 \lambda \sqrt{q} |\Phi^+\kt+\lambda \sum_{j=1}^{q^2-1} c_j |a_j\kt $. Taking an inner-product with $|\Phi^+\kt$ and observing that $\br \Phi^+|M_+[U]=\br \Phi^+|$, that is the left eigenvector of the trivial eigenvalue
%is also the adjoint of the right eigenvector, we conclude that $c_0=0$ unless $\lambda=1$, as $\sqrt{q}\, \br \Phi^+|a_j\kt=\tr(a_j)=0$ for $j \neq 0$.
%Thus it follows that $\tilde{M}_+[U]|\lambda\kt=\lambda|\lambda\kt$ for $\lambda \neq 1$. If $\lambda=1$ is not the trivial eigenvalue, but is part
%of a degenerate manifold, this can be spanned by an orthogonal set with $|\Phi^+\kt$ as one of the states, and hence $|\lambda\kt$ is also an eigenvector of $\tilde{M}_+[U]$ with eigenvalue $1$.

Focusing on $M_+[U]$, the non-trivial eigenvalues determine the decay rates
\begin{equation} 
\mu_k=-\ln|\lambda_k|, \; k\geq 1.
\label{eq:decay}
\end{equation}
 If any $|\lambda_k|=1$, these correspond to non-decaying modes,   on the other hand there is instant mixing along a mode if the
corresponding $\lambda_k=0$. Note that $|\lambda_k|\leq 1$ and hence $\mu_k \geq 0$.  
Arranging the decay rates in an increasing order $\mu_1 \leq \cdots \leq \mu_{q^2-1}$, ignoring the trivial mode as $\mu_0=0$, we may define $\mu_1$ as a degree of mixing. It is derived from the largest magnitude eigenvalue (the spectral radius $|\lambda_1|$) of $\tilde{M}_+[U]$
and vanishes when any one of the non-trivial modes does not decay.
This is infinite if all the $\mu_k=\infty$ for $k>0$,
which is the case if the circuit is Bernoulli. The converse is also true: if $\mu_1=\infty$, then the circuit is Bernoulli. 
For the most part, we will concentrate on estimating the largest nontrivial eigenvalues $|\lambda_1|$ which controls these rates
as well as the spectral gap $1-|\lambda_1|$.

If the local transformed unitary $U^\prime = (u_1 \otimes u_2 ) U (v_1 \otimes v_2 )$
is, as in Eq.~(\ref{eq:LU}), 
using Eq.~(\ref{eq:LocalIdentities}) we get
\beq
\label{eq:UT2p}
U'^{T_2} U'^{T_2 \dagger}= (u_1\otimes v_2^T)\utb \utbd (u_1\otimes v_2^T)^{\dagger}. 
\eeq
%and therefore $M_+[U]$ (and hence $\nu_+[U]$) does not depend on $u_2$ and $v_1$. It is easy to see that $M_{-}[U]$ does not depend on 
%$u_1$ and $v_2$. 
This proves the earlier observation that the singular values of $\utb$ and the eigenvalues of $\utb \utbd$, are local unitary invariant. 
However their realignment, which results in  the CPTP map (\ref{eq:mpdia}), does not have this property, and hence the decay rates depend on the locals used. 
If $M_+[U]=(\utb \utbd)^{R_1}/q$ is modified to $M_+[U']=(\utbp \utbpd)^{R_1}/q$, on using a further identity in Eq.~(\ref{eq:LocalIdentities}) results in 
\beq
\label{eq:Mprime}
M_+[U']=(v_2^{\dagger} \otimes v_2^T)M_+[U](u_1^{\dagger} \otimes u_1^T).
\eeq
Therefore the eigenvalues of $M_+[U]$ are not LUI. We note that the eigenmode corresponding to the trivial eigenvalue of $1$ remains
$|\Phi^+\kt$ as $(u \otimes u^*)|\Phi^+\kt=|\Phi^+\kt$, where $u$ is any local unitary and $u^*$ is its element-wise complex conjugate, which follows
from the vectorization of the identity $u u^{\dagger}=\mathbb{1}$, as $|ABC\kt =(A\otimes C^T)|B\kt$.

%Note though that the singular values of $M_+[U]$, which are (square of) the eigenvalues of $M_+[U]M_+[U]^{\dagger}$  are LUI. 

%and may therefore be correlated with the entangling power which is maximized by such unitaries.
%\red{AL:Mixing rates of Markov chains are so defined, this makes sense I think, has this been studied before in the dual-unitary  quantum circuit literature?}

As $\mu_1$ is dependent on the local operators, we may define the maximal mixing rate associated with $U$ as 
\beq
\label{eq:MixingRate}
\nu_+[U] = \text{max}_{u_i,v_i} \mu_1[ (u_1\otimes u_2) U (v_1\otimes v_2)],
\eeq
which is the maximum possible mixing rate under the action of arbitrary local or single-particle unitaries. A similar maximal mixing rate can also be defined from $M_{-}[U]$.
%The assumption that this exists is dependent on the diagonalizability of the CPTP map. When this is not the case, the $\nu_+$ can be 
%arbitrarily large. 
%However this does not ensure that all correlations instantly decay. 
Another useful measure is the average mixing rate
\beq
\label{eq:MixingRateAvg}
\mu_+[U]= \mathbb{E}_{u_i,v_i} \mu_1[ (u_1\otimes u_2) U (v_1\otimes v_2)],
\eeq
where the $u_i$ and $v_i$ are independently drawn from the Haar measure on the local single-particle spaces. One may also first perform the average on the spectral radius $|\lambda_1|$ before
finding the rate. All of these quantities, measuring the mixing induced by the channel, are by construction LUI and we address their connection to LUI measures based on $U$ itself, in particular to the entangling power $e_p(U)$.

The primary connection is to the norm of the channel. The Eq.~(\ref{eq:Mprime}) implies that this norm is a LUI, as 
\beq
\label{eq:NormMplus}
\|\tilde{M}_+[U]\|^2=\|M_+[U]\|^2-1=(q^2-1)(1-e_p(U)),
\eeq
where $\|A\|= \sqrt{\tr(A A^{\dagger})}$ is the Hilbert-Schmidt or Frobenius norm. 
To see this, use Eq.~(\ref{eq:M+UT}), and the easily verified identities that 
$(A^{R_1})^{\dagger}=(A^{\dagger})^{R_2}$ and $\tr(X^{R_2} X^{R_1})=\tr(X^2)$ to get
\beq
\|M_+[U]\|^2=\frac{1}{q^2}\tr \left(U^{T_2} U^{T_2\, \dagger}\right)^2.
\eeq
As $U$ is dual-unitary,  $E(U)=E(S)$ and from Eq.~(\ref{eq:ep}) $e_p(U)=E(US)/E(S)$. Using the relation 
in Eq.~(\ref{eq:EUS}), we then arrive at the above connection, which is also the norm of $M_-[U]$. 

Let $\tilde{M}_+=VTV^{\dagger}$ be a Schur decomposition \cite{horn2012matrix} of $\tilde{M}_+$, where $V$ is unitary and $T$ is an upper-triangular 
matrix whose diagonal entries are the eigenvalues $\lambda_i$. It then follows from the unitary invariance of
the norm that $\|\tilde{M}_+\|^2=\|T\|^2$ and hence 
\beq
\sum_{i=1}^{q^2-1} |\lambda_i|^2 \leq \|\tilde{M}_+[U]\|^2=(q^2-1)(1-e_p(U)).
\eeq
These consequences are immediate:
\begin{enumerate}
\item From $|\lambda_{q^2-1}|^2 (q^2-1) \leq \sum_{i=1}^{q^2-1} |\lambda_i|^2 \leq (q^2-1)(1-e_p(U))$, it follows 
that the smallest eigenvalue is bounded from above as
\beq
\label{eq:smallest_eigval}
|\lambda_{q^2-1}| \leq \sqrt{1-e_p(U)}.
\eeq
\item 
An upper-bound for the largest nontrivial eigenvalue of $M_+$ or its spectral radius $|\lambda_1|$ immediately 
follows as $|\lambda_1|^2 \leq \sum_{i=1}^{q^2-1} |\lambda_i|^2 \leq (q^2-1)(1-e_p(U))$
gives
\beq
\label{eq:Inequality}
|\lambda_1|\leq \sqrt{1-e_p(U)} \sqrt{q^2-1},
\eeq
which also gives a lower-bound on the spectral gap $1-|\lambda_1|$, and the rate of mixing.
\item The above two are subsumed in a series of inequalities which follow from the observation that
$k |\lambda_k|^2\leq \sum_{i=1}^k |\lambda_i|^2 \leq  \sum_{i=1}^{q^2-1} |\lambda_i|^2 \leq (q^2-1)(1-e_p(U))$. This yields
\beq
\label{eq:kInequality}
|\lambda_k|\leq \sqrt{1-e_p(U)} \sqrt{q^2-1}/\sqrt{k}.
\eeq

\end{enumerate}
  The upper-bound in Eq.~(\ref{eq:smallest_eigval}) implies that whenever $e_p(U)>0$,  $|\lambda_{q^2-1}|<1$ and there will be at least one mixing 
mode and hence iff $U$ is the swap or locally equivalent to it, all the $|\lambda_k|=1$  as $e_p(U)=0$. Thus the presence of mixing modes is generic.
On the other extreme, if $e_p(U)=1$,  $|\lambda_1|=0$ and hence all nontrivial $\lambda_k=0$ and we are in the Bernoulli circuit case, consistent 
with our previous observations.

We observe that the upper-bound for $|\lambda_1|$  in Eq.~(\ref{eq:Inequality}) is trivial, as it will not be less than $1$, unless
\beq
\label{eq:NontrivialityCondition}
e_p(U) > e^*_p= \frac{q^2-2}{q^2-1}.
\eeq
This is then a sufficiency condition for the dual-unitary  circuit to be mixing as it
ensures that $|\lambda_1|<1$, and hence $|\lambda_k|<1$ for all $k>0$, and for both $M_+$ and $M_-$. The inequality in Eq.~(\ref{eq:Inequality}) is a tight one,
that is there exists dual-unitary gates $U$ such that $e_p(U)=e_p^*=(q^2-2)/(q^2-1)$ and $|\lambda_1|=1$, in fact $\lambda_1=1$,
as we show explicitly via examples when discussing specific instances. The qualitative content of the sufficiency
condition is plain: if the interaction as measured by the entangling power of the unitary bricks of the quantum circuit
is sufficiently large, the resultant circuit is always mixing, whatever be the local fields. For the case of qubits, $q=2$, the sufficiency condition $e_p(U)>2/3$ is never met, as the maximum 
possible entangling power is $2/3$ and 2-unitaries do not exist, as we have already noted. Gates such as {\sc dcnot} \cite{Kus2013,Bhargavi2017,Bhargavi2019}  have the maximum 
entangling power of $2/3$, and circuits built of them need not be mixing depending on the local operations used.

\begin{figure}[h]
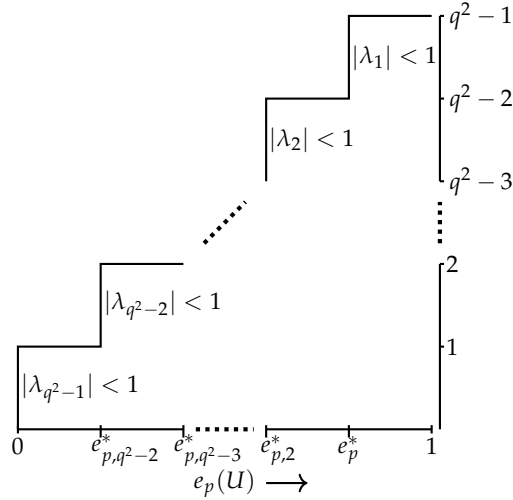

\btp[baseline=(current  bounding  box.center),scale=0.55,font=\small]
\draw [ thick] (0,0) -- (4,0);
\draw [ thick] (6,0) -- (10,0);
\draw [ultra thick,dotted] (4.3,0) -- (5.7,0);

\draw [thick] (0,-0.1) -- (0,0.1);
\draw [thick] (2,-0.2) -- (2,0.1);
\draw [thick] (4,-0.2) -- (4,0.1);
\draw [thick] (6,-0.2) -- (6,0.1);
\draw [thick] (8,-0.2) -- (8,0.1);
\draw [thick] (10,-0.1) -- (10,0.1); 

%\draw [thick,dotted] (2.3,-0.5) -- (5.7,-0.5);

\draw [thick] (0,0) -- (0,2) -- (2,2) -- (2,4) -- (4,4) ; 
\draw [thick] (6,6) -- (6,8) -- (8,8) -- (8,10) -- (10,10);
\draw [ultra thick, dotted] (4.5,4.5) -- (5.5,5.5); 

\node at (10,-0.4) {\small{$1$}};
\node at (0,-0.4) {\small{$0$}};
%\node at (8.8,-1) {\small{$e_p^* = \frac{q^2-2}{q^2-1}$}};
%\node at (6.7,-1) {\small{$e_{p,2}^* = \frac{q^2-3}{q^2-1}$}};
\node at (8,-0.5) {$e_p^*$};
\node at (6.3,-0.5) {$e_{p,2}^*$};
\node at (2.6,-0.5) {$e_{p,q^2-2}^*$};
\node at (4.6,-0.5) {$e_{p,q^2-3}^*$};
\node at (9.1,9) {$\abs{\lambda_1} <1$};
\node at (7.1,7) {$\abs{\lambda_2} <1$};
\node at (3.5,3) {$|\lambda_{q^2-2}| <1$};
\node at (1.5,1) {$|\lambda_{q^2-1}| <1$};
\node at (5,-1.3) {$e_p(U)$};
\draw [thick, ->] (6,-1.3) -- (7,-1.3);
\draw [thick] (10.2,0) -- (10.2,4);
\draw [thick] (10.2,6) -- (10.2,10);

\draw [thick] (10.2,2) -- ((10.3,2);
\draw [thick] (10.2,4) -- ((10.3,4);
\draw [thick] (10.2,6) -- ((10.3,6);
\draw [thick] (10.2,8) -- ((10.3,8);
\draw [thick] (10.2,10) -- ((10.3,10);

\node at (10.5,2) {$1$};
\node at (10.5,4) {$2$};
\node at (11.2,6) {$q^2-3$};
\node at (11.2,8) {$q^2-2$};
\node at (11.2,10) {$q^2-1$};

\draw [ultra thick, dotted] (10.2,4.5) -- (10.2,5.5);
\etp 
\caption{As the entangling power of the two-particle unitary $U$ is increased, by multiples of $1/(q^2-1)$, beyond $e_{p,k}^*$, this ensures that $q^2 -k$ modes of the dual-unitary  circuit are mixing. At least one mode is mixing if  $e_p(U)>e_{p,q^2-1}^*=0$ and all the $q^2-1$ nontrivial modes are mixing if $e_p(U)>e_{p,1}=e_p^*$.}
\label{fig:heira}
\end{figure}

While the smallest eigenvalue $|\lambda_{q^2-1}|<1$ for any entangling unitary $U$, and sufficiently entangling ones
ensure mixing, there is a hierarchy of intermediate cases. It follows from Eq.~(\ref{eq:kInequality}) that for $1 \leq k \leq q^2-1$,
\beq
\begin{split}
&e_p(U)>e_{p,k}^*=1-\frac{k}{q^2-1}\; \text{implies that} \\
&|\lambda_{q^2-1}|\leq \cdots \leq |\lambda_{k}|<1,
\end{split}
\eeq
where $e_{p,1}^* \equiv e_p^*$ and $e_{p,q^2-1}^*=0$. Thus the influence of the increasing entangling power of the two-particle unitary $U$ on the mixing properties of the many-body circuit is encapsulated by this periodic gradation: if $e_p(U)>e_{p,k}^*$ there are at least
$q^2-k$ mixing modes. This is illustrated in Fig. (\ref{fig:heira}).

The entangling power does encode at all levels, the degree of mixing in the dual-unitary circuits.
Further, as $e_p^*$ is a tight bound, so also the others maybe, in the sense that there exist $U$ such that 
$e_p(U)=e_{p,k}^*$ and $\lambda_1=\cdots =\lambda_k=1$.
The rather large upper-bound on the spectral radius Eq.~(\ref{eq:Inequality})
is due to the factor $\sqrt{q^2-1}$, which we now go on to whittle down.

\subsection{Single-particle averaged spectral radius and rates \label{ref:genavg}}

We wish to go beyond the bounds such as in Eq.~(\ref{eq:Inequality}) to find approximations.
Local unitary operations $U\mapsto U'=(u_1\otimes u_2) U (v_1\otimes v_2)$ change the mixing rate $\mu_1$ in Eq.~(\ref{eq:decay}) and can substantially lower it as well. We will motivate, both numerically and analytically that  
$\mu_+[U]$, the average mixing rate, is an universal function of the entangling power $e_p(U)$, independent of the local dimension for $q>2$. Analytical results include the case of $q=2$, which is different from the rest, as well as the approximation that arise from $\mathbb{E}_{u_i,v_i} |\lambda_1|[U'] \approx \sqrt{1-e_p(U)}$ which is seen to be qualitatively valid and quantitatively so near $e_p(U)=1$. Thus the averaging effectively removes the $\sqrt{q^2-1}$ factor from the upper bound. It is quite remarkable, that while $e_p(U)$ is just one of many local unitary invariants, it plays a central role in the mixing properties of the quantum circuit built of $U$.

The spectral radius of an arbitrary matrix $A$ is the largest absolute value of its eigenvalues.  A formula of Gelfand is well known for the spectral radius and is given as the limit $\|A^k\|^{1/k}$ as $k \rightarrow \infty$, where $\|A\|$ is  any norm \cite{horn2012matrix}, but we will continue to imply the
Hilbert-Schmidt norm. Thus we have  $|\lambda_1|=\lim_{k \rightarrow \infty}\|\tilde{M}_{+}^k\|^{1/k}$. For simplicity we will write $\tilde{M}_{+}$ for what we call $\tilde{M}_{+}[U]$. We have that the norm $\|\tilde{M}_+\|$ is a LUI and is given by Eq.~(\ref{eq:NormMplus}). However the norm of $M_+^k$
for $k>1$ is not LUI. 

Consider Eq.~(\ref{eq:Mprime}), and identify $u \equiv v_2^{\dagger}$ and $v \equiv u_1^{\dagger}$.
We then have that the local-unitary Haar averaged spectral radius of $\tilde{M}_+$ is
\beq
\begin{split}
\mathbb{E}_{u_i,v_i}|\lambda_1|=&\int_{\text{Haar}} du dv \,\,|\lambda_1|\left[ (u \otimes u^*) \tilde{M}_{+}(v \otimes v^*) \right]\\=
&\int_{\text{Haar}} du dv \,\,|\lambda_1|\left[ [(vu) \otimes (vu)^*] \tilde{M}_{+} \right]\\=
&\int_{\text{Haar}} dv \int_{\text{Haar}} dw  \,\,|\lambda_1|\left[ (w \otimes w^*) \tilde{M}_{+} \right]\\=
&\int_{\text{Haar}} du\,\, |\lambda_1|\left[ (u \otimes u^*) \tilde{M}_{+}\right]\\
&=\lim_{k \rightarrow \infty} \int_{\text{Haar}} du \|\left[(u \otimes u^*) \tilde{M}_{+}\right]^k\|^{1/k}.
\end{split}
\eeq
The invariance of the Haar measure under multiplication by a fixed unitary is used, as well as 
the limit and the integration is interchanged. 

To begin with consider
\beq
%\begin{multline}
\overline{\|\tilde{M}_{+}^2[U']\|^2} = \int_{\text{Haar}}du \|\left[(u \otimes u^*) \tilde{M}_{+}\right]^2\|^2
=\int_{\text{Haar}}du \|\tilde{M}_{+}(u \otimes u^*) \tilde{M}_{+} \|^2.
%\end{multline}
\eeq
To perform the Haar average  $\overline{\|\left(A (u \otimes u^*)B \right) \|^2}$ 
use the identity   (see Appendix \ref{app:haarid} for a derivation) which uses the results from Ref.  \cite{collins2006integration,hiai2000semicircle,Pucha_a_2017} for calculating Haar average of monomials over the unitary group  \bla 
\beq 
\label{eq:Haar}
\begin{split}
&\int_{\text{Haar}} du \tr \left[ X (u \otimes u^*) Y (u^{\dagger} \otimes u^T) \right]=
\frac{1}{q^2-1}\left(\tr X^{R_2} \tr Y^{R_2}+\tr X \tr Y\right) \\
&-\frac{1}{q(q^2-1)}\left( \tr X^{R_2} \tr Y +\tr X \tr Y^{R_2}\right).
\end{split}
\eeq
With $X=A^{\dagger} A$ and $Y=BB^{\dagger}$ and the identification of $A=B=\tilde{M}_+$, 
we have $\tr X^{R_2}=\tr Y^{R_2}=0$. This follows as $\tr X^{R_2}=q \br \Phi^{+}|A^{\dagger} A|\Phi^{+}\kt$ and 
$\tilde{M}_{+}|\Phi^+\kt=0$. We have then 
\beq
\label{eq:AvgAB}
\overline{\|\left(A (u \otimes u^*)B \right) \|^2}=\frac{1}{q^2-1}\|A\|^2 \|B\|^2, 
\eeq
and hence an exact evaluation, 
\beq
\overline{\|\tilde{M}_{+}^2[U']\|^2} =\frac{1}{q^2-1}\|\tilde{M}_+[U]\|^4 =(q^2-1)(1-e_p(U))^2,
\eeq
where we have used Eq.~(\ref{eq:NormMplus}). 

This then implies in the spirit of Eq.~(\ref{eq:Inequality}),
a local unitary averaged inequality:
\beq
\mathbb{E}_{u_i,v_i}(|\lambda_1|^4) \leq (q^2-1) (1-e_p(U))^2,
\eeq
which implies
\beq
\label{eq:AvgLambda1Inequality}
\mathbb{E}_{u_i,v_i}(|\lambda_1|) \leq (q^2-1)^{1/4} \sqrt{(1-e_p(U))}.
\eeq
For the mixing rate,
\beq
\label{eq:MuInequality}
\begin{split}
\mu_+[U]=&\mathbb{E}_{u_i,v_i}(-\ln |\lambda_1|)\geq -\ln (\mathbb{E}_{u_i,v_i} |\lambda_1|)\\& \geq -\frac{1}{2}\ln(1-e_p(U)) -\frac{1}{4}\ln(q^2-1). 
\end{split}
\eeq
This gives a sufficiency condition for the circuit to be mixing as 
\beq
e_p(U)>1-\frac{1}{\sqrt{q^2-1}},
\eeq
in the sense that there will exist locals that will make the circuit mixing. However, these are weak conditions and it appears
from numerical results that only $e_p(U)>0$ is required for suitable local unitaries to make the circuit mixing. More importantly,
a comparison of the inequalities in Eqs.~(\ref{eq:Inequality}) and ~(\ref{eq:AvgLambda1Inequality}), shows how
the local unitary averaging reduces the effect of the factor $(q^2-1)$ at even $k=2$.

\begin{figure}[htbp]
\includegraphics[scale=.42]{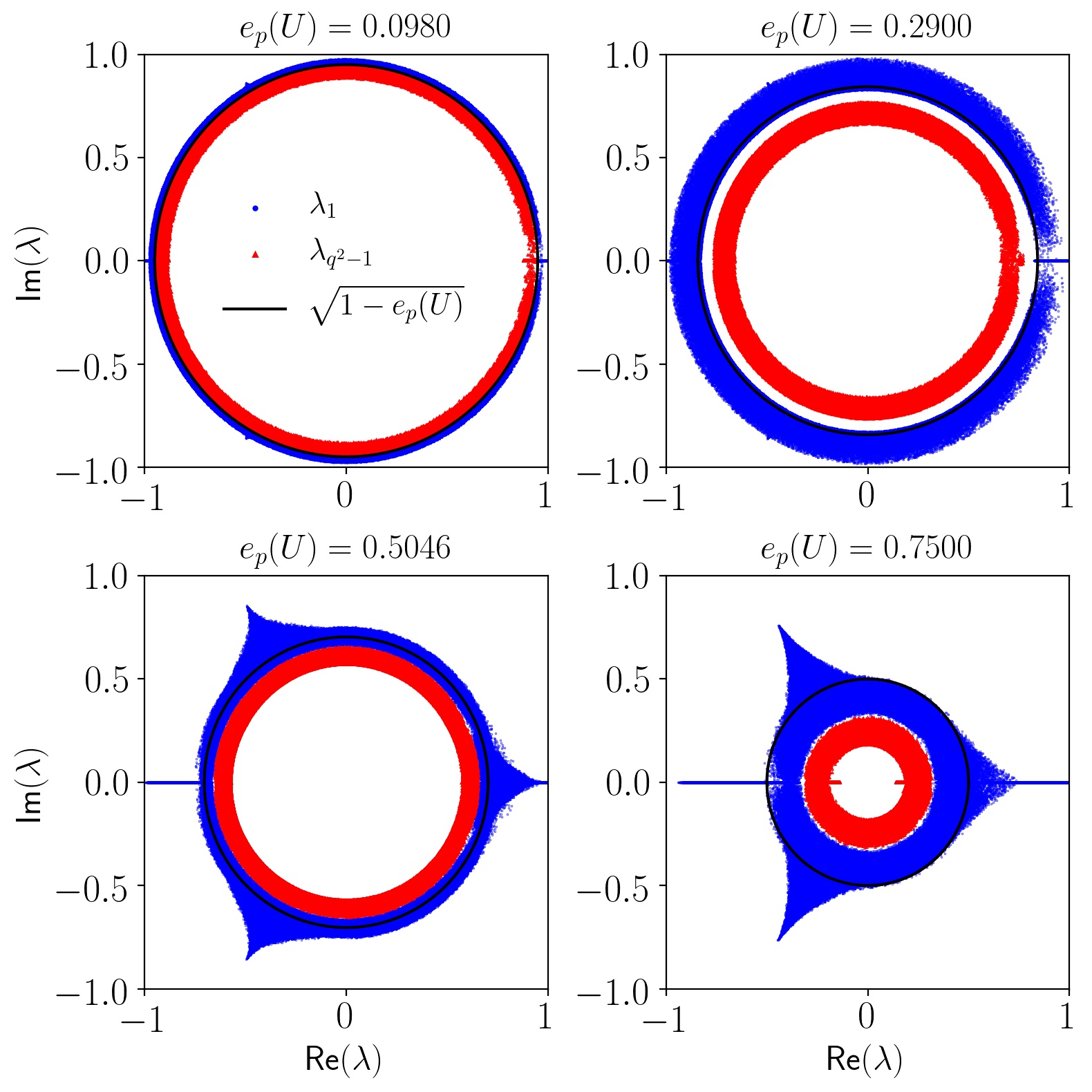} 
\caption{The real and imaginary parts of the nontrivial largest and smallest magnitude eigenvalues $\lambda_1$ and $\lambda_{q^2-1}$ 
of $M_+[U']$ for 4 dual-unitary $U$, the first three being of the form $D_1 S$, 
where $D_1$ is a diagonal unitary and $S$ the swap and  $q=3$. The $U'$ are generated in all cases with $10^4$ local, random single-particle unitaries, 
while a solid circle of radius $\sqrt{1-e_p(U)}$ is shown in each figure.}
\label{fig:EvalsM_DS}
\end{figure}   

We now indulge in an approximate evaluation for the norm of the higher powers of $\tilde{M}_{+}$. Continuing to denote Haar averages by overbars
\beq
\begin{split}
&\overline{\|\left[ A (u \otimes u^*)\right]^k \|^2}=  \overline{\|\left[A (u \otimes u^*)B \right] \|^2}\\
&=\overline{\tr\left[A^{\dagger} A (u \otimes u^*) B B^{\dagger} (u^{\dagger} \otimes u^T)\right) },
\end{split}
\eeq
where $B= [A (u \otimes u^*)]^{k-1}$. For $k>2$ the Haar average seems hard to perform exactly and we resort to treating the equality in Eq.~(\ref{eq:AvgAB}) 
as remaining valid even in this case although $B=(A (u \otimes u^*))^{k-1}$ depends on $u$. Similar reasoning as before, including the fact that $u \otimes u^*$ leaves $|\Phi^+\kt$ invariant, implies still that $\tr Y^{R_2}=0$. But now we can recursively 
separate out one more factor of $A(u \otimes u^*) B'$ from $B$ to average $\|B\|^2$. 
We get the estimate
\beq
\overline{\|\left[A (u \otimes u^*)\right]^k \|^2} \approx \frac{1}{(q^2-1)^{k-1}}\|A\|^{2k},
\eeq
and hence finally, with $A=\tilde{M}_+$ and using Eq.~(\ref{eq:NormMplus}),
\beq
\label{eq:MpluspowerkNorm}
\overline{\|\tilde{M}_+^k[U']\|^2} \approx (q^2-1)(1-e_p(U))^{k}.
\eeq

\begin{figure} 
\includegraphics[scale=.47]{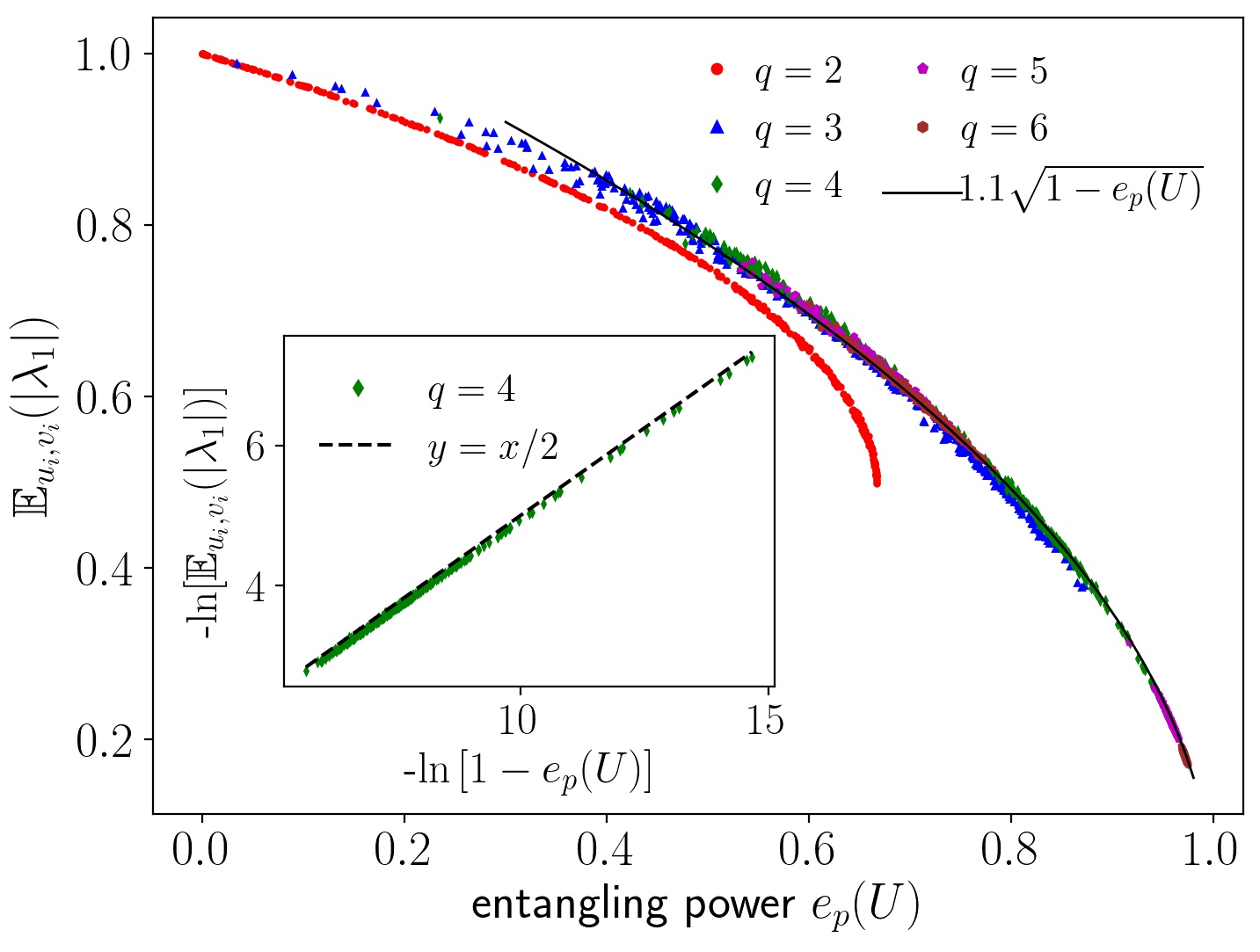} 
\caption{The single-particle averaged spectral radius $\mathbb{E}_{u_i,v_i}(|\lambda_1|)$ {\it vs} the entangling power. The dual-unitaries are  $200$ realizations each of the $D_1S$ and dual-CUE ensembles, for all the local dimensions $q$. The number of single-particle unitaries used in the averaging is  $10^4$. The analytical approximations  shown  are discussed in the text. The inset shows how well the analytical expression fits in the neighborhood of $e_p(U) =1$.} 
\label{fig:AvgLambda1}
\end{figure}

Using the Gelfand formula leads to an estimate of the local unitary average of the largest nontrivial eigenvalue as
\beq
\label{eq:AvgLambda1}
\mathbb{E}_{u_i,v_i}(|\lambda_1|)\approx f_q \sqrt{1-e_p(U)},
\eeq
where $f_q \approx 1$ is a potential ``fudge factor". Thus the $(q^2-1)$ factor in the norm of $\tilde{M}_+[U]$ is essentially eliminated
by the local unitary averaging procedure with the exactly same factor appearing in the Haar average formula
of Eq.~(\ref{eq:Haar}). 

Shown in Fig.~(\ref{fig:EvalsM_DS}) are eigenvalues $\lambda_1$ and $\lambda_{q^2-1}$ of $M_+[U']$ when an ensemble of local unitaries have been applied to fixed dual-unitary $U$, see Eq.~(\ref{eq:Mprime}). Except for the case when $e_p(U)=0.75$, these are the largest and smallest nontrivial eigenvalues for $3$ realizations of $U=D_1 S$ when $D_1$ are themselves diagonal matrices of random phases uniformly distributed on the unit circle. This is an ensemble of dual-unitary  matrices, which we discuss in the section (\ref{sec:blockdiag}) as the $D_1S$ ensemble. Its entangling power does not exceed $q/(q+1)$. The realizations have been chosen to have a representative range of values
of the entangling power, which is also displayed. The case when $e_p(U)=3/4$ is a dual-unitary obtained from an algorithm described in Sec.~\ref{sec:dualcue}. We will presently discuss the reason for this choice. 

In the figures a circle of radius $\sqrt{1-e_p(U)}$ is also marked.
For small $e_p(U)$ the $|\lambda_1|$ is large, but so is $|\lambda_{q^2-1}|$ and both nearly coincide with the $\sqrt{1-e_p(U)}$ circle.
In all cases it is clear that the inequality in Eq.~(\ref{eq:smallest_eigval}) holds.
We notice also that there are substantial number of real eigenvalues, apart from the one at $1$: this arises due to a symmetry in the spectra that if $\lambda$ is an eigenvalue then so is $\lambda^*$. There is a similar ``equatorial" enhancement for the  real Gaussian random or Ginibre ensemble \cite{edelman1994many}.  As the entangling power increases, the separation between the eigenvalues is also more prominent, but the ring of $\lambda_1$ values is essentially governed by the same circle, and indicates that that Eq.~(\ref{eq:AvgLambda1}) is valid.

This is shown in Fig.~(\ref{fig:AvgLambda1}) where quite remarkably the simple formula in Eq.~(\ref{eq:AvgLambda1}) fits reasonably well the average of $\abs{\lambda_1}$ for a  variety of ensembles of 
dual-unitaries in a large range of entangling power, with $f_q=1.1$ for  $3 \leq q\leq 6$. The dual-unitaries used are from two sources:
first is the $D_1S$ ensemble used in the last figure, whose entangling power is limited. The other is a numerical iterative procedure that generates an ensemble dubbed dual-CUE \cite{SAA2020} and which we recall in the next section. It is capable of producing high entangling power dual-unitaries including 2-unitaries. Perturbations of such 2-unitaries are used to 
generate unitaries with entangling power arbitrarily close to $1$ to populate the inset of the figure, which shows how well the Eq.~(\ref{eq:AvgLambda1})
works with $f_q=1$ near $e_p(U)=1$. Thus the factor $f_q$ appears to be a slow function of the entangling power. Note that we have made use of $q=4$ for the inset, as $q=3$ is a special case wherein
$e_p(U)=1$ appears to be isolated in the set of dual-unitary  matrices. Numerical results indicate that  for $q=3$ there is a gap whose left extreme seems larger than $e_p^*$, but is otherwise not yet known \cite{Bhargavi2019}.

\begin{figure}
\includegraphics[scale=0.47]{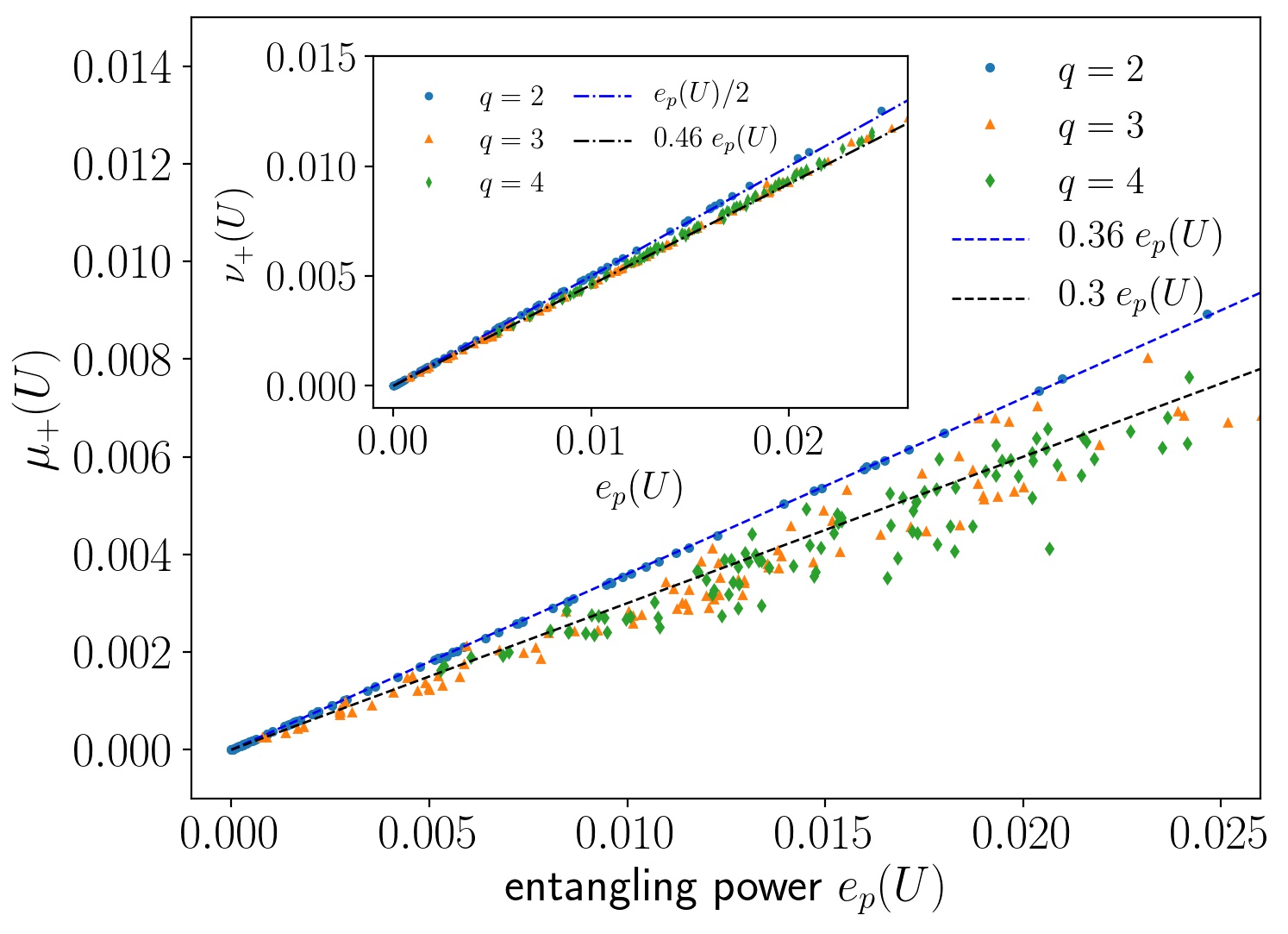}
\caption{The maximal mixing rate $\nu_+$ and the average mixing rate $\mu_+$ for dual-unitaries with entangling power $e_p(U)$ close to $e_p(U) =0$ for $q=2,3$ and $4$. $10^4$ local unitary realizations are used in the averages.}
\label{fig:nearswap}
\end{figure}

The average mixing rate $\mu_+$ and the maximal rate $\nu_+$ for dual-unitaries with entangling power close to $e_p(U) =0$ for $q=2,3$ and $4$ are shown in Fig.~(\ref{fig:nearswap}). These dual-unitaries are of the form $D_1 S$, $D_1$ is diagonal with entries $e^{i \epsilon \phi}$ where $\phi$ is uniform and random in $[-\pi,\pi)$ and $\epsilon$ controls how far from the swap the dual-unitary  is. Firstly, it is remarkable that for $q=2$, $\nu_+$ seems to lie exactly on the $e_p(U)/2$ line. We shortly derive the analytical form of $\nu_+$ for qubits, which indeed is a singular case.
For $q=3,4$ the $\nu_+$ has a slightly smaller slope than $q=2$ and both lie on the same line. The $\mu_+$
line for $q=2$ has a slope of $\approx 3/8$, while for $q=3,4$ we find a scatter of points.
% that is within sample size fluctuations of $1/\sqrt{N_s}$, we have used $N_s=10,000$.  
 We discuss complementary data and comparisons with analytics 
later in Figs.~(\ref{fig:nuplusqubits}) and ~(\ref{fig:blodia}).

The scatter of the average mixing rates  near $e_p(U)=0$ signals the involvement of other local-unitary invariants than 
the entangling power alone. That these are not fluctuations due to sample size was verified by increasing the sample size by
several orders of magnitude and observing that the scatter remained practically invariant. To examine this further, we take gates $U_1$ and $U_2$ with the same entangling powers: $e_p(U_1)=e_p(U_2)$, but which 
are not local-unitarily connected, that is there exists no $u_i$ and $v_i$ such that $(u_1\otimes  u_2)U_1(v_1\otimes v_2)= U_2$.
\begin{figure}
\includegraphics[scale=0.47]{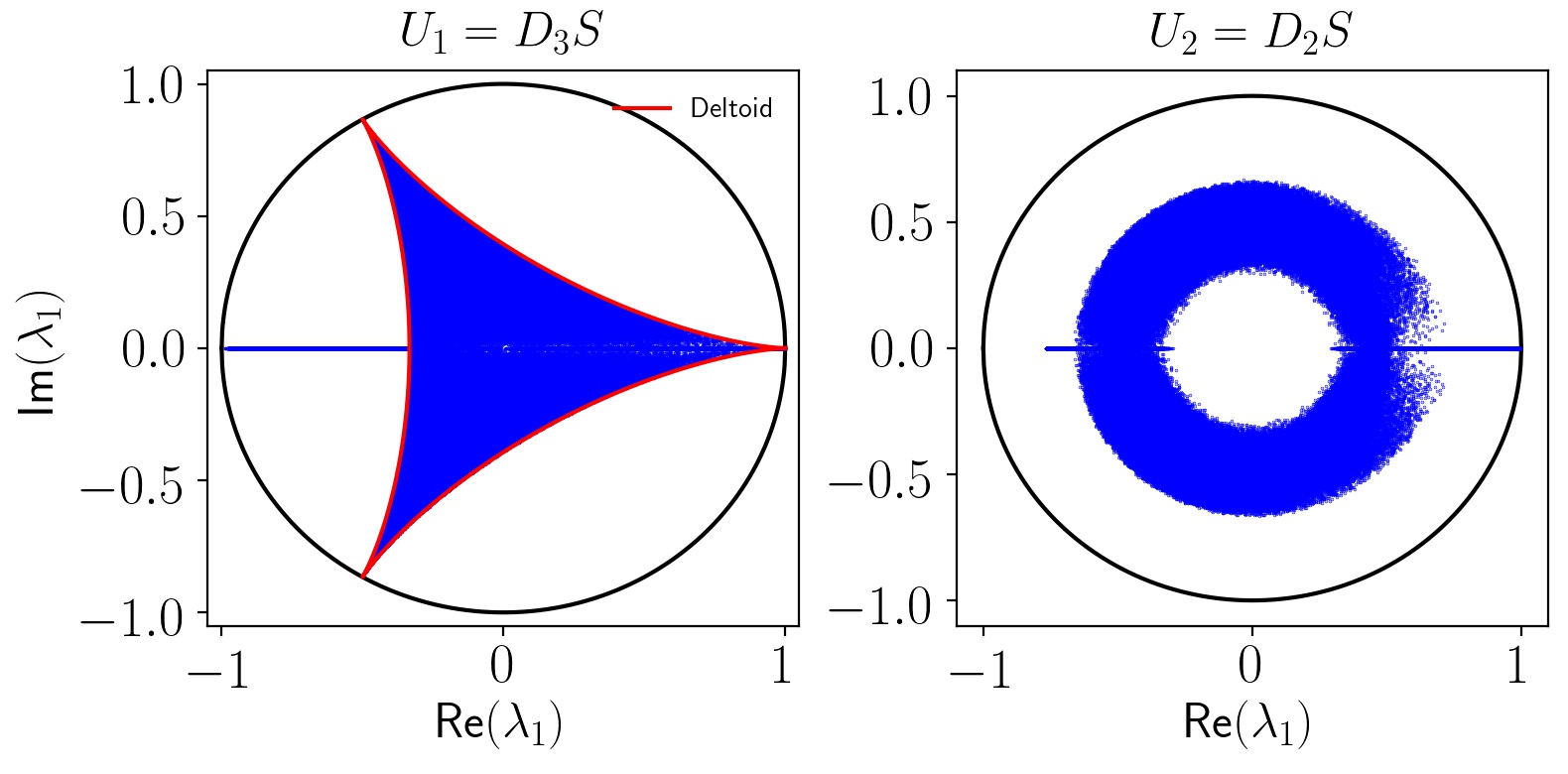}
\caption{The real and imaginary parts of the nontrivial largest eigenvalue $\lambda_1$ 
of $M_+[U']$ for two dual-unitary $U_i$ that have the same entangling power, $e_p(U_i)=3/4$. The circle shown is of unit radius, while for 
$U_1$ the complex eigenvalues are bounded by a hypocycloid called a deltoid.}
\label{fig:permlamb1}
\end{figure}
See Fig.~(\ref{fig:permlamb1}) for the largest eigenvalues $\lambda_1$ of $M_+[U'_1]$ and  $M_+[U'_2]$ when $q=3$.  Fig.~(\ref{fig:EvalsM_DS}) shows yet another instance of $e_p(U)=3/4$ but is not locally equivalent to either $U_1$ or $U_2$. For details of these matrices, see the Appendix~(\ref{app:unistoc}). 
The fact that the $\lambda_1$ is very differently distributed in the unit circle itself indicates that these operators are not locally connected. It is found that $\mathbb{E}_{u_i,v_i}(|\lambda_1|) \approx 0.48655$ for $U=U_1$ whereas it is $\approx 0.55045$ for $U=U_2$ with $N_s=10^6$ samples of local unitaries, and seem to have converged to an accuracy of at least $10^{-3}$. 

The peculiar structure seen in Fig.~(\ref{fig:permlamb1}) corresponding to $U_1$ warrants some comments,
especially as some aspects of the ``triangular features" are also evident in two of the cases in Fig.~(\ref{fig:EvalsM_DS}). Very similar figures have appeared in the study of so-called unistochastic matrices \cite{zyczkowski2003random}. The map corresponding to $M_+[U'_1]=(v_2^{\dagger} \otimes v_2^T)M_+[U_1](u_1^{\dagger} \otimes u_1^T)$ (from Eq.~(\ref{eq:Mprime})) is special in this case and 
apart from a null 6-dimensional space, it is given by the bistochastic $3 \times 3$ matrix whose elements are $|u_{ij}|^2$, where $u=u_1^{\dagger} v_2^{\dagger}$, as shown in Appendix~(\ref{app:unistoc}). A bistochastic matrix is unistochastic if its elements are derived as the absolute value squared of the elements of a unitary matrix. The eigenvalues of random unistochastic matrices for $q=3$ are conjectured to be bound by the deltoid (3-hypocycloid) \cite{zyczkowski2003random} which are the origin 
of the structure observed in the first part of Fig.~(\ref{fig:permlamb1}). It must be emphasized that the CPTP maps $M_+[U']$ are in general not related in such a manner to unistochastic matrices and hence such structures are generically absent. In particular we have found them for specific subsets
of CPTP maps derived from special dual unitaries with special values of the entangling power (such as $3/4$ in $q=3$).

\subsection{Inhomogeneous circuits}
If the two-particle unitary operators are different, but still dual-unitary, within a circuit, this represents an inhomogeneous case and in this brief interlude we use the results above to bound the correlation decay. The CPTP maps after $t$ time steps are
\beq
M_{\pm}(t)= M_{\pm}[U'_{2t}]M_{\pm}[U'_{2t-1}]\cdots M_{\pm}[U'_{1}].
\eeq
Here $U'_k=(u_{1k}\otimes u_{2k}) U_k (v_{1k} \otimes v_{2k})$ is the interaction between site $k-1$ and $k$. 
The correlation function from Eq.~(\ref{eq:Map1}) becomes in this case, considering traceless
basis operators,
%\beq
\begin{multline}
|C_{\pm}^{ij}(\pm t,t)|^2=\frac{1}{q^2} |\br a_j|M_{\pm}(t)|a_i \kt|^2 = \frac{1}{q^2} |\br a_j|\tilde{M}_{\pm}(t)|a_i \kt|^2 \leq  \frac{1}{q^2}\br a_j|a_j\kt \br a_i |a_i\kt \|\tilde{M}_{\pm}(t)\|^2
= \|\tilde{M}_{\pm}(t)\|^2,
\end{multline}
%\eeq
where we have used the Schwarz inequality and $\tr(a_j^{\dagger} a_j)=\br a_j|a_j\kt=q$. The replacement of $M_{\pm}$ by $\tilde{M}_{\pm}$, 
a product of $\tilde{M}_{\pm}[U'_k]$ is possible as the $a_j$ are traceless.

Allowing for single particle unitaries that are also inhomogeneous ($u_{1k}$ etc. are i.i.d. random) and averaging over them leads to the 
inequality
\beq
\overline{|C_{\pm}^{ij}(t,t)|^2}\leq \overline{\|\tilde{M}_{\pm}(t)\|^2} =(q^2-1) \prod_{k=1}^{2t}[1-e_p(U_k)].
\eeq
Here we note that the averaging in Eq.~(\ref{eq:MpluspowerkNorm}) is exact as the local unitaries are
independent at every stage of the averaging. As a special case, assuming single particle 
inhomogeneity, but uniform two-particle interactions, we have 
\beq
\overline{|C_{\pm}^{ij}(t,t)|^2}\leq (q^2-1)\exp(-t \gamma),
\eeq
where $\gamma=-\ln(1-e_p(U))^2$. Thus the decay of correlations in the quenched disordered dual circuits are exponentially 
fast and the rate is explicitly given in terms of the entangling powers of the two-particle unitaries used. 

\subsection{Qubit case analytics \label{sec:qubit}}

The qubit case is special in many ways including the fact that the dual-unitary family has only one free parameter.
This presents a completely solvable case to find the connections to the entangling power when a restricted volume of 
local unitary operations are used. All local unitary inequivalent classes of two-qubit operators are characterized by their Cartan decomposition that is a  three parameter family \cite{KBG01,KC01,Zhang2003}. For dual-unitary  subclass, this is reduced to the one parameter family
\begin{equation}
\begin{split}
U(J) & = \exp\left[ i \left(\frac{\pi}{4}\sigma_x \otimes \sigma_x+ \frac{\pi}{4}\sigma_y \otimes \sigma_y+ J \sigma_z \otimes \sigma_z\right)\right] \\
& = \begin{pmatrix}
e^{-iJ} & 0 & 0 & 0 \\ 
0 & 0 & -i e^{iJ} & 0 \\
0 & -i e^{iJ} & 0 & 0 \\
0 & 0 & 0 & e^{-iJ} 
\end{pmatrix},
\end{split} 
\label{eq:cartan}
\end{equation}  
where $\sigma_j$ are Pauli matrices and $0 \leq J \leq \pi/4$. Note that 
$$U(J) = SD(J) = D(J)S, $$ 
where $S$ is the  swap operator and $D(J) = \text{diag} (e^{-iJ},-ie^{iJ},-ie^{iJ},e^{-iJ}) $, and is thus a member of the $D_1S$ ensemble 
we have used in earlier figures. 

The  corresponding channel is simply a diagonal matrix: $M_+[U(J)] = \text{diag}(1,\sin(2J), \sin(2J), 1)$, with $|\lambda_1|=1$. The entangling power is  \cite{ep} 
\beq 
e_p(U) = \frac{2}{3} \cos^2 (2J),
\label{eq:epqubit}
\eeq
and the case $J=\pi/4$ corresponds to a {\sc swap} gate while $J=0$ corresponds to the {\sc dcnot} \cite{Zhang2003}.

A general local or single qubit gate $u \in SU(2)$ can be parametrized as,
\beq 
u=\begin{pmatrix}
\cos\frac{\theta}{2} e^{i\frac{\phi}{2}} & -e^{i \frac{\psi}{2}} \sin \frac{\theta}{2} \\
e^{-i \frac{\psi}{2}} \sin\frac{\theta}{2} & \cos\frac{\theta}{2} e^{-i \frac{\phi}{2}}
\end{pmatrix}, 
\label{eq:genlocal}
\eeq 
where, $\theta \in [0,\pi]$, and $\phi,\, \psi \in [0,4\pi]$. We wish to find the eigenvalues of the channel corresponding to a locally equivalent $U'$: $M_+[U'(J)]=(u^{\dagger} \otimes u^{T})M_+[U(J)]$.
This turns out to be analytically difficult for us and therefore we take single qubit gates from two different subsets. These are found to be useful,
as the first gives an approximation to the average mixing rate $\mu_+$, while the second leads to, what appears numerically to be the exact analytical expression for the maximum mixing rate $\nu_+$.

\begin{figure}[h]
 \includegraphics[scale=0.65]{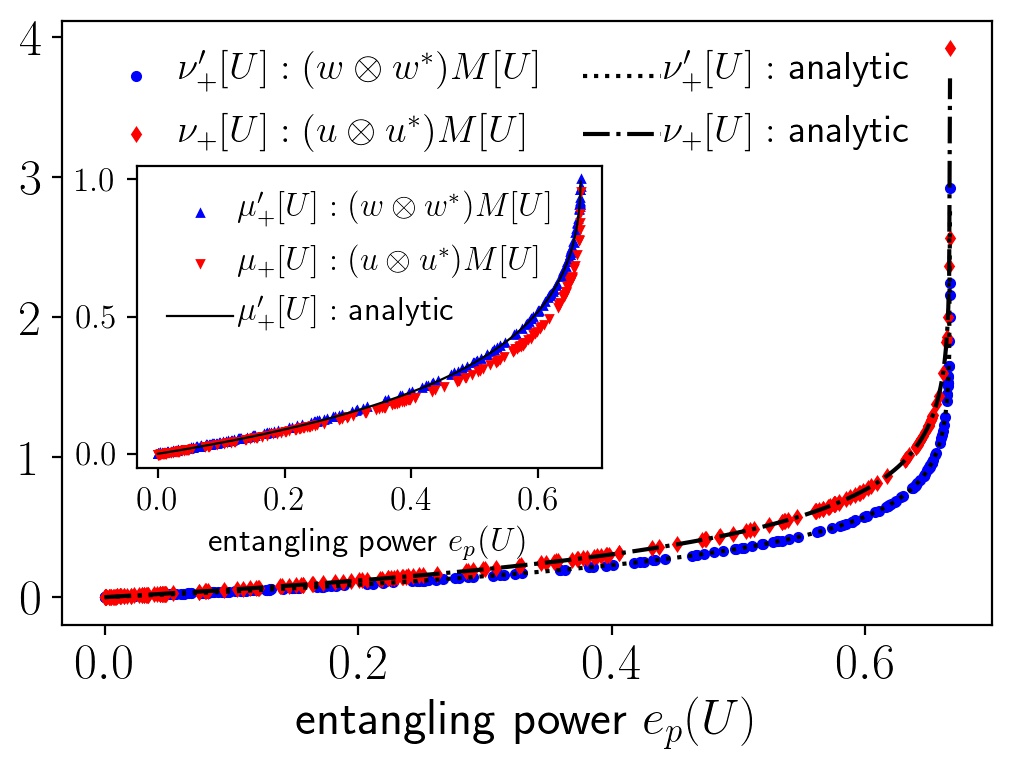} 
 \caption{The maximal mixing rate $\nu_+[U]$ for $q=2$ as function of the entangling power. Shown are the results of maximization over the restricted
 subset of local unitaries represented by $w$ ($\nu_+'[U]$ with its analytical evaluation in Eq.~(\ref{eq:nuplusep})) as well as over the unrestricted set comprising $u \in SU(2)$. The analytical results for the restricted set $v$ in Eq.~(\ref{eq:nuplusfixedr}) is seen to fit exactly the data from the optimization over the unrestricted set. The inset shows the corresponding average mixing rates $\mu'_+[U]$, for which we compare with the analytical expression in Eq.~(\ref{eq:epmuplus}).}
 \label{fig:nuplusqubits}
\end{figure}

The first subset is parametrized as,
\beq 
w=\begin{pmatrix}
\cos\frac{\theta}{2} & -e^{i \frac{\psi}{2}} \sin \frac{\theta}{2} \\
e^{-i \frac{\psi}{2}} \sin\frac{\theta}{2} & \cos\frac{\theta}{2}
\end{pmatrix},
\label{eq:local}
\eeq 
obtained by setting $\phi=0$ in Eq.~(\ref{eq:genlocal}). The maximal mixing rate   is 
\beq
\nu_+^{\prime}[U]=-\frac{1}{2}\ln(\sin 2J)  =  -\frac{1}{4} \ln \left(1-\frac{e_p(U)}{e_p^\text{max}}\right),
\label{eq:nuplusep}
\eeq 
where $e_p^\text{max}=2/3$, the maximum possible value for qubits (See appendix \ref{app:qubitw}). The prime on $\nu_+$ is to indicate that the maximum rate
is found within a subset of all possible single-particle unitary gates.  The connection to the entangling power is via 
Eq.~(\ref{eq:epqubit}) and displays explicitly its monotonic relation to the maximum mixing rate.
Thus $\nu_+[U]\geq \nu_+'[U]$, and comparisons with numerical results are shown in Fig. ~(\ref{fig:nuplusqubits}).

The average mixing rate $\mu_+^{\prime}$, again with the restricted set of local unitaries drawn from $w$, with $\cos\theta$ being uniformly distributed, is 
\beq
\begin{split}
\mu_+^{\prime}[U] & = -\frac{1}{2}\int_{0}^{\pi} \ln (|\lambda_1[\theta,J]|) \sin \theta d{\theta}\\
&=\frac{1-\sin 2J}{1+\sin 2J}= \frac{1-\sqrt{1-\frac{e_p(U)}{e_p^{\text{max}}}}}{1+\sqrt{1-\frac{e_p(U)}{e_p^{\text{max}}}}},
\end{split}
\label{eq:epmuplus}
\eeq
after taking into account the non-smooth changes at $\theta_c$ and $\pi -\theta_c$.
Again the monotonic relation between $\mu_+^{\prime}[U]$ and $e_p[U]$ is made explicit,
and agrees with numerical results as shown in Fig.~(\ref{fig:nuplusqubits}).
%%
%\begin{figure}[h]
% \includegraphics[scale=0.25]{mu_plus_ep_qubits.jpg}
% \caption{$\mu_+$ for 200 dual-cue gates with 1000 locals parametrized by $(\theta,\phi)$. \red{merge nu and mu figures}}
% \label{fig:muplusqubits}
%\end{figure}
We reiterate that the expressions for $\mu_+^{\prime}$ and $\nu_+^{\prime}$ are valid only for the given parametrization of the locals (Eq.~\ref{eq:local}), and in particular do not agree when all the unitaries are sampled using the CUE. The latter sampling that uses general locals $u \in SU(2)$ of the 
form in Eq.~(\ref{eq:genlocal}) is also shown in Fig.~(\ref{fig:nuplusqubits}). It is interesting that the average mixing rates are approximately the same from both sampling. For the case when $J=0$, when the entangling power of $U$ is the maximum possible value of $2/3$,
$\mu_+'[U(J=0)]=1$. The CUE average in this case is also easy to find and $\mu_+[U(J=0)]=1$ as well. This reflects the fact that 
there are no 2-unitary gates for qubits, as for $q>2$, $\mu_+[U]=\infty$ when $e_p(U)=1$.

For small $e_p(U)$, where from Eq.~(\ref{eq:epmuplus}) we infer that $\mu_+'[U] \approx 3 e_p(U)/8$. From Fig.~(\ref{fig:nearswap}) we see that with CUE sampling, $\mu_+[U] \sim 0.36 e_p(U)$ which is indeed close to this.
However, Eq.~(\ref{eq:nuplusep}) gives $\nu_+'[U]\sim 3 e_p(U)/8$ which is quite different from the numerically obtained estimate
of $e_p(U)/2$ from Fig.~(\ref{fig:nearswap}). A better bound on $\nu_+[U]$, which we believe to be the exact value, is attained with the second subset of 
local gates of the form, 
\beq 
v= \frac{1}{\sqrt{2}} \begin{pmatrix}
e^{i \frac{\phi}{2}}  & -e^{i \frac{\psi}{2}} \\
e^{-i \frac{\psi}{2}} & e^{-i \frac{\phi}{2}}
\end{pmatrix},
\label{eq:localfixedr}
\eeq 
obtained by setting $\theta=\pi/2$ in Eq.~(\ref{eq:genlocal}). The maximal mixing rate,  
\beq
\nu_+[U(J)]=-\ln \sin^{\frac{2}{3}}(2J)= -\frac{1}{3}\ln\left[1-\frac{e_p(U)}{e_p^{\text{max}}}\right]
\label{eq:nuplusfixedr}
\eeq
where the maximum mixing rate is found over single or local qubit gates of the form Eq.~(\ref{eq:localfixedr}), and is also written in terms of $e_p(U)=\frac{2}{3}\cos^2(2J)$ (See appendix \ref{app:qubitv}).

Indeed this is a better bound than the one in Eq.~(\ref{eq:nuplusep}), however as this is still based on a subset of local
unitaries it is not a proof that this is indeed $\nu_+$. In Fig.~(\ref{fig:nuplusqubits}), Eq.~(\ref{eq:nuplusfixedr}) is plotted and fits exactly the $\nu_+[U]$ calculated from general single qubits sampled from the CUE. Further, calculations and numerical results shown in appendix  \ref{app:genSU2fig2} strongly suggest 
that Eq.~(\ref{eq:nuplusfixedr}) is a global optimization using single qubit gates of the general form Eq.~(\ref{eq:genlocal}) from the entire $SU(2)$ space.
Notice also that Eq.~(\ref{eq:nuplusfixedr}) gives for small $e_p(U)$ that $\nu_+[U]\sim e_p(U)/2$, agreeing with the corresponding numerical result displayed in Fig.~(\ref{fig:nearswap}). Note that the results for qubits are not surprising as there is only one parameter $J$ and the 
invariants are all function of these. However, the monotonic dependence of the mixing rates with the entangling power does not follow from
this and is a forerunner of the predominantly monotonic dependence for $q>2$.

We  now construct dual-unitary,  and 2-unitary, operators in a variety of different ways, keeping in view the resulting entangling
power which we have just demonstrated plays a key role in the ergodic life of a dual-unitary  circuit.

\section{Dual-unitary  constructions and their associated CPTP maps} 
\label{sec:DualConstructions}

The channel $M_{\pm}[U]$ and the influence of the entangling power of $U$ on its mixing properties
have just been discussed in some generality. We now step back to describe families of dual-unitary 
operators which are the building blocks of the circuits in the first place.  There have been few constructions in
the literature already \cite{gutkin2020local,claeys2020ergodic},  we augment them considerably, especially with the understanding that their entangling power
forms a crucial aspect to control. dual-unitary  operators are also those with maximum operator entanglement and 
while there exists no general parametrization of these,  there are few concrete families   including a numerical 
algorithm constructed by the current authors, which results in the ``dual-CUE" ensemble \cite{SAA2020}. However, the dual-CUE have typically large entangling power and it may be good to construct analytically families that may have smaller entangling power but are parameterized well.   For large entangling power and 2-unitaries there are also permutation matrices and complex Hadamard matrices that are quantizations of classical cat maps: paradigms of classical chaos. 

\subsection{From block-diagonal unitaries \label{sec:blockdiag} }

Dual-unitary  matrices are those that remain unitary under realignment.
However, of the four permutations involving the partial transpose and realignment, the simplest to visualize is $T_2$. Therefore we may construct first 
T-dual matrices and left or right multiply these by the swap operator $S$ to get dual matrices.
For the symmetric $q\times q$ 
spaces, $A^{T_2}$ amounts to partitioning $A$ into $q^2$ $q\times q$ matrices, the first block starting at the first element,
and taking the transpose within these blocks. As the transpose of a unitary matrix remains unitary, this implies that any matrix comprising uniformly of only diagonal, unitary, $q\times q$ blocks is
$\text{T-dual}$. Denoting this as $D_q$, we have 
\beq
D_q = \bigoplus_{j=1}^q U_j \equiv 
\btp[baseline=(current  bounding  box.center),scale=0.5] 
\draw [thick] (0.2,-0.2) -- (-0.2,-0.2) -- (-0.2,8.2) -- (0.2,8.2);
\draw [thick] (7.9,-0.2) -- (8.2,-0.2) -- (8.2,8.2) -- (7.9,8.2) ;
\draw (0,6) rectangle (2,8);
\draw (2,4) rectangle (4,6);
\draw (6,0) rectangle (8,2);
\draw [ultra thick, dotted] (4.5,3.5) -- (5.5,2.5);
\node at (1,7) {$U_1$};
\node at (3,5) {$U_2$};
\node at (7,1) {$U_q$};
\node at (2.45,6.3) {\tiny{$q\times q$}};
\node at (4.45,4.3) {\tiny{$q\times q$}};
%\node at (8.7,0.3) {\tiny{$q\times q$}};
\etp ,
\eeq
where $\{U_j\}$ is any set of $q\times q$ unitary matrices. Hence
\beq
U=SD_q, \;\; U=D_qS,
\eeq
are two families of dual-unitaries, parametrized by $q$ matrices $\in \mathcal{U}(q)$. 

\subsubsection{Diagonal unitary constructions}

The blocks $U_j$ may in turn have a block structure,
an extreme case being when they are all diagonal. In this case, we refer to the diagonal matrix as $D_1$ and the ensemble $SD_1$ was studied recently in this context in \cite{claeys2020ergodic},
but was studied from the point of view of entanglement content in \cite{Lakshminarayan2014}. The channel $M_+[U]$ depends on which side the swap operator 
is multiplying the block diagonal matrices, but the entangling power does not. 
The entangling power of $e_p(U=D_1S)$ is essentially the 
operator entanglement of diagonal gates. If $D_1$ is the uniform ensemble of
diagonal unitary gates (the phase in the diagonal elements is i.i.d. random, uniform in $[-\pi, \pi)$)  
moments of the singular values of $D_1^{R_2}$ maybe derived \cite{Lakshminarayan2014}. For example we find from there that the average
entangling power is 
\beq
\label{eq:avg_ep_for_DS}
\mathbb{E}_D[e_p(D_1S)]=\frac{q-1}{q+1}.
\eeq
This compensates for an additional factor of $(q-1)/(q+1)$ present in \cite{Lakshminarayan2014} due to different
scaling used in the definition of entangling power. Recall that in the present convention, the maximum value
of the entangling power is $1$. However, we show below that the maximum value of $e_p(U=D_1S)$ is only $q/(q+1)$, see Eq.~(\ref{eq:Ep_bound_BD}).
Hence this is $\leq e_p^*$, for $q>2$ and $=e_p^*$ for $q=2$, where $e_p^*$ is the border for ensuring mixing in Eq.~(\ref{eq:NontrivialityCondition}).  
\begin{figure}[h]
\includegraphics[scale=0.6]{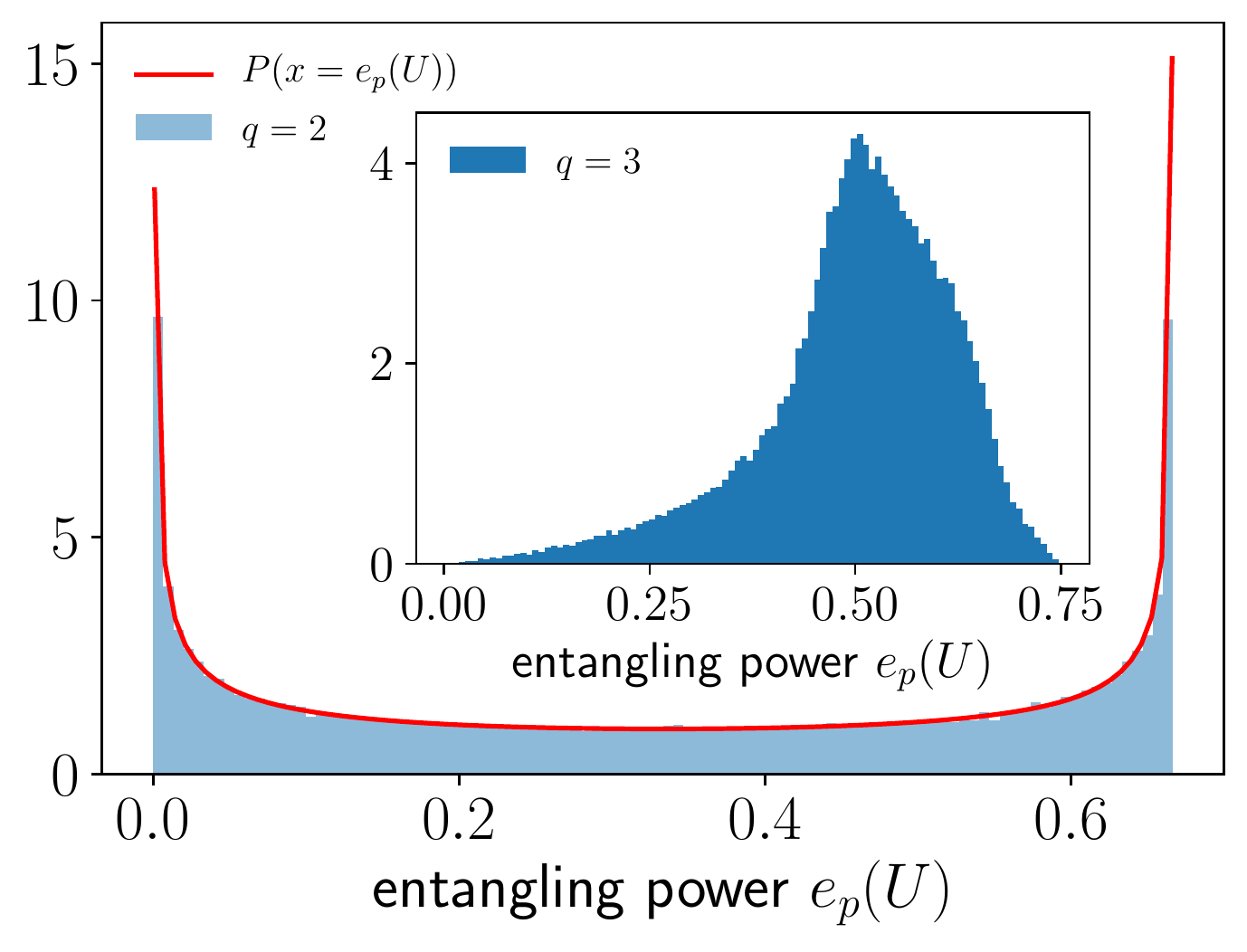}
\caption{The distribution of entangling power $e_p(U)$ of the dual-unitary $SD_1$ ensemble, consisting of the {\sc swap} gate multiplied by diagonal unitaries, for $q=2$ and $q=3$, the former is an arcsin law given in Eq. (\ref{eq:distsd}). 
}
\label{fig:emphas2and3} 
\end{figure} 

For  $q=2$, the distribution of the entangling power may also be explicitly found, as it is observed in \cite{Lakshminarayan2014} that the
distribution of the squared singular values of $U^{R_2}$ follows the arcsin law. This implies the distribution
\beq
P(x=e_p(D_1S))=\frac{1}{\pi}\frac{1}{\sqrt{x \left({2 \over 3} -x\right)}},
\label{eq:distsd}
\eeq
as seen in  Fig. (\ref{fig:emphas2and3}), where for comparison the case $q=3$ is shown as well.
These show the members of these dual-unitary  family can have a fair spread of the entangling power
from nearly $0$ to $q/(q+1)$. 
%\red For the family of dual unitaries $U(J)$ in Eq. (\ref{eq:cartan}), the distribution of entangling power $e_p(U(J))$ is 
%\begin{equation}
%P(e_p(U(J)) = \frac{1}{\pi} \frac{1}{e_p^{\text{max}} \cos 2J \sin 2J}. 
%\end{equation} \bla 
%With enphasing, $e_p(DS)$ modifies from the zero of swap, providing an ensemble of locally inequivalent dual unitary operators.  

\subsubsection{General block diagonal constructions}

A more general ensemble is constructed from 
\beq
\label{eq:DqK}
D_{q_1, \cdots, q_K}=\bigoplus_{j=1}^K U_j \equiv \btp[baseline=(current  bounding  box.center),scale=0.5] 
\draw [thick] (0.2,-0.2) -- (-0.2,-0.2) -- (-0.2,8.2) -- (0.2,8.2);
\draw [thick] (7.9,-0.2) -- (8.2,-0.2) -- (8.2,8.2) -- (7.9,8.2) ;
\draw (0,5) rectangle (3,8);
\draw (3,3) rectangle (5,5);
\draw (6.5,0) rectangle (8,1.5);
\draw [ultra thick, dotted] (5.3,2.7) -- (6.3,1.7);
\node at (1.5,6.5) {{\large$U_1$}};
\node at (4,4) {$U_2$};
\node at (7.3,0.7) {\small{$U_K$}};
\node at (4.3,5.3) {\scriptsize{$m_1q\times m_1q$}};
\node at (6,3.3) {\tiny{$m_2q\times m_2q$}};
%\node at (9.4,0.3) {\tiny{$m_Kq\times m_Kq$}};
\etp ,  
\eeq
where $1 < K\leq q$ and $U_j$ are special classes of $q_j\times q_j$  unitary matrices, where $q_j$ are multiples of $q$ such that $\sum_{j=1}^K q_j=q^2$.
For example when $q=3$, we could have $K=2$ blocks with $q_1=6$ and $q_2=3$. The extreme case of $K=1$ simply corresponds to a full $\text{T-dual}$
matrix of $q^2$ dimension, there is no block structure and is avoided here. The case when $K=q$ is the uniform one discussed above. 

Only when $q_j=q$, $U_j$ is an arbitrary unitary. If $q_j=m_j q$, with $1<m_j<q$, $U_j$ comprise of $m_j^2$ $q\times q$ sub-blocks.
These $U_j$ may be thought to belong to a tensor product space $\mathcal{H}^{m_j} \otimes \mathcal{H}^q$, of a $m_j$ dimensional system with a $q$ dimensional one. If all such $U_j$ are $\text{$T_2$-dual}$, that is remain unitary under partial transpose of the $q-$dimensional subsystem then $D_{q_1, \cdots, q_K}$ is $\text{T-dual}$. Of course this is recursive process and the difficulty is now transferred to finding such special unitaries in lower dimensions. 

However, we note that even if one such set of T-dual matrices are known then local unitary transformations on the $\mathcal{H}^{m_j} \otimes \mathcal{H}^q$ provide a parameterization of T-dual $q^2 \times q^2$ matrices of differing nonlocal content in the $\mathcal{H}^q \otimes \mathcal{H}^q$ space. The simplest T-dual matrix is the identity and hence we may take 
\beq
U_j= u_{m_j}\otimes v_{q},\;\text{where}\; u_{m_j} \in \mathcal{U}(m_j), \; v_{q} \in \mathcal{U}(q)
\nonumber 
\eeq
are arbitrary $m_j$ and $q$ dimensional unitary matrices. This provides one way of constructing the blocks $U_i$ of $D_{q_1, \cdots,q_K}$, such that it is T-dual.
Once these are constructed $U=S D_{q_1, \cdots, q_K}$ and $U=D_{q_1, \cdots, q_K} S$ provide ensembles of dual-unitaries generalizing the easy uniform case.
As $K\neq 1$ (if $K=1$, $m_1=q$ and such operators are locally equivalent to swap $S$), they result in nontrivial dual-unitary  families.

The entangling power of dual-unitaries depends only on $E(US)=E(SU)$ and these
in turn follow from the singular values of the partial transpose such as $U^{T_2}$. From Eq.~(\ref{eq:IdentityR1T2}) we get that  
$U^{T_2}=(S D_{q_1, \cdots, q_K})^{T_2}=S (D_{q_1, \cdots, q_K})^{R_1}$. The action of the realignment $R_1$ on a $q^2\times q^2$ matrix again
proceeds by uniformly dividing it into $q^2$ $q\times q$ sub-blocks, column vectorize the sub-blocks starting from the top left corner, and going along the blocks column-wise. Thus $(D_{q_1, \cdots, q_K})^{R_1}$ contains as many columns as there are $q\times q$ blocks in it and we get that
\beq
\text{Rank} \left[(D_{q_1, \cdots, q_K})^{R_1} \right]=\sum_{j=1}^K m_j^2,
\eeq
as each $q_j-$dimensional sub-block contains $m_j^2$ $q-$ dimensional ones. Along with $\sum_{j=1}^K m_j=q$, this renders
$(D_{q_1, \cdots, q_K})^{R_1}$ rank-deficient (rank $<q^2$). In the extreme case when $K=q$ blocks are present, $m_j=1$ uniformly and the rank is lowest at $q$, which includes the case of $D$ being a diagonal unitary. 

This implies, as we show in Appendix \ref{app:BlockMaxEp}, that for these dual-unitary  ensembles constructed from block-diagonal T-dual matrices,
\beq
\begin{split}
\label{eq:maxEPblockD}
E(SU)&\leq 1- \frac{K}{q^2}, \\
e_p(U)=&\frac{1}{E(S)}E(SU) \leq \frac{q^2-K}{q^2-1}\leq e^*_p,
\end{split}
\eeq
where $e^*_p$ is the bound from Eq.~(\ref{eq:NontrivialityCondition}).
The smallest range of the entangling power occurs for the uniform
case when $K=q$, 
\beq
\label{eq:Ep_bound_BD}
E(SU)\leq 1-\frac{1}{q},\\\ e_p(U)\leq \frac{q}{q+1},
\eeq
Hence, for $q=2$ the maximum value $e_p(U)=2/3$  is already reached by the block-diagonal cases above and can be achieved even with diagonal
2-qubit gates. The largest range occurs for the case $K=2$, which interestingly coincides with the bound $e^*_p$ derived from the 
norm of the channel as a sufficiency condition for mixing. Thus none of the dual-unitary  constructions based on diagonal blocks, satisfy the sufficiency condition, Eq.~(\ref{eq:NontrivialityCondition}) for mixing: there are always members that do not mix. 

Whenever $(D_{q_1, \cdots, q_K})^{R_1}$ is rank-deficient the entangling power cannot be maximized as $K>1$.  As all block-diagonal based 
constructions are rank-deficient these will not reach the maximum entangling power, with the exception of $q=2$. Quite remarkably, the maximum depends only the number of blocks and not on the block sizes $m_j$.
For the case of qutrits, the uniform block-diagonal based ensemble can only reach a maximum entangling power of $3/4$. 
If we take the non-uniform case when $K=2$, $m_1=2$ and $m_2=1$, the maximum according to Eq.~(\ref{eq:maxEPblockD}) is $7/8=0.875$. 
Interestingly, in this case, the bound is not tight, and is not achieved. For ququads with $q=4$, if $K=2$, $m_1=m_2=2$ the bound on $e_p(U)$ is 
$14/15$ and is achieved. The distinction is that there are $8$ dimensional 2-unitary matrices in the bipartite qubit-ququad split: $\mathcal{H}^2 \otimes \mathcal{H}^4$ \cite{USinpreparation}, however there are no $6$ dimensional 2-unitary matrices with the   qubit-qutrit bipartition \cite{huber2018bounds}.

\subsubsection{Channels from block-diagonal based constructions}
 
We will study the CPTP maps $M_+[U]$ when $U=D_{q_1, \cdots, q_K} S$ and $U= S D_{q_1, \cdots, q_K}$ and denote them simply as $M_+[DS]$, and $M_+[SD]$.
Note from Eq.~(\ref{eq:M_pm_relation}) that $M_{\pm}[DS]=M_{\mp}[SD]$. Recall that $D_{q_1, \cdots, q_K}$ consists of $K$ unitary blocks, denoted $U_k$, on the diagonal of sizes $q_j=m_j q$, with $\sum_j m_j=q$.

From Eq.~(\ref{eq:M+UT}) and using Eq.~(\ref{eq:IdenityAST2}) we get 
\beq
\label{eq:M+DS}
M_+[DS]=\frac{1}{q}[(DS)^{T_2} (DS)^{T_2\dagger}]^{R_1}=\frac{1}{q}[D^{R_2} D^{R_2 \dagger}]^{R_1}.
\nonumber 
\eeq
%The block-diagonal nature of $D=D_{q_1, \cdots,q_K}$ implies that the nonzero elements are:
%\beq
%\br ij|D^{R_2} D^{R_2 \dagger}|i'j'\kt = \br ij|U_k^{R_2} U_l^{R_2 \dagger}|i'j'\kt,
%\eeq
%where
%\beq
%\begin{split}
%\label{eq:ij_range}
%i,j \in \left[ \sum_{r=1}^{k-1}m_r, \sum_{r=1}^k m_r -1 \right],
%\;i',j' \in \left[ \sum_{r=1}^{l-1}m_r, \sum_{r=1}^l m_r -1 \right].
%\end{split}
%\eeq
%Here $U_k^{R_2}$ is the realignment of the block $U_k$ which is considered as acting on $\mathcal{H}^{m_k} \otimes \mathcal{H}^q$,
%and hence is in general rectangular of shape $m_k^2\times q^2$, and $U_k^{R_2} U_l^{R_2 \dagger}$ are of shape $m_k^2 \times m_l^2$. The further realignment $R_1$ for $M_+$
%implies that 
%\beq
%\br j' j| M_+[DS] |i' i \kt = \frac{1}{q}\br j'j|[U_k^{R_2} U_l^{R_2 \dagger}]^{R_1}|i'i\kt.
%\eeq
%As the pairs $(i'i)$ and $(j'j)$ range over the same set of values, 
It is not hard to see that the CPTP map $M_+[DS]$ consists of 
diagonal (square) blocks $[U_k^{R_2} U_l^{R_2 \dagger}]^{R_1}$. There are in all $K^2$ such blocks with sizes in the set $\{ m_k m_l|1\leq k,l \leq K \}$, that is
\beq
\label{eq:MplusDS}
M_+[DS]=\frac{1}{q}\bigoplus_{k,l=1}^K [U_k^{R_2} U_l^{R_2 \dagger}]^{R_1}.
\eeq

In the simplest uniform case, there are $K=q$ blocks, all the $q_j=q$ and $m_j=1$, $M_+$ is diagonal. In this case $U_{k,l}^{R_2}$ are row vectors of length $q^2$ and hence the product $U_k^{R_2} U_l^{R_2 \dagger}$ is simply an inner product, which in terms of the original matrices is the Hilbert-Schmidt inner product. Hence the (unordered) spectrum of $M_+[DS]$ is 
\beq
\lambda_{kl}=\frac{1}{q}\tr(U_k U_l^{ \dagger}).
\eeq  
Hence there are $q$ eigenvalues $\lambda_{kk}=1$, and local unitary operations are needed to render the circuit mixing.
Also interestingly, if the blocks contain orthonormal unitary matrices, then all the other eigenvalues are $0$, equivalently $\lambda_{kl}=\delta_{kl}$, and represent the most mixing that can be obtained from this type of dual-unitaries. These continue to hold for the special case when $U_k$ are themselves diagonal matrices, and was considered recently in \cite{claeys2020ergodic}. In the uniform blocks case, the entangling power cannot exceed $q/(q+1)$ as shown in
Eq.~(\ref{eq:Ep_bound_BD}).

The maximum possible entangling power is when $K=2$, when $e_p(U)\le e^*_p$. It was
mentioned that there are gates $U$ such that $e_p(U)=e^*_p=(q^2-2)/(q^2-1)$ and $\lambda_1=1$.
It is shown in App.~(\ref{app:BlockMaxEp}) that if $U_k$ are 2-unitaries in mixed dimensions $m_k \times q$,
that is they satisfy the conditions of Eq.~(\ref{eq:2uni_conditions}), then the entangling power is $e_p^*$. 
However, these conditions imply that there are $K$ blocks in $M_+[DS]$ which are 
\beq
\frac{1}{q} [U_k^{R_2} U_k^{R_2 \dagger}]^{R_1}=\frac{1}{m_k}\mathbb{1}_{m_k^2}^{R_1}=|\phi^+_{m_k}\kt \br \phi^+_{m_k}|, 
\eeq
where $|\phi^+_{m_k}\kt$ is the maximally entangled state $\sum_{i=1}^{m_k}|ii \kt/\sqrt{m_k}$. Thus in general there are
$K$ eigenvalues that are $1$ and $\sum_k(m_k-1)=q-K$ zero eigenvalues. Hence $|\lambda_1|=1$, also in the particular case of $K=2$. As we have previously noted, there exists no matrices that satisfy these conditions in $q=3$ and $K=2$, as there exists no $2\times 3$ 2-unitary matrix. However there do exist such block matrices for $q=4$ when we could have two $8 \times 8$ blocks, each satisfying the conditions of Eq.~(\ref{eq:2uni_conditions}) as there exists 2-unitary matrices in $\mathcal{H}^2 \otimes \mathcal{H}^4$. This follows from the existence of a 4-party AME state of two qubits and 2 ququads, which is a corollary of the fact that 6 qubit  AME states exist \cite{huber2018bounds}. We show below that the quantum cat map is another example wherein
$e_p(U)=e_p^*$ but can be non-mixing.

Turning to $M_+[SD]$, which is also $M_-[DS]$, we get 
\beq
M_+[SD]=\frac{1}{q}[(SD)^{T_2} SD)^{T_2 \dagger}]= \frac{1}{q}[S D^{R_1} D^{R_1 \dagger}S]^{R_1},
\nonumber 
\eeq
where an identity in Eq.~(\ref{eq:IdentityR1T2}) is used. The difference with the previous
case that involved $R_2$, Eq.~(\ref{eq:M+DS}), is that here there no ``interference" terms between the blocks as a simple schematic of 
multiplying such realigned matrices will convince a reader. This results in 
\beq
\label{eq:M+SD}
M_+[SD]=M_-[DS]=\frac{1}{q}\sum_{j=1}^K[S U_j^{R_1} U_j^{R_1 \dagger}S]^{R_1}.
\eeq
This simplifies for $K=q$, when $U_j$ are uniformly $q\times q$ unitary matrices as
\beq
\label{eq:M+SDuniform}
M_+[SD]=M_-[DS]=\frac{1}{q}\sum_{j=1}^q U_j ^{\dagger} \otimes U_j^T.
\eeq
Thus unlike $M_+[DS]$ this is not diagonal and generically there are no other eigenvalue that is $1$ apart from the trivial one. 
Thus the same circuit could be chiral in the nature of their mixing. Also, note that in this case, using Eq.~(\ref{eq:Mprime})
$M_+[(u_1 \otimes u_2) DS (v_1 \otimes v_2)]=M_+[SD']$ where $D'=\bigoplus_j u_1 U_j v_2$. Thus the channels of locally transformed
operators are equivalent to the channels of other members of dual-unitary  matrices of the same form.

To construct dual-unitary  operator $U$ with an entangling power in excess of the bound $e^*_p$, which assures mixing, and
to construct Bernoulli circuits from 2-unitaries, we need to
look beyond the block-diagonal constructions. A few possibilities are outlined below.

\subsection{Dual-CUE \label{sec:dualcue}}
As shown above, the entangling power $e_p(U)$ of block diagonal dual-unitaries  is bounded by the number of blocks, 
and cannot exceed the sufficiency bound in Eq.~(\ref{eq:NontrivialityCondition}) for a mixing circuit. They certainly cannot result in Bernoulli circuits. Permutations and 
decorations thereof may exceed it and even reach the maximum possible, and for odd dimensions there exist cat maps that can do this.
However, as these are special constructions, an ensemble of dual-unitary  matrices with large entangling power still seems inaccessible. However,
in Ref. \cite{SAA2020}, we constructed a nonlinear dynamical map on the space of unitary operators, which acting iteratively, limit to dual operators and which have large entangling power.

We recall this procedure here for completeness. Define the map $\mathcal{M}_R : \mathcal{U} (q^2) \rightarrow \mathcal{U} (q^2)$, which is schematically represented as follows: 
\beq 
\mathcal{M}_R (U) : U \xrightarrow{\mathcal{R}} U^{R_2} \xrightarrow{\mathcal{PD}} V.
\label{eq: MRmap}
\eeq 
$\mathcal{R}$ is a linear map, $\mathcal{R} (U) = U^{R_2}$, whereas the map $\mathcal{PD}$ is nonlinear, such that $\mathcal{PD}(X) = V$, where $V$ is the closest unitary operator to $X$ (under any unitarily invariant norm), which is given by the polar decomposition $X = V\sqrt{XX^\dagger}$. It was observed  in \cite{SAA2020} that, starting from almost any unitary operator $U_0$, an iterative application of $\mathcal{M}_R$, $U_n = \mathcal{M}_R^n (U_0)$, produces unitary $U_n$ with an operator entanglement $E(U)$ monotonically approaching the maximum value $1- 1/q^2$ as $n \rightarrow \infty$, so that $\mathcal{M}_R^{\infty} (U_0)$ is dual-unitary .

\begin{figure}[h]
\includegraphics[scale=0.42]{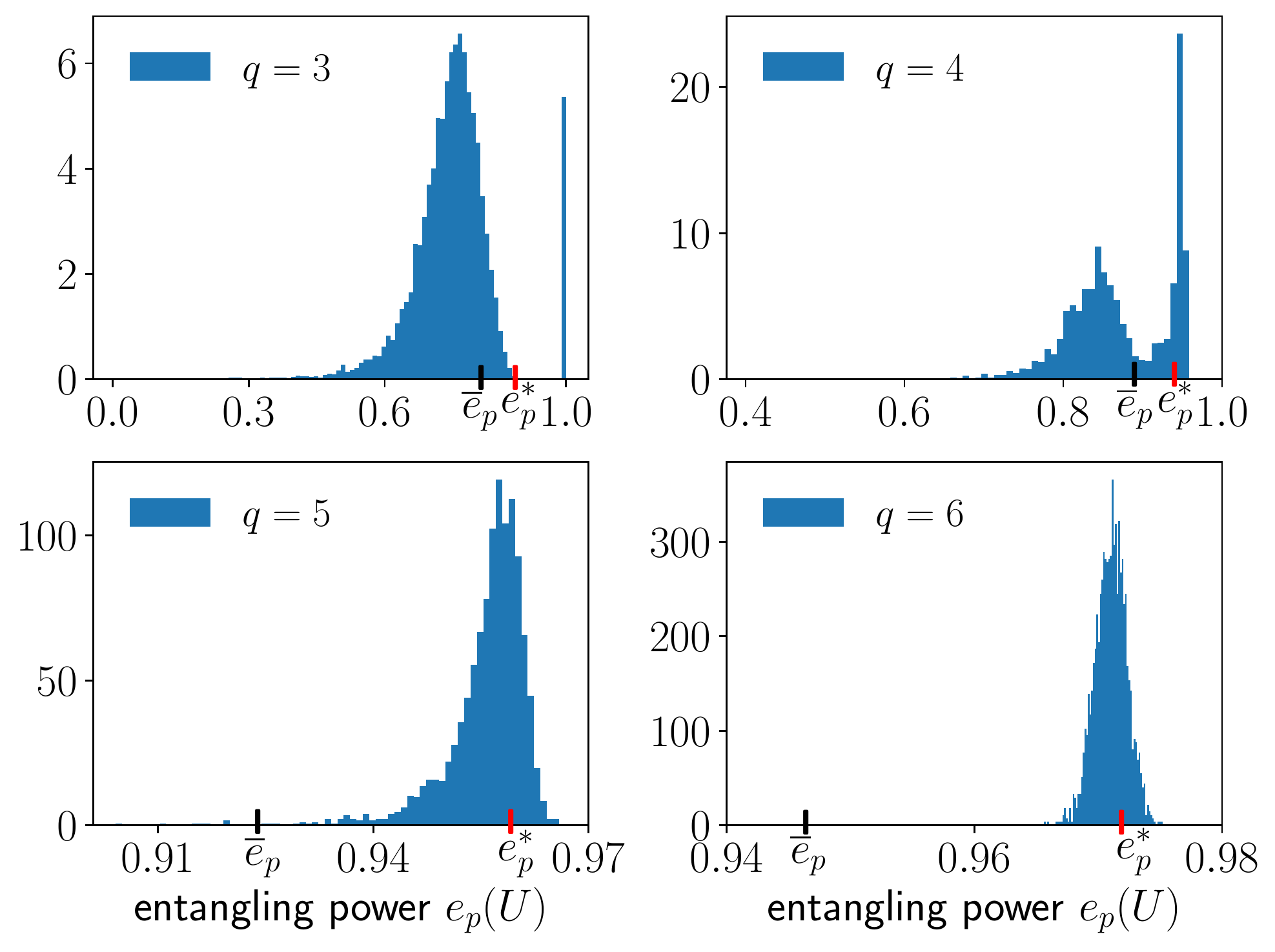}
\caption{Distribution of $e_p(U)$ of dual-CUE ensemble generated by applying the $M_R$ map on the CUE of ensemble.  It can be seen that for $q=3$ the $M_R$ map also produces a fraction  of 2-unitaries whose $e_p(U) =1$. If  $ e_p(U) > e_p^*$, the dual-unitary  circuit is guaranteed to be mixing. For comparison, the average entangling power of CUE $\bar{e}_p $ is also indicated.}
\label{fig:cuedist} 
\end{figure}

Furthermore, defining the Tsallis entropies,
\beq 
S_\alpha (U) = \frac{1 - \sum_j^{q^2} p_j^\alpha}{\alpha -1},
\nonumber 
\eeq  
where $p_j = \gamma_j/q^2$ (see Eq. (\ref{eq:sch})), the $\alpha = 2$ case corresponds to the operator entanglement $E(U)$ defined in  Eq. (\ref{eq:EUdefn}). Under the $\mathcal{M}_R$ map, it was proved in \cite{SAA2020} that
\beq 
S_{1/2} (U_{n+1}) \geq S_{1/2} (U_n),   
\nonumber 
\eeq 
showing a nondecreasing trend towards an entropy increase and making plausible that the limiting process results in 
a maximum entropy state.
From a dynamical systems perspective, the dual-unitaries are fixed points of the $\mathcal{M}_R^2$ map. To see this, let $U$ be a dual-unitary operator to begin with, then action of $\mathcal{M}_R^2$ is as follows: 
\beq 
U \xrightarrow{\mathcal{R}} U^{R_2} \xrightarrow{\mathcal{PD}} U^{R_2}  \xrightarrow{\mathcal{R}} U \xrightarrow{\mathcal{PD}} U. 
\nonumber 
\eeq 
Since $U$ is dual-unitary,  so that $U^{R_2}$ is unitary and it remains invariant under the $\mathcal{PD}$ map. As shown in \cite{SAA2020} there is overwhelming numerical evidence that  almost any unitary is in the basin of attraction of the dual-unitary fixed point set, and is reached under the $\mathcal{M}_R$ map. Thus if the starting unitaries are sampled from the Haar measure, the resultant ensemble of dual-unitary matrices, $\mathcal{M}_R^{\infty}(\text{CUE})$ was dubbed ``dual-CUE". The distribution of entangling power $e_p(U)$ for dual-CUE is given in Fig. (\ref{fig:cuedist}) in which the average entangling power of CUE, $\bar{e}_p = (q^2-1)/(q^2+1) $ \cite{Zanardi2001,Bhargavi2019} and $e_p^*$  are both indicated for comparison. 

Additionally, it is possible to define other maps wherein the partial transpose is added as a step in the protocol
so that the resultant matrices are 2-unitary. It was shown in \cite{SAA2020} that especially for $q=3$ and $q=4$, this was successful
in producing ensembles of 2-unitary matrices, with $e_p(U)=1$, from which Bernoulli circuits may be constructed. Thus, although these maps
result typically in numerically close to dual and 2-unitary matrices, they are for practical purposes as good as analytical ones 
and crucially enable us to explore gates with large entangling powers.  In Appendix (\ref{sec:appnice}), we provide examples of a dual-unitary operator with entangling power $e_p(U) = 8/9 > e_p^* = 7/8$ and 2-unitary operator obtained via this numerical procedure for $q=3$.  Note that this procedure is capable of identifying exact operators if 
the elements are simple enough to be recognized, as in these examples.

\begin{figure}[h] 
\includegraphics[scale=0.43]{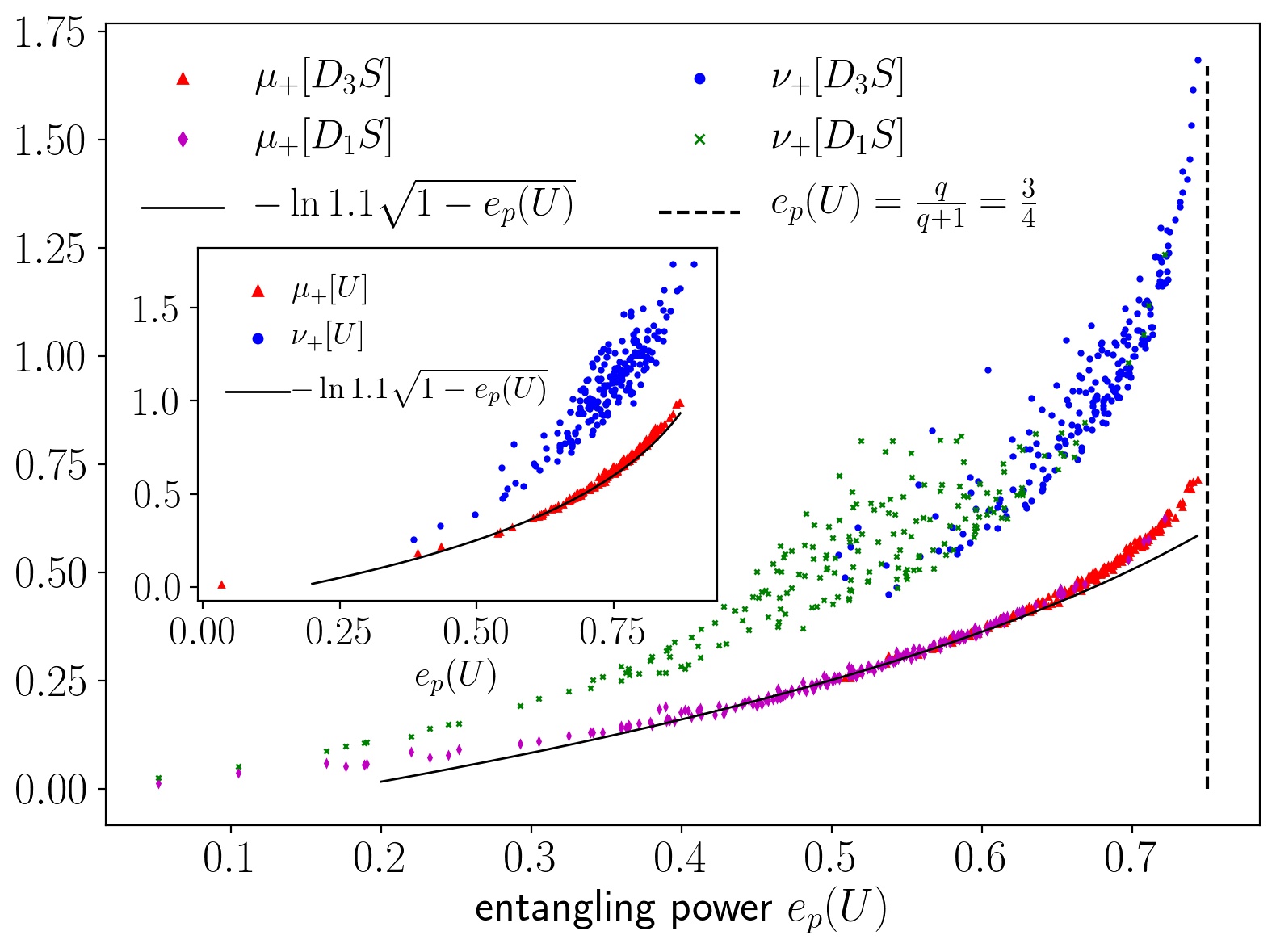} 
\caption{The maximal mixing rate $\nu_+$ and the average mixing rate $\mu_+$ for $q=3$ and $K=1, 3$  cases of dual-unitary  operators of the form $D_3S$,
and $D_1S$. In the former case the 3 blocks are random unitaries drawn from the CUE, while in the latter the diagonal entries are random 
phases drawn uniformly from the unit circle. The maximum possible entangling power $3/4$ is  indicated by dashed vertical line. The inset shows
the case of dual-CUE, again with $q=3$, which can explore a larger range of entangling power. }
\label{fig:blodia}
\end{figure}

The maximum and the average mixing rates $\nu_+$ and $\mu_+$ of Eqs.~(\ref{eq:MixingRate}) and ~(\ref{eq:MixingRateAvg}) are shown for $q=3$ in Fig.~(\ref{fig:blodia}), using three types of dual-unitary constructions. Two of these are based on block diagonal constructions, while  the $D_1S$
uses diagonal unitary matrices,  the $D_3S$ uses random $3 \times 3$ CUE blocks on the diagonals. While both of these have the same range of possible entangling 
power, namely $[0, q/(q+1)]$, it is seen that $D_3S$ typically has larger entangling power. The third ensemble used is the dual-CUE, which 
can have as large as $e_p(U)=1$, but $q=3$ is a peculiar case, where there is a gap at $e_p(U)=1$. Nevertheless we can see that this 
realizes gates of larger entangling power than the diagonal blocks based ensembles. For larger entangling power $\nu_+$ does not appear to be 
a function of the entangling power alone and these numerical results indicate that it is a more complex quantity. For smaller values of $e_p(U)$
it has been noted in Fig.~(\ref{fig:nearswap}) that it is simple function of the entangling power. On the other hand the single-particle averaged
$\mu_+$ seems to be determined by the entangling power and is close to the estimation from $|\lambda_1|$.

\subsection{From permutations \label{sec:permu}}

Permutation matrices form a source of unitary gates with a wide range of entangling powers \cite{Clarisse2005}. They have been used to construct gates with $e_p(U)=1$ for all dimensions except $q=2$ and $q=6$. The {\sc swap} and identity operators are two permutations with vanishing entangling power but have different nonlocal contents. The characterization of permutation matrices that are dual and T-dual does not seem to have been systematically exposed, although the ideas are present in \cite{Clarisse2005}.

If $[q]=\{1,\cdots,q\}$, a permutation of $q^2$ elements $[q] \times [q] $, written as $P(i,j) = (k_{ij},l_{ij})$, is represented by the pair of matrices $K = (k_{ij}) $ and $L = (l_{ij})$. Corresponding permutation operator $P$ acts on the product basis as $P(\ket{i}\ket{j}) = \ket{k_{ij}} \ket{l_{ij}}$, and represented as 
\beq
P = \sum_{i,j} \ket{k_{ij}} \ket{l_{ij}} \bra{i}\bra{j}.
\label{eq:permu}
\eeq
For example, for $q=3$, if $P$ permutes the elements of  $ [3] \times [3] $ as 
\beq
\begin{bmatrix}
11 & 12 & 13 \\
21 & 22 & 23 \\
31 & 32 & 33 
\end{bmatrix}
\xrightarrow{P} \, \begin{bmatrix}
11 & 23 & 31 \\
32 & 21 & 13 \\
33 & 12 & 22 
\end{bmatrix}, 
\nonumber 
\eeq
the matrices $K$ and $L$ are recognized as
\beq 
K=\begin{bmatrix}
{\bf 1} & {\bf 2} & {\bf 3} \\
3 & 2 & 1 \\
3 & 1 & 2 
\end{bmatrix},\; L=\begin{bmatrix}
{\bf 1} & 3 & 1 \\
{\bf 2} & 1 & 3 \\
{\bf 3} & 2 & 2
\end{bmatrix}.
\nonumber 
\eeq

As $P^{R_2} = \sum_{i,j} \ket{k_{ij}} \ket{i}  \bra{l_{ij}} \bra{j}$, if $\braket{k_{ij}}{k_{in}} = \delta_{jn}$, then $P^{R_2\, \dagger} P^{R_2}=I$.
It is also easy to see that the converse is true. Similarly considering conditions under which $P^{R_2} P^{R_2\, \dagger}=I$, one concludes that 
the permutation operator $P$ is dual-unitary,  that is $P^{R_2}$ is unitary, if and only if $\braket{k_{ij}}{k_{in}} = \delta_{jn}$ 
 and $\braket{l_{ij}}{l_{mj}} = \delta_{im}$. This imposes a constraint on the $K$ and $L$ matrices. The elements along the each row of the  $K$ matrix and the elements along the column of the $L$  matrix must be distinct. Thus the example above does represent a dual-unitary  permutation.

\begin{figure}[h] 
\includegraphics[scale=0.48]{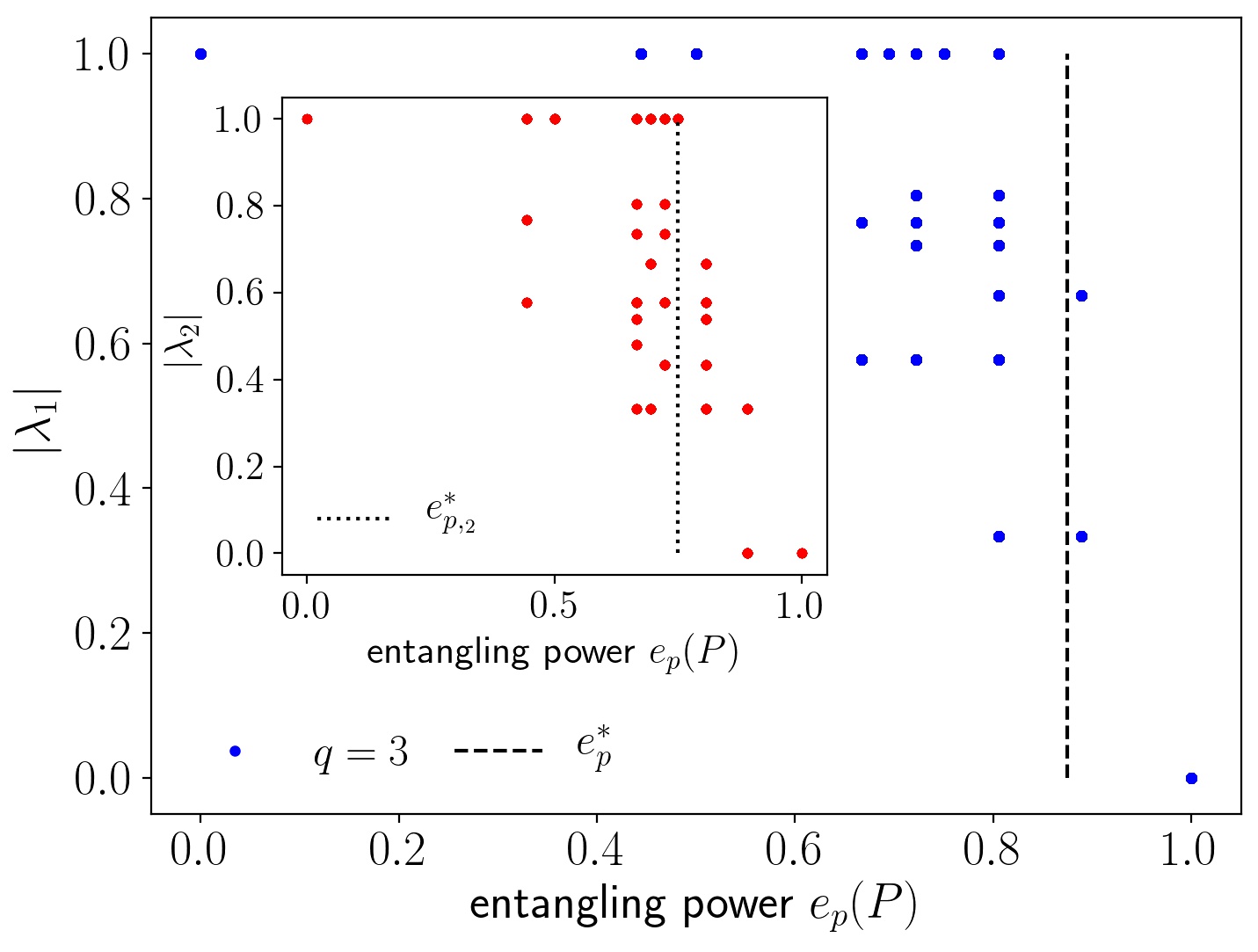} 
\caption{The largest and second (inset) largest eigenvalues $|\lambda_{1,2}|$  of the channel $M_+$ for all possible dual-unitary  permutations for qutrits, $q=3$. The figure also illustrates the bounds of the entangling power, beyond which the corresponding modes are definitely mixing.}
\label{fig:lambdualperm}
\end{figure}

 Similarly, the permutation operator $P$ is T-dual if and only if 
 $\braket{k_{ij}}{k_{mj}} = \delta_{im}$ and $\braket{l_{ij}}{l_{in}} = \delta_{jn}$. The elements in each column of the $K$ matrix and the elements in each row of the $L$ matrix must be distinct.
The conditions for dual and T-dual together defines the conditions for $P$ to be 2-unitary. This requires that both $K$ and $L$ be Latin squares, with unique entries along any row or column. In addition for the combined $KL$ to be a permutation, we must require that these Latin squares be orthogonal:
overlaying the squares result in no repetition of the order pairs in $[q]\times [q]$. 
Thus orthogonal Latin squares give unitary gates with $e_p(U)=1$.
The above example fails this condition, whereas the following satisfies it:  
\begin{align}
\begin{bmatrix}
11 & 12 & 13 \\
21 & 22 & 23 \\
31 & 32 & 33 
\end{bmatrix}
&\xrightarrow{P}  \begin{bmatrix}
11 & 23 & 32  \\
22 & 31 & 13 \\
33 & 12 & 21 
\end{bmatrix} \nonumber \\
& \equiv   \begin{bmatrix}
{\bf 1} & {\bf 2} & {\bf 3} \\
{\bf 2} & 3 & 1 \\
{\bf 3} & 1 & 2 
\end{bmatrix}\begin{bmatrix}
{\bf 1} & {\bf 3} & {\bf 2} \\
{\bf 2} & 1 & 3 \\
{\bf 3} & 2 & 1 
\end{bmatrix}.
\nonumber 
\end{align}	 
Figure~(\ref{fig:lambdualperm}) shows the largest and second largest eigenvalues of the channels $M_+[P]$ from all the possible dual-unitary 
permutations for $q=3$ as function of their entangling power. Many permutations share the same entangling power and interestingly many also have 
$|\lambda_1|=1$, unless $e_p(P)>e_p^*$ which is indicated by a vertical dashed line in the main figure. Note that 2-unitary permutations have $e_p(P)=1$
and $|\lambda_1|=0$. The inset shows the same for the second largest 
eigenvalue  and shows that if the entangling power exceeds $e_{p,2}^*$, indeed $|\lambda_2|<1$. \bla

While permutations provide a discrete set of dual, T-dual, and 2-unitaries, one may use a kind of gauge-freedom they enjoy to define continuous families that are not local unitarily connected. Multiplying by diagonal unitaries on either sides does not change the dual nature of the permutations, we define generically locally inequivalent ensembles of unitaries that are ``enphased" permutations as    
\beq 
P_{\theta} = \sum_{i,j} e^{i\theta_{ij}} \ket{k_{ij}} \ket{l_{ij}} \bra{i}\bra{j},
\label{eq:emphas}
\eeq     where $ -\pi \leq \theta_{ij} < \pi $. It is easy see that the same condition for duality, T-duality and 2-unitary  given above for $P$ continues to hold. However, if we take a single dual permutation,  the members of its enphased family will have different entangling powers.  The exceptions are 2-unitaries, which remain 2-unitary under any enphasing. This follows as 2-unitaries are both dual and T-dual, and enphasing does not change either of these properties.

\subsection{From cat maps \label{sec:cat}} 

It seems natural to look at quantized classically chaotic systems for dual and 2-unitary operators.   Quantum maps
have long been studied whose classical counterparts are dynamical systems with a high degree of chaoticity \cite{berry1979quantum},   in particular
the baker \cite{balazs1989quantized,saraceno1994towards,schack2000shifts}   and  the cat maps \cite{hannay1980quantization,Ford1991}.  The cat maps are linear automorphisms of the 2-torus that are hyperbolic, Anasov or C-systems.
Their quantization was first studied by Berry and Hannay \cite{hannay1980quantization} and provides a rich source of models
for which semiclassical methods are exact. But they also have peculiar properties such as periodicity in time, whose
origins lie in number-theoretic aspects of the classical cat itself \cite{hannay1980quantization,keating1991cat,dyson1992period}.
 As our concentration in this work
is not primarily with these interesting models, but rather use them as examples for the class of unitary operators to use
in quantum circuits, we restrict ourselves to a minimal description. Coupled cat maps from the point of view of duality have been studied in \cite{gutkin2016classical} and coupled quantum cat maps
have also been studied in \cite{mantica2019many}.

The Arnold cat map is area-preserving and defined on the torus $[x,p] \in [0,1)^2$ by
\beq
\begin{bmatrix}
x'\\p'
\end{bmatrix}
=
\begin{bmatrix}
1 & 1\\1&2 
\end{bmatrix}
\begin{bmatrix}
x\\p
\end{bmatrix} \text{mod\; 1}.
\eeq
As we are interested in bipartite unitary operators, two coupled cat maps are considered in which case the automorphism is of the 4-dimensional 
torus and the map on it is symplectic. There have been previous studies involving such coupled maps, such as in   \cite{rivas2000quantization}.  However, we  define the quantum propagator via the classical generating function, a route taken in the absence of a Hamiltonian for example
in the construction of the quantum bakers map \cite{balazs1989quantized}. The linearity of the map implies a quadratic 
generating function. Considering the generating function on $\mathbb{R}^4$
\beq
\label{eq:Cat_action}
F_1(x_1,x_2,x_1',x_2')=a x_1x_2+b x'_1x'_2+c(x_1x'_2+x'_1x_2),
\eeq
a symplectic transformation is obtained from $p_i=\partial F_1/\partial x_i$ and $p'_i=-\partial F_1/\partial x'_i$ \cite{goldstein2002classical}. This linear map is a differentiable 
one on the compactified unit torus $\mathbb{T}^4$ if $a$, $b$, $c$ are integers and $c|a,\, c|b$. We take $a=-1$, $b=-2$ and $c=1$ to get 
the 4-dimensional symplectic map
\beq
\label{eq:4DcatmapClassical}
\begin{bmatrix}
x_1'\\p'_1\\x_2'\\p_2'
\end{bmatrix}
=
\begin{bmatrix}
1 & 0&0&1\\0&1&1&0\\
0&1&2&0\\
1&0&0&2 
\end{bmatrix}
\begin{bmatrix}
x_1\\p_1\\x_2\\p_2
\end{bmatrix} \text{mod\; 1},
\eeq
which we recognize as Arnold cat maps on the $(x_1,p_2)$ and $(x_2,p_1)$ subspaces.
The corresponding quantum propagator is given by the prescription of Jordan and Dirac
$\br x'_1 x'_2|U|x_1 x_2\kt = e^{i F_1/\hbar}$ \cite{jordan1926kanonische,dirac1945analogy}.   On the unit torus with $q$ states in each degree of freedom $h=1/q$
and the position and momentum take discrete values $k/q$ with $0 \leq k \leq q-1$. This leads to the quantum cat map
corresponding to the classical map in Eq.~(\ref{eq:4DcatmapClassical}) acting on the bipartite system in 
$\mathcal{H}^q \otimes \mathcal{H}^q$:
\begin{equation}
\matrixel{k\alpha}{U_C}{j\beta} = \frac{1}{q} \exp(-\frac{2\pi i}{q} [k\alpha + 2 j\beta -( k\beta + j\alpha) ]),
\label{eq:cat}
\end{equation}
where $\alpha, \beta, j, k \in \{0,1, \cdots, q-1\}$ label position eigenkets.

The operator $U_C$ is dual-unitary for all $q$ and it is 2-unitary operator for all odd local dimension $q$ (See appendix \ref{app:cat}).  
To our knowledge, this is the first time a simple non-permutation class of unitaries have been 
shown to be 2-unitary. Note that this is a complex Hadamard matrix.  Examples of complex Hadamard matrices are known to be 2-unitary for $q=3$   \cite{Goyeneche2015}. 
Besides, being related to dynamical systems, and their similarity to Fourier transforms, they may be implementable in efficient quantum circuits or other physical setups \cite{weinstein2001implementation,wang2011simple}. Thus while $e_p(U_C)=1$ for $q$ odd, it is amusing that for $q$ even, \beq
\label{eq:cat_ep}
e_p(U_C)=e_p^*=\frac{q^2-2}{q^2-1}
\eeq
is at the border of the sufficiency condition for mixing.

While the unitary $U_C$ does not provide a family or ensemble of dual operators, a one parameter family may be found by setting 
$a=0$ and $c=1$ for arbitrary $b$ in the generating function Eq.~(\ref{eq:Cat_action}). The corresponding quantum maps 
are
\begin{equation}
\matrixel{k\alpha}{U_C(b)}{j\beta} = \frac{1}{q} \exp(\frac{2\pi i}{q} [b j\beta +( k\beta + j\alpha) ]). 
\label{eq:cat_0b1}
\end{equation}
This forms a dual-unitary family for all $b \in \mathbb{R}$. However its entangling power is no larger than $q/(q+1)$ (reached when $b=1$), which one recognizes
from Eq.~(\ref{eq:Ep_bound_BD}) as that of a uniform block-diagonal case. Indeed, it is not hard to see that $U_C(b)=(F_q \otimes F_q) S D_{q^2}(b)$
where $F_q$ is the $q-$ dimensional discrete Fourier transform and $\br k \alpha |D_{q^2}(b)|j \beta\kt=e^{2 \pi i b(j \beta)/q}\delta_{kj}\delta_{\alpha \beta}$ is a diagonal unitary. Hence this family of cats is locally equivalent to a swap-diagonal family such as we considered earlier, irrespective of the parity of $q$. We note that $U_C$ in Eq.~(\ref{eq:cat}) also has a decomposition of the form $S D_{q^2} (F_q \otimes F_q) D_{q^2}'$ where $D$ are diagonal matrices, but the fact that the local product is sandwiched between two nonlocal ones makes the structure special and require that the parameters $a,b$ be fine-tuned. Complex Hadmard matrices have been used in similar constructions by \cite{gutkin2020local} and it will be interesting to compare them and 
find their entangling power.

\subsubsection{Cat map channels}
The channel $M_+$ for the quantum cat map in Eq.~(\ref{eq:cat}) is given by 
\beq
\begin{split}
\mel{k\alpha}{M_+[U_C]}{j\beta} =& \frac{1}{q}  \mel{k\alpha}{(U_C^{T_2}U_C^{T_2\dagger})^{R_1}}{j\beta}\\ 
%=& \frac{1}{q^3} \sum_{mn} e^{-\frac{2\pi i}{q}(\beta - j)n} e^{-\frac{2\pi i}{q}2(\alpha - k)m}  e^{\frac{2\pi i}{q} (\alpha \beta - jk)} \nonumber \\ 
=& \frac{1}{q^2} e^{\frac{2\pi i}{q} (\alpha \beta - jk)} \delta_{j\beta} \sum_{m=0}^{q-1} e^{-\frac{2\pi i}{q}2(\alpha - k)m}\\
=&  \frac{1}{q} e^{\frac{2\pi i}{q} (\alpha \beta - jk)} \delta_{j\beta} \delta_{2(\alpha-k) \mod q ,0}. 
\end{split}
\label{eq:catmp}
\eeq 
For $q$ odd, $\delta_{2(\alpha-k) \mod q ,0}=\delta_{\alpha k}$ and $M_+[U_C]=|\Phi^+\kt \br \Phi^+|$ as behoves a 2-unitary cat, and the circuit is Bernoulli. 
For $q$ even, the cat map $U_C$ is dual but not 2-unitary. The $M_+$ in Eq. (\ref{eq:catmp}) takes a simple form as 
\begin{equation}
\label{eq:M+Cat}
\begin{split}
\mel{k\alpha}{M_+[U_C]}{j\beta} &= \frac{1}{q} \delta_{j,\beta}\delta_{k,\alpha} + \frac{1}{q} (-1)^\beta \delta_{j,\beta} \delta_{k, \alpha \pm \frac{q}{2}},\\
M_+[U_C]& = \op{\Phi^+}{\Phi^+} + \op{\Psi}{\overline{\Psi}} ,
\end{split}
\end{equation}
where $\ket{\Phi^+}$ is the maximally entangled state in Eq.~(\ref{eq:iso}) and 
\beq
\label{eq:psi_psibar}
\begin{split}
\ket{\Psi} &= \frac{1}{\sqrt{q}} \sum_{k=0}^{q-1} \ket{k, \left(k+\frac{q}{2}\right)\mod (q)}, 
 \\ \ket{\overline{\Psi}}  &= \frac{1}{\sqrt{q}} \sum_{k=0}^{q-1} (-1)^k \ket{kk},
\end{split}
%\nonumber 
\eeq 
are two other maximally entangled states.
The operator $\op{\Phi^+}{\Phi^+}$ as always corresponds to the trivial eigenvalue, and  $\op{\Psi}{\overline{\Psi}}$ is non-diagonalizable, as $\ket{\Psi}$ and $\ket{\overline{\Psi}}$ are orthogonal. However, it is easy to see that, 
\begin{equation}
M_+^n[U_C] = \op{\Phi^+}{\Phi^+}, \qquad \forall n \geq 2, 
\end{equation}
and hence the cat map for $q$ is such that correlations vanish after the second time step. We also note that $M_-[U_C]=M_+[U_C]$
as $U_C S=S U_C$, the cat is parity or exchange symmetric.

Under local single-particle unitaries though the cat map channel changes as Eq.~(\ref{eq:Mprime}). It is sufficient to consider
one local unitary to get $M_+[U_C']=(u \otimes u^*) M_+[U_C]$, where $u\in \mathcal{U}(q)$. Then in general $M^2_+[U_C'] \neq \op{\Phi^+}{\Phi^+}$. 
From Eq.~(\ref{eq:M+Cat}) $ M_+[U_C']=\op{\Phi^+}{\Phi^+}+ (u \otimes u^*)  \op{\Psi}{\overline{\Psi}}$. Therefore any state that is orthogonal to 
both $|\Phi^+\kt$ and $|\overline{\Psi}\kt$ is a (right) eigenstate of $M_+[U_C']$ with eigenvalue $0$, and hence the channel is rank-2. Ignoring the trivial
eigenvalue $1$, we get 
\beq
\lambda_1(U_C')=\br \overline{\Psi}|(u \otimes u^*) |\Psi\kt=\frac{1}{q}\sum_{k,l=0}^{q-1}(-1)^k u_{lk} u^*_{(l+\frac{q}{2})\, k}. 
\eeq 
If $u$ is the discrete Fourier transform on $q$ sites,
\[
u_{kl}=\frac{1}{\sqrt{q}}\exp[2 \pi i (l+\varphi_1)(k+\varphi_2)/q],
\]
with $0 \leq \varphi_{1,2}<1$ being phases, then $\lambda_1(U_C')=\cos(\pi \varphi_2)$. Thus the dual circuit
can be rendered non-ergodic if $\varphi_2=0$ or ergodic if $\varphi_2 \neq 0$.This shows that the cat map for $q$ even, with suitable
local unitaries, is an example of a case when $e_p(U)=e_p^*$, but the circuit is non-mixing. 

 A generic choice of $u$ will however render the circuit mixing.
Indeed, if $u$ is chosen from the Haar measure, it is a straightforward application of Eq.~(\ref{eq:appid2}) to get the local-unitary averaged
\beq
\label{eq:Lambd1AvgCat}
\overline{|\lambda_1(U_C')|^2}=\frac{1}{q^2-1}=1-e_p(U_C).
\eeq
This quantum cat map is therefore a solvable case wherein the general relation between the largest nontrivial 
local single-particle averaged eigenvalue and the entangling power, Eq.~(\ref{eq:AvgLambda1}) takes this exact and simple form.

%The $\mu_+(U)$ Vs $e_p(U)$ for $q=3$ and $q=4$ for both $DS$ ensemble and Dual-CUE ensemble is shwon in Fig. (\ref{fig:dualcue}). The dependece of $\mu_+(U)$ on $e_p(U)$ can be clearly seen from this plots. The maximum $\mu_+(U)$ for $DS$ ensemble for $q = 3$ and $q=4$ is given as 0.62 and 0. 7  \blu (Suhail: give exact numbers)\bla   where as for Dual-CUE it can be 1.0 and 1.5 for $q = 3$ and $q=4$ respectively.  \blu (Suhail: give exact numbers) \bla 
%
%\begin{figure}
%\includegraphics[scale=0.6]{average_mu_1_ep_d_3_4_DS_dual_cue.png}
%\label{fig:mup1}
%\caption{Plots of  $\mu_+(U)$ Vs $e_p(U)$ for $q=3$ and $q=4$. Left panel shows tha variation 
%$\mu_+(U)$ Vs $e_p(U)$ for ensemble DS given in Eq. and the left panel is for dual-CUE.}
%\label{fig:dualcue}
%\end{figure}

\section{Summary and discussions \label{sec:discu}}

This paper has considered the recently introduced dual-unitary  circuits that are models of many-body quantum 
chaotic systems \cite{Bertini2019}. We showed that the partial-transpose and realignment operations central to the problem of
separability of quantum states \cite{peres1996separability,chen2002matrix,rudolph2003some}, play a central role in these circuits as well. Defining T-dual gates then, we showed 
how the CPTP maps that govern the decay of correlations along the lightcone is related to the partial-transpose
operation. This leads naturally to Bernoulli circuits that are at the apex of a putative quantum ergodic hierarchy.
These are made of 2-unitaries or perfect tensors which remain unitary under both realignment and partial transpose.
Determining  necessary and sufficient conditions for the circuit to be Bernoulli, we showed how two-site observables
retain the property of having Bernoulli correlations. 

A simple observation that the norm of the CPTP maps determining the decay of correlations is directly related to 
the entangling power of the two-particle unitary building the circuit, enables us to derive a sufficiency
condition for mixing. There are unitaries $U$ such that $e_p(U)=e_p^*$ and the circuit is not 
mixing, and not even ergodic. For the case of qubits it is also true that if $e_p(U)\leq e_p^*$, there 
exists local unitary transformations such that the circuit can be made non-ergodic. This follows
from the results in the Supplementary material of \cite{Bertini2019}. It is unclear if this 
generalizes to $q>2$, but we believe that this true. However, increasing entangling power does
ensure that an increasing number of modes are rendered mixing, whatever the local unitaries maybe.

While the entangling power are local (single-site or single-particle)
unitary invariants represent local fields, the mixing properties of the circuit are not.
We show however that local unitary averaged mixing rates depend simply and universally on the entangling power. 
That these averages are in fact typical values is shown by numerical results. This paper has also systematically 
enumerated several methods of constructing dual-unitary  operators. Importantly, 
we also calculate the entangling power in these constructions as indeed these play a central part in the ergodic
properties of the quantum circuit.

There are very many natural directions that this work throws up. The use of T-duality by itself or in conjunction with 
realignment, such that the unitary remains unitary under serial applications of these rearangements, in quantum circuits maybe interesting. This is related to using operators other than dual-unitary  ones. The rudimentary phase-diagram where 
the ergodicity properties depend on the entangling power alone, will likely expand to include at least two LUI
quantities such as the operator entanglements $E(U)$ and $E(US)$ or the entangling power and the so-called gate typicality \cite{Bhargavi2019}.
Another direction is the use of dual-unitary  operators with particles with different local dimensions, such as a mixture of 
qubits and qutrits \cite{USinpreparation}. A closely related possibility is to have multi-particle interactions, rather than only nearest neighbor
ones. For example, a tripartite 3-body interaction can enable Bernoulli or perfect tensor circuits even with qubits as there
are AME states of 6 qubits \cite{huber2018bounds,USinpreparation}. 

Other directions that seem intriguing are comparisons of the ergodic properties as discussed via
correlation functions and those from standard quantum chaos literature, such as spectral statistics. 
Finite systems, with recurrences are natural in this context, and the nature of eigenstates of the Floquet 
operator and the development of complexity in nonequilibrium states are of interest. Also, we have 
used a cat operator that is capable of being a 2-unitary. It will be interesting to study other physical
systems whose Floquet operators are indeed 2-unitary or in general dual operators. 
It is unclear to us what quantity would play the role of entangling power in organizing the ergodic
hierarchy of possible classical limits of such dual-unitary  circuits.
\bla

\begin{acknowledgments}
AL thanks Toma{\v{z}} Prosen for introducing  dual-unitary circuits during a seminar and for subsequent discussions in the summer of 2019.
He is also grateful for the hospitality of the MPIPKS, Dresden, that made this possible. We are happy to acknowledge a detailed feedback 
from  Karol \.{Z}yczkowski of a preliminary version of this paper.  Email clarifications from Boris Gutkin are much appreciated. SA and SAR acknowledge support by the International Centre for Theoretical Sciences (ICTS), Bangalore, during a visit for participating in the program  (Thermalization, Many body localization and Hydrodynamics (Code: ICTS/hydrodynamics2019/11)) relevant to this work.    

\end{acknowledgments}
\bibliography{ent_local}

\appendix 

\section{Bernoulli circuits: Correlation between extended operators \label{app:extende} }
In this appendix, we consider other cases of two-site observables in  Bernoulli circuits than treated in the main text. For the case $x_1,x_2 \neq t-\frac{1}{2}, t$, the repeated application of duality relation (\ref{eq:dual2}) imply the vanishing correlation function which can be seen from the following diagrammatical evaluation, $C^{ijkl}(x_1,x_2,t) = $
\begin{equation} 
\frac{1}{q^{4t}}
\btp [baseline=(current  bounding  box.center),scale=0.28] 

\foreach \i in {0,...,2} %left v shaped 
{
\draw [thick,fill=purple,rounded corners=2.2pt] (2*\i,5 + 2*\i) rectangle (1+2*\i,6 + 2*\i);
}

\foreach \i in {0,...,1} %left v shaped 
{
\draw [thick,fill=purple,rounded corners=2.2pt] (-2-2*\i,7 + 2*\i) rectangle (-1-2*\i,8 + 2*\i);
}

\draw [thick,fill=purple,rounded corners=2.2pt] (0,9) rectangle (1,10);

%lower figure 

\foreach \i in {0,...,2} %left v shaped 
{
\draw [thick,fill=teal,rounded corners=2.2pt] (2*\i,-5 - 2*\i) rectangle (1+2*\i,-4 - 2*\i);
}

\foreach \i in {0,...,1} %left v shaped 
{
\draw [thick,fill=teal,rounded corners=2.2pt] (-2-2*\i,-7 - 2*\i) rectangle (-1-2*\i,-6 - 2*\i);
}

\draw [thick,fill=teal,rounded corners=2.2pt] (0,-9) rectangle (1,-8);

%lines now 
\foreach \i in {0,...,3}
{
\draw [thick] (3+2*\i, -2 -2*\i) -- (3.5 + 2*\i, -1.5 - 2*\i) -- (3.5+2*\i,2.5+2*\i) -- (3+2*\i,3+2*\i);
\draw [thick] (-2-2*\i, -2 -2*\i) -- (-2.5 - 2*\i, -1.5 - 2*\i) -- (-2.5-2*\i,2.5+2*\i) -- (-2-2*\i,3+2*\i);
\draw [thick] (1+2*\i, 2+2*\i) -- (2+2*\i,3+2*\i);
\draw [thick] (1+2*\i, -1-2*\i) -- (2+2*\i,-2-2*\i);
\draw [thick] (-2*\i, 2+2*\i) -- (-1-2*\i,3+2*\i);
\draw [thick] (-2*\i, -1-2*\i) -- (-1-2*\i,-2-2*\i);
}

\foreach \i in {0,...,2}
{
\draw [thick] (2 + 2*\i, 4 +2*\i) -- (1+2*\i,5+2*\i);
\draw [thick] (-1 - 2*\i, 4 +2*\i) -- (-2*\i,5+2*\i);
\draw [thick] (2+2*\i, -3 -2*\i) -- (1+2*\i,-4-2*\i);
\draw [thick] (-1-2*\i, -3 -2*\i) -- (-2*\i,-4-2*\i); 
}

\foreach \i in {0,...,1}
{
\draw [thick] (1 + 2*\i, 6 +2*\i) -- (2+2*\i,7+2*\i);
\draw [thick] ( -2*\i, 6 +2*\i) -- (-1-2*\i,7+2*\i);
\draw [thick] (1 + 2*\i, -5 -2*\i) -- (2+2*\i,-6-2*\i);
\draw [thick] ( -2*\i, -5 -2*\i) -- (-1-2*\i,-6-2*\i);
}

\draw [thick] (2,8) -- (1,9);
\draw [thick] (-1,8) -- (0,9); 
\draw [thick] (2,-7) -- (1,-8);
\draw [thick] (-1,-7) -- (0,-8); 

\foreach \i in {0,...,4}
{
\draw [thick] (-8+4*\i,10) -- (-8.5+4*\i,10.5);
\draw [thick] (-7+4*\i,10) -- (-6.5+4*\i,10.5);
\draw [thick] (-8+4*\i,-9) -- (-8.5+4*\i,-9.5);
\draw [thick] (-7+4*\i,-9) -- (-6.5+4*\i,-9.5);
}

\draw [thick] (0,0) to[out =110, in = -110] (0,1);
\draw [thick] (1,0) to[out=70, in=-70] (1,1); 

\draw[thick, fill=black] (-0.1,0.5) circle (0.1cm);
\draw[thick, fill=black] (1.1,0.5) circle (0.1cm);
\Text[x=-0.8,y=0.5]{$a_i$}
\Text[x=1.8,y=0.5]{$a_j$}

\draw[thick, fill=black] (3.5,10.5) circle (0.1cm);
\draw[thick, fill=black] (5.5,10.5) circle (0.1cm);
\Text[x=3.5,y=11.2]{$a_k$}
\Text[x=5.5,y=11.2]{$a_l$}

\draw [thick,fill=purple,rounded corners=2.2pt] (0,1) rectangle (1,2);
\draw [thick,fill=teal,rounded corners=2.2pt] (0,-1) rectangle (1,0);

\foreach \i in {0,...,3}
{
\draw [thick] (2+2*\i,4+2*\i) -- (3+2*\i,4+2*\i);
\draw [thick] (2+2*\i,3+2*\i) -- (3+2*\i,3+2*\i);
\draw [thick] (2+2*\i,-2-2*\i) -- (3+2*\i,-2-2*\i);
\draw [thick] (2+2*\i,-3-2*\i) -- (3+2*\i,-3-2*\i);
\draw [thick] (-1-2*\i,4+2*\i) -- (-2-2*\i,4+2*\i);
\draw [thick] (-1-2*\i,3+2*\i) -- (-2-2*\i,3+2*\i);
\draw [thick] (-1-2*\i,-2-2*\i) -- (-2-2*\i,-2-2*\i);
\draw [thick] (-1-2*\i,-3-2*\i) -- (-2-2*\i,-3-2*\i);
}
\etp .
\nonumber 
\end{equation} 

For $x_1,x_2 = t-\frac{1}{2}$, by again using duality conditions  (\ref{eq:dual2}),  the vanishing correlation function can be seen from the correlation function $C^{ijkl}(x_1,x_2,t) = $
\begin{equation} 
\frac{1}{q^{4t}}
\btp [baseline=(current  bounding  box.center),scale=0.28] 

\foreach \i in {0,...,2} %left v shaped 
{
\draw [thick,fill=purple,rounded corners=2.2pt] (2*\i,5 + 2*\i) rectangle (1+2*\i,6 + 2*\i);
}

\foreach \i in {0,...,1} %left v shaped 
{
\draw [thick,fill=purple,rounded corners=2.2pt] (-2-2*\i,7 + 2*\i) rectangle (-1-2*\i,8 + 2*\i);
}

\draw [thick,fill=purple,rounded corners=2.2pt] (0,9) rectangle (1,10);

%lower figure 

\foreach \i in {0,...,2} %left v shaped 
{
\draw [thick,fill=teal,rounded corners=2.2pt] (2*\i,-5 - 2*\i) rectangle (1+2*\i,-4 - 2*\i);
}

\foreach \i in {0,...,1} %left v shaped 
{
\draw [thick,fill=teal,rounded corners=2.2pt] (-2-2*\i,-7 - 2*\i) rectangle (-1-2*\i,-6 - 2*\i);
}

\draw [thick,fill=teal,rounded corners=2.2pt] (0,-9) rectangle (1,-8);

%lines now 
\foreach \i in {0,...,3}
{
\draw [thick] (3+2*\i, -2 -2*\i) -- (3.5 + 2*\i, -1.5 - 2*\i) -- (3.5+2*\i,2.5+2*\i) -- (3+2*\i,3+2*\i);
\draw [thick] (-2-2*\i, -2 -2*\i) -- (-2.5 - 2*\i, -1.5 - 2*\i) -- (-2.5-2*\i,2.5+2*\i) -- (-2-2*\i,3+2*\i);
\draw [thick] (1+2*\i, 2+2*\i) -- (2+2*\i,3+2*\i);
\draw [thick] (1+2*\i, -1-2*\i) -- (2+2*\i,-2-2*\i);
\draw [thick] (-2*\i, 2+2*\i) -- (-1-2*\i,3+2*\i);
\draw [thick] (-2*\i, -1-2*\i) -- (-1-2*\i,-2-2*\i);
}

\foreach \i in {0,...,2}
{
\draw [thick] (2 + 2*\i, 4 +2*\i) -- (1+2*\i,5+2*\i);
\draw [thick] (-1 - 2*\i, 4 +2*\i) -- (-2*\i,5+2*\i);
\draw [thick] (2+2*\i, -3 -2*\i) -- (1+2*\i,-4-2*\i);
\draw [thick] (-1-2*\i, -3 -2*\i) -- (-2*\i,-4-2*\i); 
}

\foreach \i in {0,...,1}
{
\draw [thick] (1 + 2*\i, 6 +2*\i) -- (2+2*\i,7+2*\i);
\draw [thick] ( -2*\i, 6 +2*\i) -- (-1-2*\i,7+2*\i);
\draw [thick] (1 + 2*\i, -5 -2*\i) -- (2+2*\i,-6-2*\i);
\draw [thick] ( -2*\i, -5 -2*\i) -- (-1-2*\i,-6-2*\i);
}

\draw [thick] (2,8) -- (1,9);
\draw [thick] (-1,8) -- (0,9); 
\draw [thick] (2,-7) -- (1,-8);
\draw [thick] (-1,-7) -- (0,-8); 

\foreach \i in {0,...,4}
{
\draw [thick] (-8+4*\i,10) -- (-8.5+4*\i,10.5);
\draw [thick] (-7+4*\i,10) -- (-6.5+4*\i,10.5);
\draw [thick] (-8+4*\i,-9) -- (-8.5+4*\i,-9.5);
\draw [thick] (-7+4*\i,-9) -- (-6.5+4*\i,-9.5);
}

\draw [thick] (0,0) to[out =110, in = -110] (0,1);
\draw [thick] (1,0) to[out=70, in=-70] (1,1); 

\draw[thick, fill=black] (-0.1,0.5) circle (0.1cm);
\draw[thick, fill=black] (1.1,0.5) circle (0.1cm);
\Text[x=-0.8,y=0.5]{$a_i$}
\Text[x=1.8,y=0.5]{$a_j$}

\draw[thick, fill=black] (7.5,10.5) circle (0.1cm);
\draw[thick, fill=black] (-6.5,10.5) circle (0.1cm);
\Text[x=7.5,y=11.2]{$a_k$}
\Text[x=-6.5,y=11.2]{$a_l$}

\draw [thick,fill=purple,rounded corners=2.2pt] (0,1) rectangle (1,2);
\draw [thick,fill=teal,rounded corners=2.2pt] (0,-1) rectangle (1,0);

\foreach \i in {0,...,3}
{
\draw [thick] (2+2*\i,4+2*\i) -- (3+2*\i,4+2*\i);
\draw [thick] (2+2*\i,3+2*\i) -- (3+2*\i,3+2*\i);
\draw [thick] (2+2*\i,-2-2*\i) -- (3+2*\i,-2-2*\i);
\draw [thick] (2+2*\i,-3-2*\i) -- (3+2*\i,-3-2*\i);
\draw [thick] (-1-2*\i,4+2*\i) -- (-2-2*\i,4+2*\i);
\draw [thick] (-1-2*\i,3+2*\i) -- (-2-2*\i,3+2*\i);
\draw [thick] (-1-2*\i,-2-2*\i) -- (-2-2*\i,-2-2*\i);
\draw [thick] (-1-2*\i,-3-2*\i) -- (-2-2*\i,-3-2*\i);
}
\etp .
\nonumber 
\end{equation} 

The vanishing correlation function along the lightcone, $x_1, x_2 = t-\frac{1}{2}$, due to the application of T-duality condition (\ref{eq:tdual2}), and $C^{ijkl}(x_1,x_2,t) = $  
\begin{equation} 
\frac{1}{q^{4t}}
\btp [baseline=(current  bounding  box.center),scale=0.28] 

\foreach \i in {0,...,2} %left v shaped 
{
\draw [thick,fill=purple,rounded corners=2.2pt] (2*\i,5 + 2*\i) rectangle (1+2*\i,6 + 2*\i);
}

\foreach \i in {0,...,1} %left v shaped 
{
\draw [thick,fill=purple,rounded corners=2.2pt] (-2-2*\i,7 + 2*\i) rectangle (-1-2*\i,8 + 2*\i);
}

\draw [thick,fill=purple,rounded corners=2.2pt] (0,9) rectangle (1,10);

%lower figure 

\foreach \i in {0,...,2} %left v shaped 
{
\draw [thick,fill=teal,rounded corners=2.2pt] (2*\i,-5 - 2*\i) rectangle (1+2*\i,-4 - 2*\i);
}

\foreach \i in {0,...,1} %left v shaped 
{
\draw [thick,fill=teal,rounded corners=2.2pt] (-2-2*\i,-7 - 2*\i) rectangle (-1-2*\i,-6 - 2*\i);
}

\draw [thick,fill=teal,rounded corners=2.2pt] (0,-9) rectangle (1,-8);

%lines now 
\foreach \i in {0,...,3}
{
\draw [thick] (3+2*\i, -2 -2*\i) -- (3.5 + 2*\i, -1.5 - 2*\i) -- (3.5+2*\i,2.5+2*\i) -- (3+2*\i,3+2*\i);
\draw [thick] (-2-2*\i, -2 -2*\i) -- (-2.5 - 2*\i, -1.5 - 2*\i) -- (-2.5-2*\i,2.5+2*\i) -- (-2-2*\i,3+2*\i);
\draw [thick] (1+2*\i, 2+2*\i) -- (2+2*\i,3+2*\i);
\draw [thick] (1+2*\i, -1-2*\i) -- (2+2*\i,-2-2*\i);
\draw [thick] (-2*\i, 2+2*\i) -- (-1-2*\i,3+2*\i);
\draw [thick] (-2*\i, -1-2*\i) -- (-1-2*\i,-2-2*\i);
}

\foreach \i in {0,...,2}
{
\draw [thick] (2 + 2*\i, 4 +2*\i) -- (1+2*\i,5+2*\i);
\draw [thick] (-1 - 2*\i, 4 +2*\i) -- (-2*\i,5+2*\i);
\draw [thick] (2+2*\i, -3 -2*\i) -- (1+2*\i,-4-2*\i);
\draw [thick] (-1-2*\i, -3 -2*\i) -- (-2*\i,-4-2*\i); 
}

\foreach \i in {0,...,1}
{
\draw [thick] (1 + 2*\i, 6 +2*\i) -- (2+2*\i,7+2*\i);
\draw [thick] ( -2*\i, 6 +2*\i) -- (-1-2*\i,7+2*\i);
\draw [thick] (1 + 2*\i, -5 -2*\i) -- (2+2*\i,-6-2*\i);
\draw [thick] ( -2*\i, -5 -2*\i) -- (-1-2*\i,-6-2*\i);
}

\draw [thick] (2,8) -- (1,9);
\draw [thick] (-1,8) -- (0,9); 
\draw [thick] (2,-7) -- (1,-8);
\draw [thick] (-1,-7) -- (0,-8); 

\foreach \i in {0,...,4}
{
\draw [thick] (-8+4*\i,10) -- (-8.5+4*\i,10.5);
\draw [thick] (-7+4*\i,10) -- (-6.5+4*\i,10.5);
\draw [thick] (-8+4*\i,-9) -- (-8.5+4*\i,-9.5);
\draw [thick] (-7+4*\i,-9) -- (-6.5+4*\i,-9.5);
}

\draw [thick] (0,0) to[out =110, in = -110] (0,1);
\draw [thick] (1,0) to[out=70, in=-70] (1,1); 

\draw[thick, fill=black] (-0.1,0.5) circle (0.1cm);
\draw[thick, fill=black] (1.1,0.5) circle (0.1cm);
\Text[x=-0.8,y=0.5]{$a_i$}
\Text[x=1.8,y=0.5]{$a_j$}

\draw[thick, fill=black] (9.5,10.5) circle (0.1cm);
\draw[thick, fill=black] (-8.5,10.5) circle (0.1cm);
\Text[x=9.5,y=11.2]{$a_k$}
\Text[x=-8.5,y=11.2]{$a_l$}

\draw [thick,fill=purple,rounded corners=2.2pt] (0,1) rectangle (1,2);
\draw [thick,fill=teal,rounded corners=2.2pt] (0,-1) rectangle (1,0);

\foreach \i in {0,...,3}
{
\draw [thick] (2+2*\i,4+2*\i) -- (3+2*\i,4+2*\i);
\draw [thick] (2+2*\i,3+2*\i) -- (3+2*\i,3+2*\i);
\draw [thick] (2+2*\i,-2-2*\i) -- (3+2*\i,-2-2*\i);
\draw [thick] (2+2*\i,-3-2*\i) -- (3+2*\i,-3-2*\i);
\draw [thick] (-1-2*\i,4+2*\i) -- (-2-2*\i,4+2*\i);
\draw [thick] (-1-2*\i,3+2*\i) -- (-2-2*\i,3+2*\i);
\draw [thick] (-1-2*\i,-2-2*\i) -- (-2-2*\i,-2-2*\i);
\draw [thick] (-1-2*\i,-3-2*\i) -- (-2-2*\i,-3-2*\i);
}
\etp  . 
\nonumber 
\end{equation} 

\section{Integration with respect to Haar measure \label{app:haarid}}
Here we prove the identity (\ref{eq:Haar}) used in the Section (\ref{ref:genavg}) of the main text. Consider 
\begin{equation}
\begin{split}
& \int_{\text{Haar}}  du \tr \left[ X (u\otimes u^*) Y (u^{\dagger} \otimes u^T)\right]   \\ & =  \int_{\text{Haar}}  du \sum_{i\alpha j\beta k\gamma m\tau} \mel{i\alpha}{X}{j\beta} \mel{j\beta}{u\otimes u^*}{k\gamma} \\ & \mel{k\gamma}{Y}{m\tau} \mel{m\tau}{u^\dagger \otimes u^T}{i\alpha}   \\
&  =   \sum_{i\alpha j\beta k\gamma m\tau} \mel{i\alpha}{X}{j\beta}\mel{k\gamma}{Y}{m\tau} \int_{\text{Haar}} du \; u_{jk} u^*_{\beta \gamma} u^*_{im} u_{\alpha \tau}  
\end{split}
\label{eq:appid1}
\end{equation}
Using the results from \cite{collins2006integration,hiai2000semicircle,Pucha_a_2017} for calculating the monomial integrals on the unitary group with respect to Haar measure, above integral can be evaluated as 
\begin{equation}
\begin{split}
& \int_{\text{Haar}} du \; u_{jk} u^*_{\beta \gamma}  u_{\alpha \tau}u^*_{im} \\
& =  \frac{1}{q^2 - 1} \left( \delta_{j\beta} \delta_{k\gamma} \delta_{\alpha i} \delta_{\tau m} + \delta_{ji} \delta_{km} \delta_{\beta \alpha } \delta_{\gamma \tau } \right)  \\ 
& - \frac{1}{q(q^2-1)} \left( \delta_{j\beta}\delta_{km} \delta_{\alpha i } \delta_{ \tau \gamma } + \delta_{ji} \delta_{k\gamma} \delta_{\beta \alpha } \delta_{\tau m} \right).    
\end{split}
\label{eq:appid2}
\end{equation}

By substituting Eq. (\ref{eq:appid2}) into Eq. (\ref{eq:appid1}), we get the following identity:
\beq 
\begin{split}
&\int_{\text{Haar}} du \tr \left[ X (u \otimes u^*) Y (u^{\dagger} \otimes u^T) \right]=\\
&\frac{1}{q^2-1}\left(\tr X^{R_2} \tr Y^{R_2}+\tr X \tr Y\right) \\
&-\frac{1}{q(q^2-1)}\left( \tr X^{R_2} \tr Y +\tr X \tr Y^{R_2}\right).
\end{split}
\eeq

\section{Unistochastic matrices and CPTP map $M_+[U^\prime]$ \label{app:unistoc}} 

The nontrivial largest and smallest magnitude eigenvalues $\lambda_1$ and $\lambda_{q^2-1}$ of $M_+[U^\prime]$ is plotted in the Fig.~(\ref{fig:EvalsM_DS}), where $U^\prime = (u_1 \otimes u_2) U (v_1 \otimes v_2)$. The following dual-unitary $U$ with $e_p(U) = 3/4$ is used to generate $\lambda_1$ and $\lambda_{q^2-1}$ of $M_+[U^\prime]$ as shown in the lower right figure in Fig.~(\ref{fig:EvalsM_DS}). This unitary is obtained by iterative application of the nonlinear map $\mathcal{M}_R (U)$ to a permutation operator. Note the elegant form the dual-unitary operator obtained from a numerical algorithm, and is given as : 
\begin{equation}
U=\begin{pmatrix}
0 & 0 & 0 & 0 & 1 & 0 & 0 & 0 & 0 \\
0 & 1 & 0 & 0 & 0 & 0 & 0 & 0 & 0 \\
0 & 0 & 0 & 0 & 0 & 0 & 0 & 1 & 0 \\
0 & 0 & 0 & 0 & 0 & 0 & 0 & 0 & 1 \\
-\frac{1}{2} & 0 & \frac{1}{2} & \frac{1}{2} & 0 & -\frac{1}{2} & 0 & 0 & 0 \\
-\frac{1}{2} & 0 & \frac{1}{2} & -\frac{1}{2} & 0 & \frac{1}{2} & 0 & 0 & 0 \\
0 & 0 & 0 & 0 & 0 & 0 & 1 & 0 & 0 \\
-\frac{1}{2} & 0 & -\frac{1}{2} & -\frac{1}{2} & 0 & -\frac{1}{2} & 0 & 0 & 0 \\
-\frac{1}{2} & 0 & -\frac{1}{2} & \frac{1}{2} & 0 & \frac{1}{2} & 0 & 0 & 0 
\end{pmatrix}
\end{equation} 

The structure of the real and imaginary parts of the nontrivial largest eigenvalue $\lambda_1$ of $M_+[U_i^\prime]$ in Fig.~(\ref{fig:permlamb1}) deserves an explanatory note. First note that the dual-unitary operator $U_1 = D_3S $ and $U_2 = D_2S$ are having same entangling power $e_p(U_i) = 3/4$. The local unitary inequivalence between $U_1$ and $U_2$ is evident from the different  $\lambda_1$ distribution  seen in the Fig.~(\ref{fig:permlamb1}). The rank of $M_+[U_1^\prime]$ is $3$ and $M_+[U_2^\prime]$ is full.    The block diagonal unitary operators used to construct these dual-unitary operators are,   
\begin{equation}
D_3 = \begin{pmatrix}[ccc|ccc|ccc]
1 & 0 & 0 & 0 & 0 & 0 & 0 & 0 & 0 \\
0 & 1 & 0 & 0 & 0 & 0 & 0 & 0 & 0 \\ 
0 & 0 & 1 & 0 & 0 & 0 & 0 & 0 & 0 \\ 
\hline 
0 & 0 & 0 & 0 & 0 & 1 & 0 & 0 & 0 \\
0 & 0 & 0 & 1 & 0 & 0 & 0 & 0 & 0 \\
0 & 0 & 0 & 0 & 1 & 0 & 0 & 0 & 0 \\
\hline 
0 & 0 & 0 & 0 & 0 & 0 & 0 & 1 & 0 \\
0 & 0 & 0 & 0 & 0 & 0 & 0 & 0 & 1 \\
0 & 0 & 0 & 0 & 0 & 0 & 1 & 0 & 0 
\end{pmatrix}, 
\end{equation} 
\begin{equation} 
D_2=\begin{pmatrix}[cccccc|ccc]
1 & 0 & 0 & 0 & 0 & 0 & 0 & 0 & 0 \\
0 & 1 & 0 & 0 & 0 & 0 & 0 & 0 & 0 \\
0 & 0 & 0 & 0 & 0 & 1 & 0 & 0 & 0 \\
0 & 0 & 0 & 0 & 1 & 0 & 0 & 0 & 0 \\
0 & 0 & 0 & 1 & 0 & 0 & 0 & 0 & 0 \\
0 & 0 & 1 & 0 & 0 & 0 & 0 & 0 & 0 \\
\hline
0 & 0 & 0 & 0 & 0 & 0 & 1 & 0 & 0 \\
0 & 0 & 0 & 0 & 0 & 0 & 0 & \cos(a \pi) &  \sin(a \pi) \\
0 & 0 & 0 & 0 & 0 & 0 & 0 & - \sin(a \pi) & \cos(a \pi) 
\end{pmatrix},
\end{equation} 
where $a \approx 0.315167$. 

The geometric structure in the  Fig.~(\ref{fig:permlamb1}) corresponding to the statistics of $\lambda_1$ of $M_+[U_1^\prime]$ is called deltoid, a 3-hypocycloid. Here we briefly explain the reason behind this structure as these structures were also reported in Ref.~(\cite{zyczkowski2003random}) in the context of random unistochastic matrices.  
 By using Eq.~(\ref{eq:M+UT}), the operator  $M_+[U_1]$is given by its matrix elements as 
\begin{equation}
\label{eq:mpuni}
\mel{i\alpha}{M_+[U_1]}{j\beta} = \delta_{ij} \delta_{\alpha\beta} \delta_{i\alpha}.
\end{equation} 
 With $U_1^\prime = (u_1 \otimes u_2 ) U_1 (v_1 \otimes v_2)$, $M_+[U_1^\prime]=   
(v_2^\dagger \otimes v_2^T) M_+[U_1] (u_1^\dagger \otimes u_1^T)$ (see Eq.~(\ref{eq:Mprime}). The spectrum  $\text{Spec} ( M_+[U_1^\prime])=   
\text{Spec} ((v_2^\dagger \otimes v_2^T) M_+[U_1] (u_1^\dagger \otimes u_1^T)) = \text{Spec}  ((u_1^\dagger v_2^\dagger \otimes u_1^T v_2^T) M_+[U_1]) $, by identifying $u = u_1^\dagger v_2^\dagger$, it is sufficient to study the spectrum of  $M_+[U_1^\prime] = (u \otimes u^*)  M_+[U_1]$. By using Eq.~(\ref{eq:mpuni}) for $M_+[U_1]$, 
\begin{equation}
 \begin{split}
\mel{i\alpha}{M_+[U_1^\prime]}{j\beta} & =  \mel{i\alpha}{u \otimes u^*}{j\beta} \delta_{j\beta}\\ & = \mel{i}{u}{j} \mel{\alpha}{u^*}{\beta} \delta_{j\beta}  
 \end{split}. 
\end{equation}
 Here $M_+[U_1] \ket{j\beta} = \delta_{j\beta} \ket{j\beta}$ is used. The subspace contributing the three eigenvalues are spanned by the vectors $\{\ket{ii}\}$. The corresponding matrix elements are $\mel{ii}{M_+[U_1^\prime]}{jj} = \mel{i}{u}{j}\mel{i}{u^*}{j} = \abs{u_{ij}}^2$. Thus the eigenvalue of  $M_+[U_1^\prime]$ is same as that of the unistochastic matrix $B = \abs{u_{ij}}^2$ of dimension $3 \times 3 $ generated from the local unitary operators $u = u_{ij}$.  In ref.~(\cite{zyczkowski2003random}), the authors constructed an ensemble of unistochastic matrices of dimension $3 \times 3$ starting from the nontrivial permutation operators  and showed that the spectrum is bounded by a 3-hypocycloid. They conjecture that the spectrum of any random unistochastic matrices of dimension $3 \times 3$  forms a  3-hypocycloid. They also construct random unistochastic matrices for the dimension $N \times N$ and show that the spectrum forms $N$-hypocycloid for such constructions. But in general the spectrum may deviate for $N >3$. 

\begin{figure}[h]
 \includegraphics[scale =0.6]{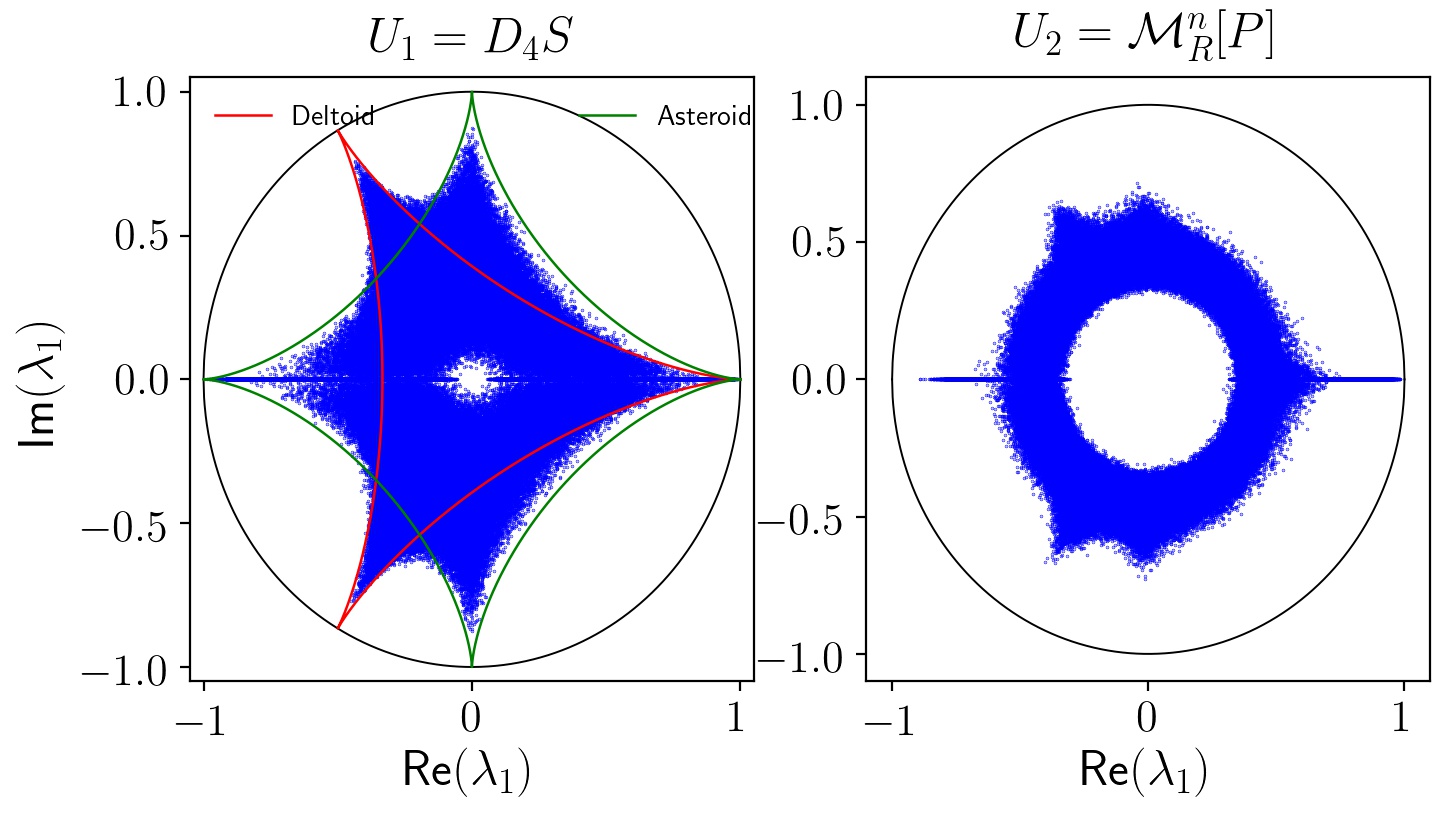}
 \caption{The real and imaginary parts of the nontrivial largest eigenvalue $\lambda_1$  of $M_+[U_i]$ for two dual-unitary $U_i$ with $e_p(U_i) = 4/5$. A finite fraction of spectrum of $M_+[U_i]$ is deviated from the 4-hypocycloid. The matrix $U_2$ converges from iterating the map $\mathcal{M}$ for a large number times $n$ to a particular permutation matrix $P$.} 
 \label{fig:hypo4}
\end{figure}

 The dual-unitary $U_1 = D_4S$ and $U_2$ with $e_p(U_i) = 4/5$ are given below. Here $U_2$ is obtained from the iterative application of $\mathcal{M}_R$ map to a permutation operator. Again note the elegant form of the dual-unitary obtained from the $\mathcal{M}_R$ map. $U_1$ and $U_2$ are not local unitarily equivalent and the rank of $M_+[U_1^\prime]$ is $4$ and $M_+[U_2^\prime]$ is of full rank. The distribution of real and imaginary values of nontrivial largest eigenvalue $\lambda_1$ of $M_+[U_i]$ is plotted in Fig.(\ref{fig:hypo4}). There is a small fraction of eigenvalues that deviates from the 4-hypocycloid. The single-particle  averages  $\mathbb{E}_{u_i,v_i}(|\lambda_1|) = 0.46170$ for $U_1$ and $\mathbb{E}_{u_i,v_i}(|\lambda_1|) = 0.4940$ for $U_2$. 

\begin{equation}
 D_4=\begin{pmatrix}[cccc|cccc|cccc|cccc]
1 & 0 & 0 & 0 & 0 & 0 & 0 & 0 & 0 & 0 & 0 & 0 & 0 & 0 & 0 & 0 \\
0 & 1 & 0 & 0 & 0 & 0 & 0 & 0 & 0 & 0 & 0 & 0 & 0 & 0 & 0 & 0 \\
0 & 0 & 1 & 0 & 0 & 0 & 0 & 0 & 0 & 0 & 0 & 0 & 0 & 0 & 0 & 0 \\
0 & 0 & 0 & 1 & 0 & 0 & 0 & 0 & 0 & 0 & 0 & 0 & 0 & 0 & 0 & 0 \\
\hline
0 & 0 & 0 & 0 & 0 & 0 & 0 & 1 & 0 & 0 & 0 & 0 & 0 & 0 & 0 & 0 \\
0 & 0 & 0 & 0 & 1 & 0 & 0 & 0 & 0 & 0 & 0 & 0 & 0 & 0 & 0 & 0 \\
0 & 0 & 0 & 0 & 0 & 1 & 0 & 0 & 0 & 0 & 0 & 0 & 0 & 0 & 0 & 0 \\
0 & 0 & 0 & 0 & 0 & 0 & 1 & 0 & 0 & 0 & 0 & 0 & 0 & 0 & 0 & 0 \\
\hline
0 & 0 & 0 & 0 & 0 & 0 & 0 & 0 & 0 & 0 & 1 & 0 & 0 & 0 & 0 & 0 \\
0 & 0 & 0 & 0 & 0 & 0 & 0 & 0 & 0 & 0 & 0 & 1 & 0 & 0 & 0 & 0 \\
0 & 0 & 0 & 0 & 0 & 0 & 0 & 0 & 1 & 0 & 0 & 0 & 0 & 0 & 0 & 0 \\
0 & 0 & 0 & 0 & 0 & 0 & 0 & 0 & 0 & 1 & 0 & 0 & 0 & 0 & 0 & 0 \\
\hline
0 & 0 & 0 & 0 & 0 & 0 & 0 & 0 & 0 & 0 & 0 & 0 & 0 & 1 & 0 & 0 \\
0 & 0 & 0 & 0 & 0 & 0 & 0 & 0 & 0 & 0 & 0 & 0 & 0 & 0 & 1 & 0 \\
0 & 0 & 0 & 0 & 0 & 0 & 0 & 0 & 0 & 0 & 0 & 0 & 0 & 0 & 0 & 1 \\
0 & 0 & 0 & 0 & 0 & 0 & 0 & 0 & 0 & 0 & 0 & 0 & 1 & 0 & 0 & 0 \\
\end{pmatrix}
\end{equation}

\begin{equation} 
U_2=\begin{pmatrix}
\begin{array}{cccccccccccccccc}
1 & 0 & 0 & 0 & 0 & 0 & 0 & 0 & 0 & 0 & 0 & 0 & 0 & 0 & 0 & 0 \\
0 & 0 & 0 & 0 & 1 & 0 & 0 & 0 & 0 & 0 & 0 & 0 & 0 & 0 & 0 & 0 \\
0 & 0 & 0 & 0 & 0 & 0 & 0 & 0 & 1 & 0 & 0 & 0 & 0 & 0 & 0 & 0 \\
0 & 0 & 0 & 0 & 0 & 0 & 0 & 0 & 0 & 0 & 0 & 0 & 1 & 0 & 0 & 0 \\
0 & 0 & 0 & 0 & 0 & 0 & 0 & 0 & 0 & \frac{1}{\sqrt{2}} & 0 & 0 & 0 & \frac{1}{\sqrt{2}} & 0 & 0 \\
0 & 1 & 0 & 0 & 0 & 0 & 0 & 0 & 0 & 0 & 0 & 0 & 0 & 0 & 0 & 0 \\
0 & 0 & 0 & 0 & 0 & 1 & 0 & 0 & 0 & 0 & 0 & 0 & 0 & 0 & 0 & 0 \\
0 & 0 & 0 & 0 & 0 & 0 & 0 & 0 & 0 & 0 & -\frac{1}{\sqrt{2}} & 0 & 0 & 0 & \frac{1}{\sqrt{2}} & 0 \\
0 & 0 & 0 & 0 & 0 & 0 & \frac{1}{\sqrt{2}} & \frac{1}{\sqrt{2}} & 0 & 0 & 0 & 0 & 0 & 0 & 0 & 0 \\
0 & 0 & \frac{1}{\sqrt{2}} & \frac{1}{\sqrt{2}} & 0 & 0 & 0 & 0 & 0 & 0 & 0 & 0 & 0 & 0 & 0 & 0 \\
0 & 0 & 0 & 0 & 0 & 0 & 0 & 0 & 0 & 0 & \frac{1}{\sqrt{2}} & 0 & 0 & 0 & \frac{1}{\sqrt{2}} & 0 \\
0 & 0 & 0 & 0 & 0 & 0 & 0 & 0 & 0 & \frac{1}{\sqrt{2}} & 0 & 0 & 0 & -\frac{1}{\sqrt{2}} & 0 & 0 \\
0 & 0 & -\frac{1}{\sqrt{2}} & \frac{1}{\sqrt{2}} & 0 & 0 & 0 & 0 & 0 & 0 & 0 & 0 & 0 & 0 & 0 & 0 \\
0 & 0 & 0 & 0 & 0 & 0 & \frac{1}{\sqrt{2}} & -\frac{1}{\sqrt{2}} & 0 & 0 & 0 & 0 & 0 & 0 & 0 & 0 \\
0 & 0 & 0 & 0 & 0 & 0 & 0 & 0 & 0 & 0 & 0 & 1 & 0 & 0 & 0 & 0 \\
0 & 0 & 0 & 0 & 0 & 0 & 0 & 0 & 0 & 0 & 0 & 0 & 0 & 0 & 0 & 1 \\
\end{array}
\end{pmatrix}
\end{equation}

\section{Maximum entangling power of block diagonal unitaries \label{app:BlockMaxEp}}
The bound in Eq.~(\ref{eq:maxEPblockD}) is attained when the blocks $U_j$ of Eq.~(\ref{eq:DqK}) are not only such that their partial transposes are
unitary, but that their realignments are \emph{also} unitary; that is they are 2-unitaries in the bipartite $\mathcal{H}^{m_j} \otimes \mathcal{H}^q$ space.
As $m_j<q$ these spaces are not symmetric under exchange and the conditions of 2-unitarity are modified accordingly \cite{USinpreparation}.
They are
\beq
\label{eq:2uni_conditions}
U_j^{T_2} U_j^{T_2\dagger}=\mathbb{1}_{q^2}, \; U_j^{R_2} U_j^{R_2\dagger}=\frac{q}{m_j}\mathbb{1}_{m_j^2},
\eeq
while $U_j^{R_1} U_j^{R_1\dagger}$ is not proportional to identity. The operator Schmidt decomposition of $U_j$ is in general
\beq
U_j =\sum_{k=1}^{m_j^2} \sqrt{\tilde{\lambda}_{j,k}} \;a_{j,k} \otimes b_{j,k},
\eeq
where $\{a_{j,k}\}$ and $\{b_{j,k}\}$ are orthonormal operator bases in $\mathcal{H}^{m_j}$ and $\mathcal{H}^q$ respectively, $\tr(a_{j,k}a_{j,l}^{\dagger})=\delta_{lk}$ and $\tr(b_{j,k}b_{j,l}^{\dagger})=\delta_{lk}$. The Schmidt values satisfy $\sum_{k=1}^{m_j^2} \tilde{\lambda}_{j,k}=m_j q$, and $U_j$ is maximally entangled if
$\tilde{\lambda}_{j,k}=q/m_j$ for all $k$, from which the second relation in Eq.~(\ref{eq:2uni_conditions}) follows. If all the $U_j$ are
maximally entangled,
\beq
(D_{q_1,\cdots,q_K})^{R_2}=\bigoplus_{j=1}^{K} \sqrt{\frac{q}{m_j}} \sum_{k=1}^{m_j^2} |a_{j,k}\kt \br b^*_{j,k}|,
\eeq
as $(A\otimes B)^{R_2}=|A\kt \br B^*|$. It follows that 
\beq
(D_{q_1,\cdots,q_K})^{R_2}(D_{q_1,\cdots,q_K})^{R_2 \dagger}=\bigoplus_{j=1}^{K} \frac{q}{m_j} \sum_{k=1}^{m_j^2} |a_{j,k}\kt \br a_{j,k}|, 
\eeq
as $|a_{j,k}\kt \br a_{j',l}|=0$ for $j \neq  j'$, their supports not overlapping. The singular values of $(D_{q_1,\cdots,q_K})^{R_2}$ are therefore $\sqrt{q/m_j}$ with a $m_j^2$ fold degeneracy. Using Eq.~(\ref{eq:EUdefn}) we get
\beq
E(D_{q_1,\cdots,q_K})=1-\frac{1}{q^4}\sum_{j=1}^K \frac{q^2}{m_j^2} m_j^2=1-\frac{K}{q^2}. 
\eeq
Hence as $E(SU)=E(D_{q_1,\cdots,q_K})$, the bound in Eq.~(\ref{eq:maxEPblockD}) arises.

\section{Qubit case analytics \label{app:qubits}} 

\subsection{For the  locals of the form $w$ \label{app:qubitw}}

\begin{figure}[h]
 \includegraphics[scale=0.65]{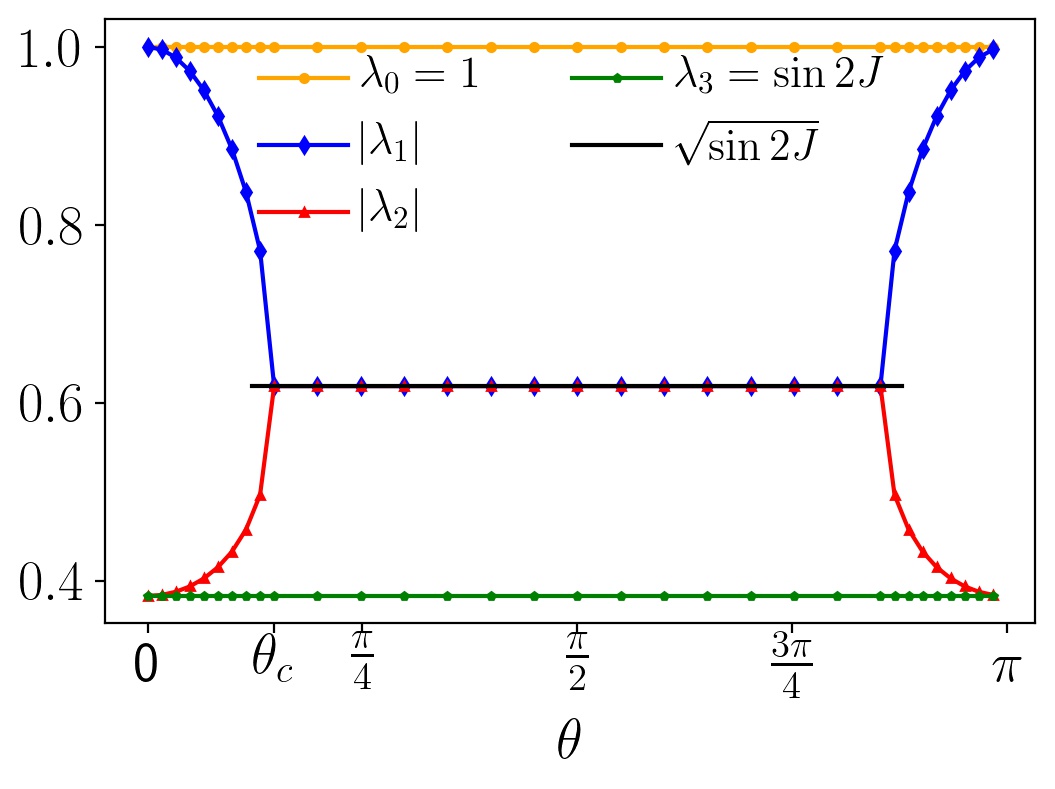}
 \caption{The ordered eigenvalues $\abs{\lambda_i}$ of the qubit channel $M_+[U'(J)]$ as a function of $\theta$ for $J=\pi/16$, when the local single particle
 gates are of the restricted form $w$.}
 \label{fig:eigvalMJprime}
\end{figure}

The eigenvalues of $M[U'(J)]=(w^{\dagger}\otimes w^T)M[U(J)]$ are found to be still analytically simple and are (apart from the trivial eigenvalue $\lambda_0'=1$)
\beq
\begin{split}
\lambda'_{1,2} & =\frac{1}{2}[(1+\sin 2J) \cos \theta \\
		  &\pm \sqrt{(1+\sin 2J)^2 \cos^{2}\theta-4 \sin 2J}]\\
\lambda'_3&=\sin 2J. 
\end{split}
\eeq
Note that these eigenvalues are independent of $\psi$, while $\lambda_3'=\sin 2J$ is always real and independent of $\theta$ as well.
The ordered eigenvalues are denoted $\lambda_i$ as above and $|\lambda_1|$ can change from $|\lambda_1'|$ to $|\lambda_2'|$.  
Figure~(\ref{fig:eigvalMJprime}) shows $\abs{\lambda_i}$ as a function of $\theta$ for a particular value of $J=\pi/16$. 

As $\theta$ increases from $0$, $\lambda'_{1,2}$ are real and $\abs{\lambda'_1}=|\lambda_1|$ decreases while $\abs{\lambda'_2}=|\lambda_2|$ increases till these values become equal at $\theta= \theta_c=\arccos[2 \sqrt{\sin 2J}/(1+\sin 2J)]$ when these eigenvalues turn complex and form a conjugate pair whose modulus $\sqrt{\sin 2J}$ remains constant till $\pi -\theta_c$ when they become real again. Thereafter $|\lambda_1|=|\lambda'_2|$ and $|\lambda_2|=|\lambda'_1|$.
Thus the smallest value of $|\lambda_1|$ for a fixed $J$ for various local $w$ is $\sqrt{\sin 2 J}$ and hence
\beq
\nu_+^{\prime}[U]=-\frac{1}{2}\ln(\sin 2J)  =  -\frac{1}{4} \ln \left(1-\frac{e_p(U)}{e_p^\text{max}}\right). 
%\label{eq:nuplusep}
\eeq

\subsection{For the  locals of the form $v$ \label{app:qubitv}}

The nontrivial eigenvalues of $M[U'(J)]$ are found to be 
roots of the cubic equation
\beq
\lambda^3-(\lambda^2- \lambda)\cos \phi\sin 2J -\sin^2 2J =0.
\label{eq:characeqnrfixed}
\eeq
Despite the restriction, explicit forms of the roots $\lambda_{1,2,3}^\prime$, 
and their ordered versions, are cumbersome.
They satisfy the following relations,
\begin{multline}
\label{eq:lambprime}
\lambda'_1 \lambda'_2  \lambda'_3 =  \sin^2 2J,
\lambda'_1 +\lambda'_2 +\lambda'_3   =  \cos \phi \sin 2J,\\
\lambda'_1 \lambda'_2 +\lambda'_2\lambda'_3 + \lambda'_1\lambda'_3  = \cos \phi \sin 2J ,
\end{multline}
%\begin{subequations}
%\label{eq:lambprime}
%\begin{align}
%\lambda_1^{'} \lambda_2^{'}  \lambda_3^{'} & =  \sin^2(2J),\\
%\lambda_1^{'} +\lambda_2^{'} +\lambda_3^{'} &  =  \cos(\phi)\sin(2J),\\
%\lambda_1^{'} \lambda_2^{'} +\lambda_2^{'}\lambda_3^{'} + \lambda_1^{'} \lambda_3^{'} & = \cos(\phi)\sin(2J),
%\end{align}
%\end{subequations}
and are independent of $\psi$. As the roots depend on $\cos \phi$ it is sufficient to restrict attention to the interval $\phi \in [0, \pi]$.
\begin{figure}[h]
 \includegraphics[scale=0.65]{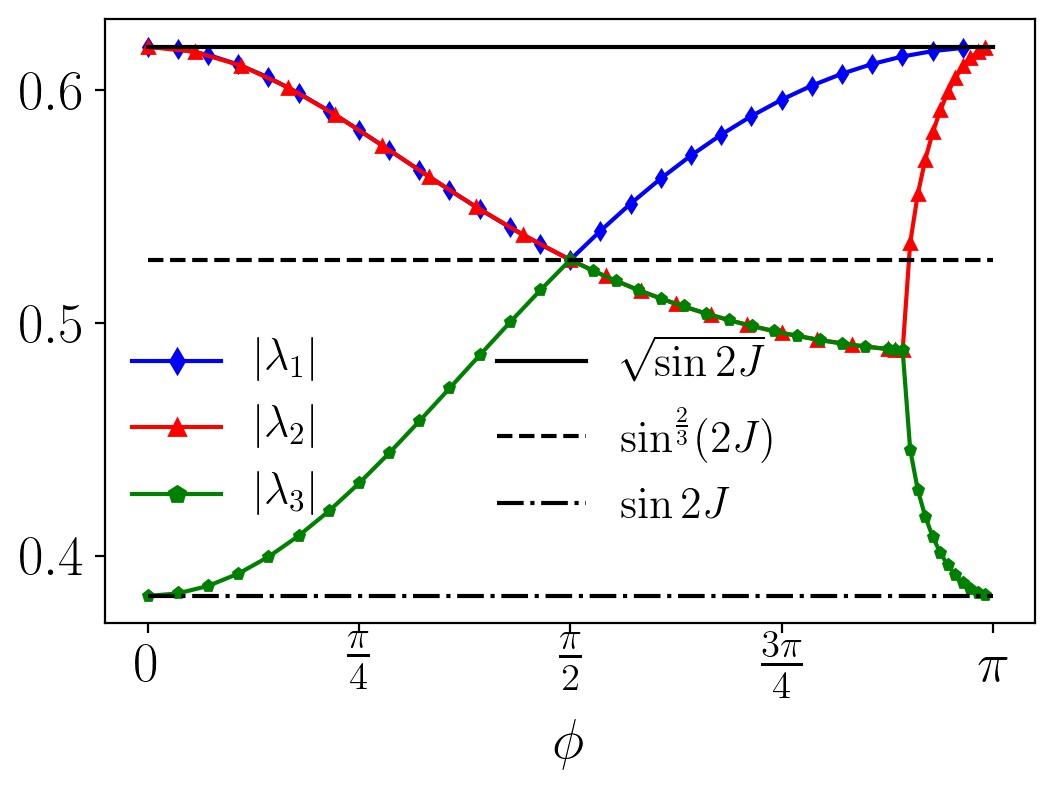}
 \caption{The ordered nontrivial eigenvalues $\abs{\lambda_i},i=1,2,3$ of $M_+[U'(J)]$ as a function of $\phi$ for $J=\pi/16$, when the locals are drawn from the restriction of local unitaries to the form $v$ in Eq.~(\ref{eq:localfixedr}). The smallest value of $|\lambda_1|$ is $(\sin 2J)^{2/3}$ at $\phi=\pi/2$ for all values of $J$.}
 \label{fig:eigvalrfixed}
\end{figure}

The absolute values of $\lambda_i^{\prime}$ are shown in Fig.~(\ref{fig:eigvalrfixed}) as a function of $\phi$ for a representative value of $J=\pi/16$. 
The phenomenology of the roots shown is independent of the value of $J$ used. For $\phi=0$, it is easy to see that $\lambda_{1,2}^{\prime}=\pm i \sqrt{\sin 2J}$, while the real root is $\lambda_3^{\prime}=\sin 2J$. At $\phi=\pi$, all three roots are real: $\lambda_3'=\sqrt{\sin 2 J}$ while $\lambda'_1=-\sqrt{\sin 2 J}$
and $\lambda_2'=-\sin 2J$. The root with the largest magnitude, $|\lambda_1|$, 
decreases from $|\lambda'_1|=|\lambda'_2|=\sqrt{\sin 2 J}$ at $\phi=0$ and then increases again to the same value at $\phi=\pi$. We are interested in its smallest value, and it turns out that this is obtained when $\phi=\pi/2$, {\it independent} of the value of $J$ used.

The root that is always real, $\lambda_3'$, increases from $\sin 2 J$ to $\sqrt{\sin 2J}$ monotonically for $\phi \in [0, \pi]$.
This follows from the non-negativity of, 
\beq
\dfrac{d \lambda_3'}{d \phi}= \left[ \frac{\lambda'_3(1-\lambda_3')  \sin \phi \sin 2J }
{3 \lambda_3 ^{\prime 2} + (1-2\lambda_3')\cos \phi \sin 2J}\right].
\label{eq:difflamb3phi}
\eeq
It is straightforward to see that this derivative is positive for $0<\phi\leq \pi/2$, and remains so for $\pi/2<\phi<\pi$
as $|\lambda_3'|<1$ \cite{Bertini2019}. Consider the case $\phi \in [0, \pi/2]$. There are two complex roots, say $\lambda'_1=\lambda'^*_2$, and 
$|\lambda'_1|$ decreases monotonically as from Eq.~(\ref{eq:lambprime}): $|\lambda_1'|^2 \lambda_3'=\sin^2 2J$.
Furthermore $|\lambda_1^{\prime}|^{2}-\lambda_3^{\prime 2} = \cos(\phi)\sin(2J)(1-\lambda_3^{\prime}) \geq 0$, the
equality is satisfied for $\phi=\pi/2$ for all $J$. As $\lambda_3'$ continues to increase, it will replace $|\lambda_1'|$
as $|\lambda_1|$ beyond $\phi=\pi/2$. Thus the minimum value of $|\lambda_1|$ is identified from $|\lambda_1|^3=\sin^2 2J$,
that is $|\lambda_1| =\sin^{\frac{2}{3}}(2J).$
%% I have removed the equation with \label{eq:lamb1rfixed}       
%%%%
Therefore,
\beq
\nu_+[U(J)]=-\ln \sin^{\frac{2}{3}}(2J)= -\frac{1}{3}\ln\left[1-\frac{e_p(U)}{e_p^{\text{max}}}\right].
%\label{eq:nuplusfixedr}
\eeq

\subsection{General $SU(2)$ as single qubit gates}
\begin{figure}[h]
 \includegraphics[scale=0.75]{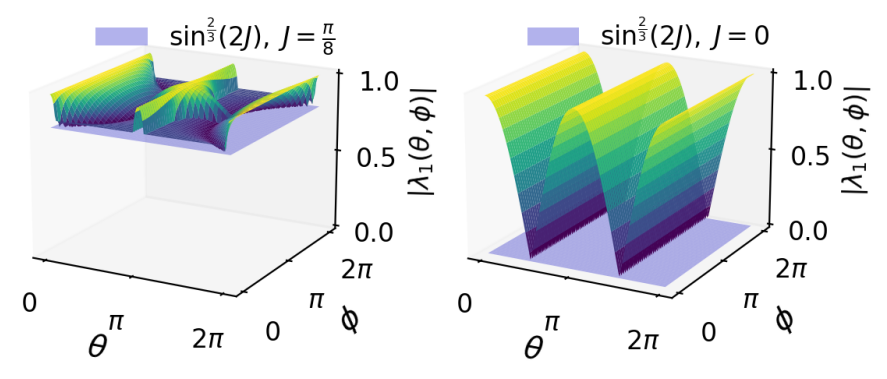}
 \caption{Absolute value of the largest non-trivial eigenvalue; $\lambda_1$, of $M[U']=(u \otimes u^*)M[J],u \in SU(2)$, is plotted as a function of $\theta, \phi$ parametrizing single qubit gate $u$. For given value of $J$, $|\lambda_1|$ is bounded below by $\sin^{\frac{2}{3}}(2J)$.}
 \label{fig:genSU2fig1}
\end{figure}

With single qubit gate $u$ of the form Eq.~(\ref{eq:genlocal}), the characteristic equation corresponding to the non-trivial eigenvalues of $M[U']=(u \otimes u^*)M[J]$ is given by,
\beq
\lambda^3+\left[1-2\cos^2\frac{\theta}{2} \left(\sin 2J \cos \phi+1 \right)\right]\lambda^2+\left[\sin 2J \left(2\cos^2\frac{\theta}{2} (\sin 2J+\cos \phi)-\sin 2J\right)\right]\lambda-\sin^2 2J=0.
\label{eq:charpolgenloc}
\eeq
This cubic equation is not easy to factorize and its roots can all be real or, a pair of complex roots with one real root depending on the values of $J,\theta,\phi$. Since getting closed form of roots is hard, we study the $\theta,\phi$ dependence of the largest root i.e., the largest non-trivial eigenvalue of $M[U']$ numerically for a given value of $J$ as shown in Fig.~(\ref{fig:genSU2fig1}). It is  seen numerically that  $|\lambda_1|$ is bounded below by $\sin^{\frac{2}{3}}(2J)$ and this minimum value is obtained for many pairs of $(\theta,\phi)$ for a given value of $J$. 

%To see it more clearly, we plot $-\log[|\lambda_1(J,\theta,\phi)|-\sin^{\frac{2}{3}}(2J)]$ as a function of $(\theta,\phi)$. Pairs of $(\theta,\phi)$ angles for which the bound is reached are seen as yellow in the colour plot Fig.~\ref{fig:modlam1su2}. For $J=0$, bound is reached at $\theta=\pi/$ for all values of $\phi$. The dependence of the minimum of $|\lambda_1(J,\theta,\phi)|$, obtained by searching over different values of $(\theta,\phi)$ pairs, on $J$ or equivalently, $e_p(U)$ is shown quantitatively for different values of $J$ in Fig.~(\ref{fig:genSU2fig2}).

\begin{figure}[h]
 \includegraphics[scale=0.4]{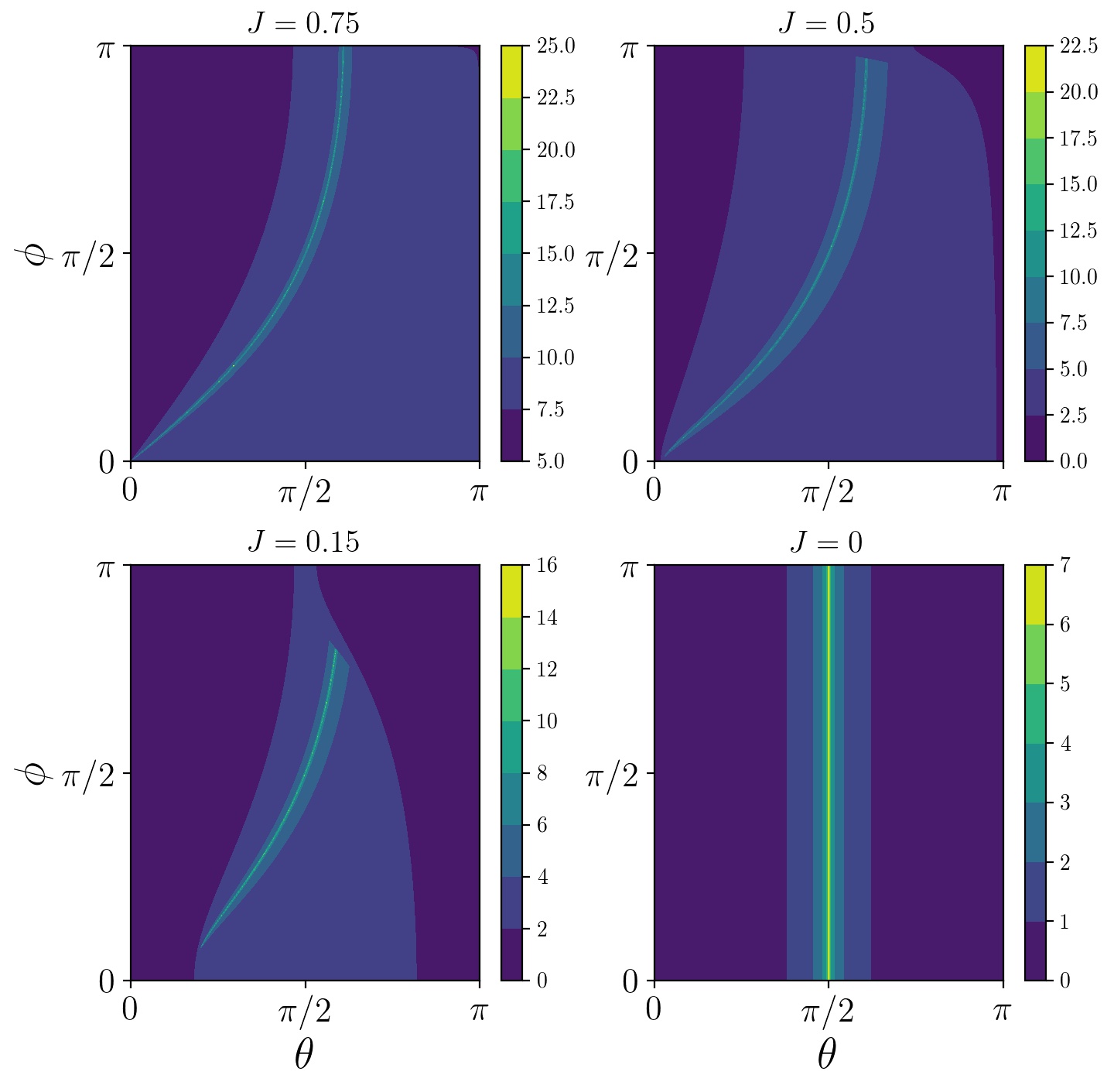}
 \caption{Function $f(J,\theta,\phi)$ is evaluated on the $(\theta,\phi)$ plane for four different values of $J$. Pairs of $(\theta,\phi)$ angles for which $f(J,\theta,\phi)$ diverges are the values at which the minimum of $|\lambda_1(J,\theta,\phi)|$ is reached.}
 \label{fig:modlam1su2}
\end{figure}

\begin{figure}[h]
 \includegraphics[scale=0.6]{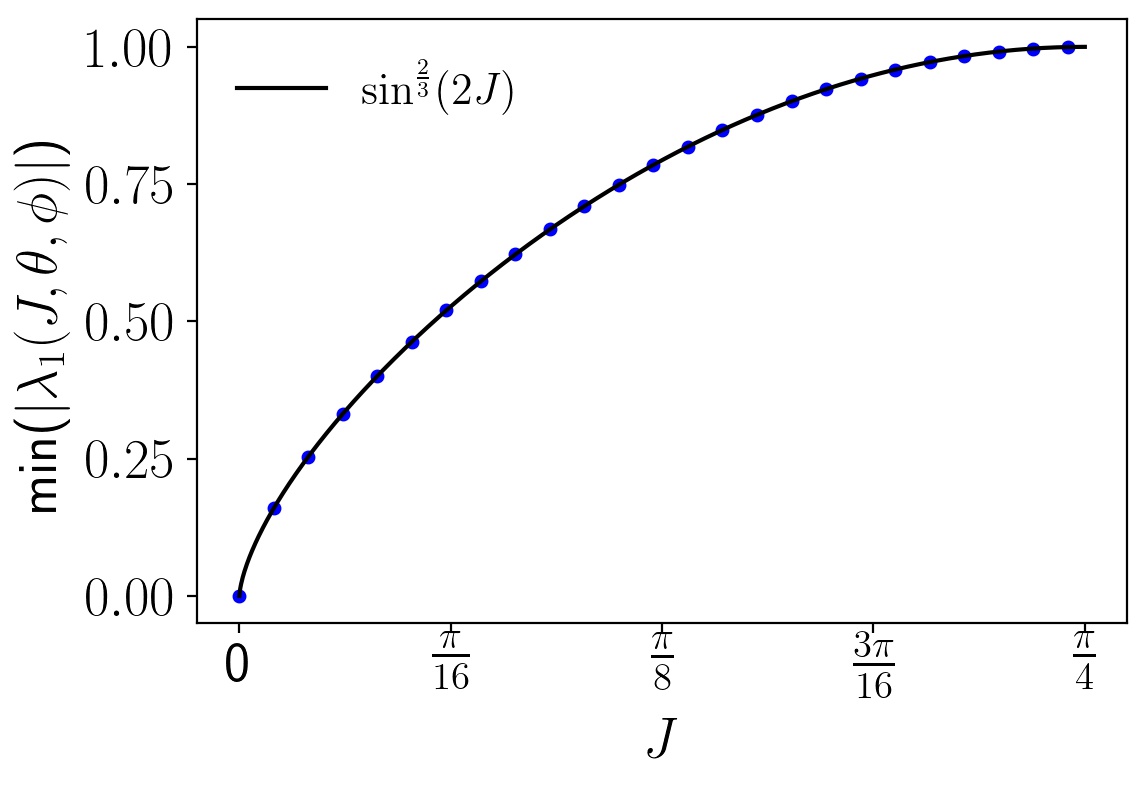}
 \caption{Minimum value of $|\lambda_1|$ (obtained numerically by searching over $\theta, \phi \in [0,2\pi]$ parametrizing single qubit gate) is plotted as a function of $J$. The minimum obtained perfectly fits the predicted theoretical value obtained using locals from a restricted set.}
 \label{fig:genSU2fig2}
\end{figure}

To see it clearly, we plot the function, 
\beq
f(J,\theta,\phi)=-\log \left| |\lambda_1(J,\theta,\phi)|-\sin^{\frac{2}{3}}(2J) \right|,
\eeq
as a function of $(\theta,\phi)$ for a fixed value of $J$. Pairs of $(\theta,\phi)$ angles for which the bound is reached are seen as lighter in the plot Fig.~\ref{fig:modlam1su2} where the function takes large values (diverges in principle). For $J=0$, bound is reached at $\theta=\pi/2$ for all values of $\phi$. The dependence of the minimum of $|\lambda_1(J,\theta,\phi)|$, obtained by searching over different values of $(\theta,\phi)$ pairs, on $J$ or equivalently, $e_p(U)$ is shown quantitatively for different values of $J$ in Fig.~(\ref{fig:genSU2fig2}).

\section{Explicit form of Dual and 2-unitaries from the $\mathcal{M_R}$ map \label{sec:appnice}} 
Here we give the explicit forms of a dual-unitary and a 2-unitary obtained  from the map $\mathcal{M}_R$ given in Eq. (\ref{eq: MRmap}). The dual-unitary for $q=3$ having entangling power $e_p(U) = 8/9 > e_p^* = 7/8$ is given as 
\begin{equation}
U = \frac{1}{2}
\begin{pmatrix}
\sqrt{3} & 0 & -1 & 0 & 0 & 0 & 0 & 0 & 0 \\
0 & 0 & 0 & 0 & 2 & 0 & 0 & 0 & 0  \\ 
0 & 0 & 0 & 0 & 0 & 0 & -\sqrt{3} & 0 & 1 \\
0 & 0 & 0 & 1 & 0 & \sqrt{3} & 0 & 0 & 0 \\
0 & 0 & 0 & 0 & 0 & 0 & 1 & 0 & \sqrt{3}\\
0 & 2 & 0 & 0 & 0 & 0 & 0 & 0 & 0\\
0 & 0 & 0 & 0 & 0 & 0 & 0 & 2 & 0 \\
0 & 0 & 0 & \sqrt{3} & 0 & -1 & 0 & 0 & 0\\
1 & 0 & \sqrt{3} & 0 & 0 & 0 & 0 & 0 & 0
\end{pmatrix},
\end{equation}
and the 2-unitary operator $e_p(U) =1$ for $q=3$ is given as 
\begin{equation}
U = \frac{1}{2}\begin{pmatrix}
\sqrt{3} & 0 & 0 & 0 & 0 & 0 & 0 & 0 & -1 \\
0 & 0 & 0 & 0 & 2 & 0 & 0 & 0 & 0\\
1 & 0 & 0 & 0 & 0 & 0 & 0 & 0 & \sqrt{3} \\
0 & -1 & 0 & 0 & 0 & \sqrt{3} & 0 & 0 & 0 \\
0 & 0 & 0 & 0 & 0 & 0 & 2 & 0 & 0 \\
0 & \sqrt{3} & 0 & 0 & 0 & 1 & 0 & 0 & 0 \\
0 & 0 & 0 & 1 & 0 & 0 & 0 & \sqrt{3} & 0 \\
0 & 0 & 2 & 0 & 0 & 0 & 0 & 0 & 0 \\
0 & 0 & 0 & -\sqrt{3} & 0 & 0 & 0 & 1 & 0
\end{pmatrix}
\label{eq:2uniq3}
\end{equation}
 This 2-unitary is different from 2-unitary operators obtained from mutually orthogonal latin squares in the following sense: considering that each row is associated with a particle, then $6$ rows with $2$ non-zeros entries each correspond to $3$ pairs of particles which are entangled, while the rest $3$ rows with only one non-zeros entry (equal to $1$) correspond to free or unentangled particles. However, under local transformations Eq.~(\ref{eq:2uniq3}) can be transformed to 2-unitary permutation obtained from mutually orthogonal latin squares \cite{Adam}.

\section{Cat map is dual for all $q$ and 2-unitary for all odd $q$ \label{app:cat}}

The unitarity of $U_C$ is straightforward to verify as 
\begin{align}
& \matrixel{k\alpha}{U_CU_C^\dagger}{j\beta}  = \sum_{m,n=0}^{q-1} \mel{k\alpha}{U_C}{mn}\mel{mn}{U_C^\dagger}{j\beta} \nonumber \\
& = \frac{1}{q^2}e^{\frac{2\pi i}{q} (j\beta - k\alpha )}\left( \sum_{m=0}^{q-1}  e^{\frac{2\pi i}{q} (\alpha - \beta )m}\right)\left( \sum_{n=0}^{q-1} e^{\frac{2\pi i}{q} (k-j )n}\right)  \nonumber \\
& =   \delta_{\alpha, \beta}  \delta_{k,j}. 
\nonumber 
\end{align}

The dual nature of $U$ can be seen as follows: 
\begin{align}
&\mel{k\alpha}{U_C^{R_1}U_C^{R_1\dagger}}{j\beta}  = \sum_{m,n=0}^{q-1} \mel{k\alpha}{U_C^{R_1}}{mn} \mel{mn}{U_C^{R_1\dagger}}{j\beta} \nonumber \\
=& \frac{1}{q^2} \left(\sum_{m=0}^{q-1} e^{\frac{2\pi i}{q} [2(j-k) + (\alpha - \beta)]m}\right)
\left(\sum_{n=0}^{q-1} e^{-\frac{2\pi i}{q} [(j-k) + (\alpha - \beta)]n}\right) \nonumber \\
=&  \delta_{\alpha, \beta}  \delta_{k,j}.
\end{align} 
To prove this, let $\sigma_1$ and $\sigma_2$ denote the $m$ and $n$  sums respectively, then for $k = j, \alpha = \beta, \sigma_1 = \sigma_2 = q$, and the diagonal elements are equal to $1$. For the off-diagonal terms such that $k=j,\, \alpha \neq \beta,\; \sigma_1 = \sigma_2 = 0$,  and similarly when $k \neq j, \, \alpha = \beta,\; \sigma_2 = 0$ $\forall q$.   Note that here we have used the fact $\sum_{m=0}^{q-1} e^{\frac{2\pi i}{q} lm} = q\delta_{l \mod q,0}$. For the other off-diagonal elements, $k \neq j, \, \alpha \neq \beta$ , the structure of elements in $\sigma_1$ and $\sigma_2$ are such that either  both $\sigma_1=\sigma_2=0$ or if $\sigma_1 \neq 0$ then $\sigma_2 = 0$ and if $\sigma_2 \neq 0$ then $\sigma_1 = 0$. This holds for all dimensions $q$, and hence the quantum cat map unitary $U$  in Eq.~(\ref{eq:cat}) is dual-unitary  in all local dimensions $q$. 

The T-dual property is restricted to only odd dimensions. This can be shown by considering the matrix elements  
\begin{align}
&\mel{k\alpha}{U_C^{T_1}U_C^{T_1\dagger}}{j\beta} =  \sum_{m,n=0}^{q-1} \mel{k\alpha}{U_C^{T_1}}{mn}\mel{mn}{U_C^{T_1 \dagger}}{j\beta} \nonumber \\
=& \frac{1}{q^2}e^{\frac{2\pi i}{q} (k\alpha - j\beta)} \left(\sum_{m=0}^{q-1} e^{\frac{2\pi i}{q} m (\beta-\alpha)}\right)\left( \sum_{n=0}^{q-1} e^{\frac{4\pi i}{q} n (j-k)} \right) \nonumber \\
=&e^{\frac{2\pi i}{q} (k\alpha - j\beta)} \delta_{\alpha,\beta} \delta_{2(j-k)\mod q,0}.
\end{align}
For $q$ odd, the second Kronecker delta product is nonzero iff $k=j$ and hence there are only diagonal entries ($=1$) in this case and T-duality follows.
If $q$  is even, there are nonzero elements ($=(-1)^{\beta})$ when $\alpha=\beta$ and $(k-j) = \pm q/2$. As the unitary cat map $U$ in Eq.~(\ref{eq:cat}) is dual for all dimensions $q$, it is 2-unitary for $q$ odd.

\end{document}